\documentclass[3p,11pt,review,preprint]{elsarticle}
 \usepackage{lineno}
\usepackage{hyperref}  
\usepackage{graphicx,subcaption}
\usepackage[labelfont=bf,figurename=Fig.,labelsep=period]{caption}
\usepackage[usenames, dvipsnames]{color}
\usepackage{bm}
\usepackage{amsmath,amssymb,amsfonts}
\usepackage{mathrsfs}
\usepackage{multirow}
\usepackage{algpseudocode}
\usepackage[inline]{enumitem}
\usepackage{soul}
\usepackage[linesnumbered,ruled,vlined]{algorithm2e}
\usepackage{bbm}
\usepackage{tabularx} % in the preamble
\begin{document}
\begin{frontmatter}



\title{Reduced-order modeling for stochastic large-scale and time-dependent problems using deep spatial and temporal convolutional autoencoders}


\author[1]{Azzedine Abdedou}
\author[1]{Azzeddine Soula\"imani\corref{cor1}}
\cortext[cor1]{Corresponding author. Tel.: +1 514 396 8977; fax: +1 514 396 8530.}
\ead{azzeddine.soulaimani@etsmtl.ca}
%\author[2]{Georges Williams Tchamen}
%-------
\address[1]{Department of Mechanical Engineering, École de Technologie Sup\'{e}rieure, 1100 Notre-Dame W., Montr\'{e}al (QC),
Canada H3C 1K3}
%\address[2]{Hydro-Qu\'{e}bec, 75 Boul. Ren\'{e}-L\'{e}vesque Ouest, Montr\'{e}al (Qc), Canada  H2Z 1A4}
%---------------

%---------------
\begin{abstract}
A non-intrusive reduced order model based on convolutional autoencoders (NIROM-CAEs) is proposed as a data-driven tool to build an efficient nonlinear reduced-order model for stochastic spatio-temporal large-scale physical problems. The method uses two 1d-convolutional autoencoders (CAEs) to reduce the spatial and temporal dimensions from a set of high-fidelity snapshots collected from the high-fidelity numerical solver. The encoded latent vectors, generated from two compression levels, are then mapped to the input parameters using a regression-based multilayer perceptron (MLP). The accuracy of the proposed approach is compared to that of the linear reduced-order technique-based artificial neural network (POD-ANN) through two benchmark tests (the one-dimensional Burgers and Stoker's solutions) and a hypothetical dam-break flow problem, with an unstructured mesh and over a complex bathymetry river. The numerical results show that the proposed nonlinear framework presents strong predictive abilities to accurately approximate the statistical moments of the outputs for complex stochastic large-scale and time-dependent problems, with low computational cost during the predictive online stage.\\  
\end{abstract}
%--------------
%--------------
\end{frontmatter}


\section{Introduction}\label{intro}
Computational science has greatly gained in efficiency due to the increase in the high-performance computational resources and the emergence of novel methods, with significant achievements in scientific and industrial areas. Most of the computational mechanics' problems are described by time-dependent and parametrized nonlinear partial differential equations.  Their discretization over fine spatial meshes and for a high number of time steps leads to the so-called high-fidelity solutions whose evaluation renders the computational techniques prohibitively expensive when dealing with multi-resolutions as in optimization and uncertainty quantification\citep{rezaeiravesh2022uncertainty, kalinina2020metamodeling}. Reduced-order modeling \citep{quarteroni2015reduced} has gained interest as a powerful methodology to reduce the cumbersome computational effort while ensuring acceptable accuracy, particularly for situations that require real-time predictions.\\

One of the most adopted approaches in ROM is the proper orthogonal decomposition (POD) \citep{berkooz1993proper, sirovich1987turbulence} which allows an approximation of the outputs using a linear combination of a limited number of basis functions \citep{chatterjee2000introduction, zokagoa2012pod}. Many non-intrusive techniques have been proposed to compute the coefficients of the POD linear approximation using the data-driven concept without any modification of the governing equations \citep{xiao2017parameterized, walton2013reduced, 12081940}. Some of these methods are based on the stochastic framework taking into account the variability stemming from the parametric domain by expressing the modal coefficients as a function of the stochastic basis functions, as in the polynomial chaos expansion method (POD-PCE) \citep{sun2021non, el2020stochastic} and the B-splines Bézier-element method (POD-BSBEM) \citep{abdedou2021non}. Recent studies have explored the combination of the POD basis with artificial neural networks (ANN) to construct an efficient regression framework for linear reduced order modeling for time-dependent problems by learning the map between the time-parameter inputs and the modal coefficients of the POD basis \citep{JACQUIER2021109854, hesthaven2018non, wang2019non}. Despite the wide adoption of the linear ROM in approximating the outputs for parametric and time-dependent problems, they still suffer to accurately capture the dynamics of some complex physical problems (with strong hyperbolic behavior or shock propagation wave) without increasing considerably the number of reduced basis functions thus compromising the low dimensionality aspect \citep{taddei2015reduced, ohlberger2015reduced}.\\

To overcome the limitations of the techniques based on linear dimensionality reduction \citep{nikolopoulos2022non}, recent methods based on nonlinear manifolds have gained interest in the dimensionality reduction area. Some of these approaches are based on the recently developed algorithms of deep learning technology which are effective in learning sophisticated abstract features \citep{lecun2015deep, zhu2020information, zhu2020deep}. Autoencoder based method (AE), a special type of neural network (NN) with two parts named encoder and decoder trained jointly,  has been successfully introduced as a nonlinear dimensionality reduction framework with an effective nonlinear relationship between variables \citep{kim2022fast, otto2019linearly}. Some limitations concerning classical autoencoders based on dense layers (AE) have been reported in the literature, due to the drastic increase in the number of trainable parameters when the dimension of the input dataset becomes large \citep{halder2022non, nikolopoulos2022non}. Convolutional autoencoders (CAEs) have emerged as an alternative to AEs in the nonlinear compression manifolds. Their structure comprises several operations such as convolution-deconvolution, pooling-upsampling, and multilayer perceptron (MLP) \citep{he2021multiblock, gonzalez2018deep}. CAEs offer the possibility of sharing coefficients and local patch connection allowing a significant reduction in the trainable parameters. Some configurations combine CAEs with LSTM and TCN \citep{xu2020multi,dutta2022reduced} to provide a tool for surrogate modeling predictions, particularly for time predictions.\\

In this paper, a non-intrusive reduced order modeling framework is proposed for parametrized and time-dependent physical problems. This data-driven approach relies on attractive features of convolutional autoencoders (CAEs) in reducing the dimensionality of high-fidelity solutions collected from numerical solvers with large-scale meshes. The method, referred to as the NIROM-CAEs, is based on two compression levels provided by 1d-autoencoders. The first 1d-encoder reduces the spatial dimension by encoding the input dataset along the spatial dimension to generate a latent space. Another 1d-encoder convolves the latent space along the temporal dimension to output the final spatio-temporal encoded latent vector which is mapped to the input parameters values through a multilayer perceptron (MLP). Therefore, for a new set of unseen parameters, the online predictive stage allows a fast and accurate reconstruction of the original spatio-temporal dynamics through the trained 1d-decoders. The method is applied for a stochastic treatment of two benchmark cases with univariate meshes, and a hypothetical dam-break flow over a real river with complex bathymetry and two-dimensional unstructured meshes which may present a challenging task for convolution autoencoders. The proposed framework presents interesting features in non-linear reduction modeling for physical problems with a high degree of complexity in their dynamics.\\

The paper is organized as follows: in Section \ref{metho} the fundamental framework of the proposed NIROM-CAEs is presented with detailed steps of the offline and online stages. Section \ref{res} provides numerical test cases for assessing the performance and accuracy of the proposed method, and a summary and concluding remarks are presented in Section \ref{conc}. 

\section{Methodology}\label{metho}
This section describes the fundamental framework of the proposed non-intrusive reduced model based on 1D convolutional autoencoders for space and time dimensions. The method belongs to the data-driven approaches where high-fidelity solutions to large-scale and time-dependent physical problems are obtained from numerical solvers and gathered for both time and parameters sequences to construct snapshot matrices.\\

\subsection{Convolutional autoencoders (CAEs)}\label{CAEs}
\begin{figure}[ht!]
 \centering
\includegraphics[width=1\textwidth]{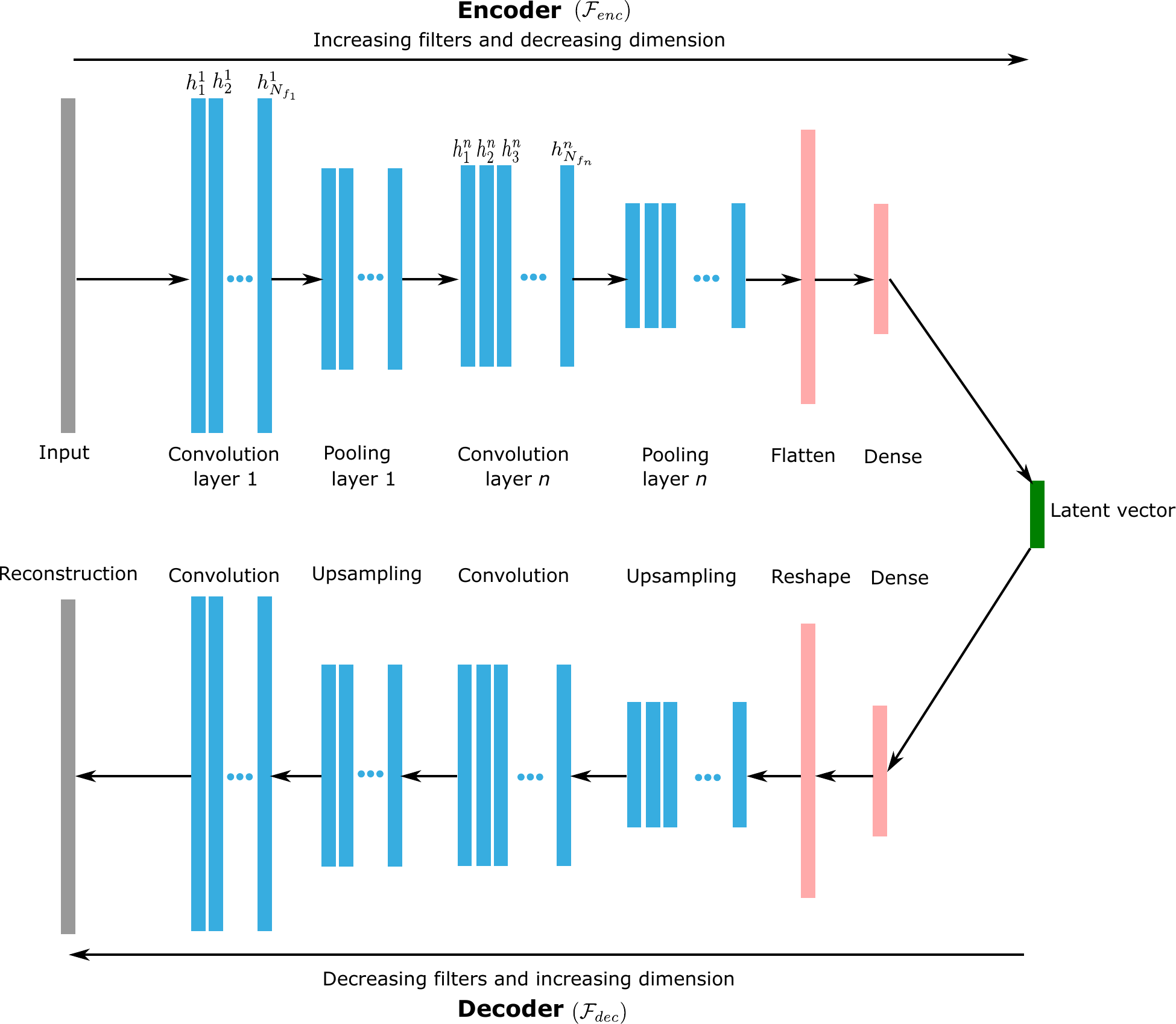}
 \caption{Schematic representation of 1D convolutional autoencoder architecture. The same structure of the 1D-CAE is adopted in the encoding and decoding processes for both space and time dimensions.}
   \label{fig:CAE_structure}
\end{figure}

Convolutional autoencoders (CAEs) gained interest as powerful non-linear reduced order modeling techniques with remarkable performances in the image recognition field. Convolutional layers have been introduced to overcome some limitations that classical autoencoders (AEs) based on dense layers may face when it comes to treating time-dependent problems with high dimensional inputs \citep{nikolopoulos2022non}. They are characterized by two properties: the local connections and shared weights, thus allowing a feature map of the input and a limitation of the number of trainable parameters. Convolutional autoencoders are constituted of two distinguished symmetrical parts; the first part called the encoder ($\mathcal{F}_{enc}$), reduces the dimension of the input matrix by mapping with a latent space through a combination of successive convolutional, pooling, and fully connected layers. The decoder part ($\mathcal{F}_{dec}$), consists of a combination of dense, upsampling, and convolutional layers; and maps the latent reduced dimension vector to a larger-dimensional reconstruction of the input. A schematic representation of a 1D convolutional autoencoder architecture is presented in Fig.\ref{fig:CAE_structure}. It should be emphasized that the same structure of the 1D CAE is adopted in this study for the encoding and decoding processes for both space and time dimensions. Thus, the presentation of the framework of the proposed approach will concern mainly the 1D convolutional autoencoders.\\ 

%========================================
%+++++++++++++++++++++++++++++++
%+++++++++++++++++++++++++++++++
%========================================

A convolution layer represents a feature map where each unit in the layer is locally connected to a selected part of the previous layer via a kernel (filter) and an activation function. This operation allows extraction of the most dominant features from the input data by applying a filter that moves with the vertical stride in the case of 1D convolution as shown in Fig.\ref{fig:1D_conv_one_multi_channel}. In this figure, a schematic representation of the 1D-convolution operation for a given layer $n$ is detailed for both space and time directions. A mathematical formulation of a typical 1d-convolutional operation denoted by the symbol ($*$), shown in Figs.\ref{fig:CAE_structure} and \ref{fig:1D_conv_one_multi_channel}, can be expressed as follows \citep{he2021multiblock,maulik2021reduced}: 
\begin{linenomath*}
\begin{equation}\label{<eq_conv1d>}
h_{i}^{\ell}=\sigma\left(\mathbf{H}^{\ell-1}*\mathbf{f}_{i}^{\ell}+b_{i}^{\ell}\right) 
\end{equation}
\end{linenomath*}
where $h_{i}^{\ell} \in \mathbb{R}^{D_{\ell}\times1}$ denotes the $i^{th}$ feature of the $\ell^{th}$ layer, $\sigma$ a non linear activation function, $\mathbf{H}^{\ell-1}=\left[h_{1}^{\ell-1},\,h_{2}^{\ell-1},\ldots,h_{N_{f_{\ell-1}}}^{\ell-1}\right]$ stands for the convolution layer $\ell-1$, $\mathbf{f}_{i}^{\ell}$ is the kernel of the layer $\ell$, and $b_{i}^{\ell}$ the bias parameter, with $i\in(1,\,N_{f_{\ell}})$ and $\ell\in(1,\,n)$. The number of feature column vectors in each layer $N_{f_{\ell}}$ corresponds to the number of kernels represented by different colors in Fig.\ref{fig:1D_conv_one_multi_channel}, and the total number of layers $n$ defines the depth of the convolution neural network (CNN). In addition to the convolutional operations, pooling layers (in this case, a max-pooling) are also integrated after each convolution layer to reduce the dimension of the convolved features by a fixed factor defined as a kernel size of the max-pooling layer, thus allowing maintaining the most dominant features of the local domain \citep{westermann2020using}. The outputs features resulting from a max-pooling operation are illustrated in the encoder part of Fig.\ref{fig:CAE_structure} for the $n^{th}$ layer, and denoted by $p_{i}^{n}$ with $i\in(1,\,N_{f_{n}})$.

\subsection{Non-intrusive reduced order model based comvolutional autoencoders (NIROM-CAEs)}\label{NIROM-CAEs}
As mentioned, the proposed approach combines the convolutional autoencoders with the reduced order modeling concept and concerns the parametrized time-dependent problems with large-scale computational domains. The framework of the non-linear surrogate model is mainly based on a succession of two compression levels through 1d-convolutions. The first level concerns the space dimension while the second deals with the compression of the time dimension of the obtained latent space from the former spatial compression.

\subsubsection{Dataset construction}\label{dataset}
The model proposed in this work belongs to the data-driven approaches where the input dataset, known as a snapshot matrix, is constructed through a collection of high-fidelity solutions from the numerical solver $\left\lbrace \mathbf{u}(\eta_{s},t_{j})\in\mathbb{R}^{N_{x}\times 1};\,j=1,\ldots,N_{t};\,s=1,\ldots,N_{s}\right\rbrace$, with $\eta_{s}$ represents the $s^{th}$ value of the random parameter $\eta$ within its generated sample set with a size of $N_{s}$, following an appropriate probability density function $ \varrho(\eta)$. $t_{j}$ denotes the $j^{th}$ time step within the temporal domain $t\in\mathcal{T}=\left[0,\;T \right]$ decomposed into $N_{t}$ time steps. The high fidelity solutions are collected from the numerical solver as a column vector $\mathbf{u}\in\mathbb{R}^{N_{x}\times1}$, where $N_{x}$ is the total number of the meshing nodes that cover the computational domain. The snapshot matrix, obtained by gathering these solution vectors, can be expressed as a 3D global matrix:
\begin{linenomath*}
\begin{equation}\label{<eq_glob_snap_mat>}
 \boldsymbol{\mathcal{U}}=\left[\;\mathcal{U}_{1}\mid\ldots\mid\mathcal{U}_{s}\mid\ldots\mid\mathcal{U}_{N_{s}} \right]\in\mathbb{R}^{N_{s}\times N_{x}\times N_{t}} 
\end{equation}
\end{linenomath*}   
with $\mathcal{U}_{s}\in\mathbb{R}^{N_{x}\times N_{t}}$ is a 2D solutions matrix corresponding to the parameter value $\eta_{s}$ over the whole time steps, expressed as:
\begin{linenomath*}
\begin{equation}\label{<eq_loc_snap_mat>}
 \mathcal{U}_{s}=\left[\mathbf{u}(\eta_{s},t_{1})\mid\ldots\mid\mathbf{u}(\eta_{s},t_{N_{t}}) \right]\in\mathbb{R}^{N_{x}\times N_{t}} 
\end{equation}
\end{linenomath*}
It should be noted that the global snapshot matrix given by Eq.\eqref{<eq_glob_snap_mat>} is divided into two sets:  a training set ($80\,\%$) and a validation set ($20\,\%$) during the learning process for the space and time autoencoders.

%========================================
%++++++++++++++++
\begin{figure}[ht!]
  \centering
    \begin{subfigure}[b]{0.43\textwidth}
      \centering
        \includegraphics[width=\textwidth]{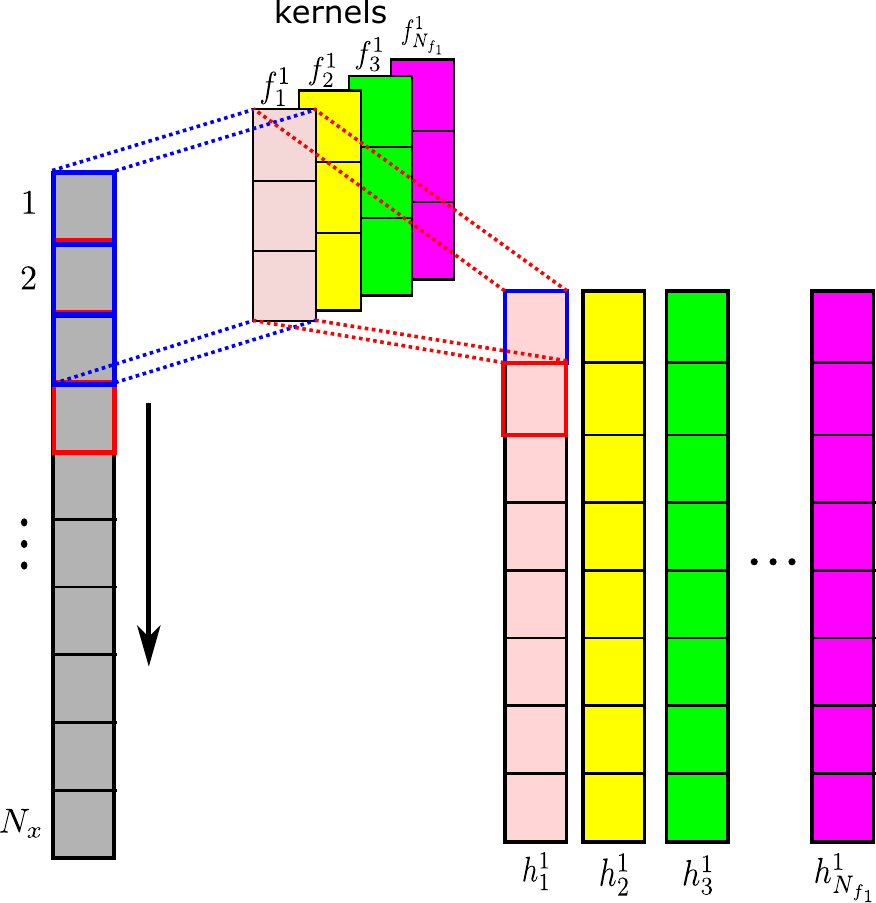}
         \caption{Convolution along the spatial dimension}
         \label{fig:1D_conv_one_channel}
    \end{subfigure}  
  \hfill
    \begin{subfigure}[b]{0.43\textwidth}
      \centering
        \includegraphics[width=\textwidth]{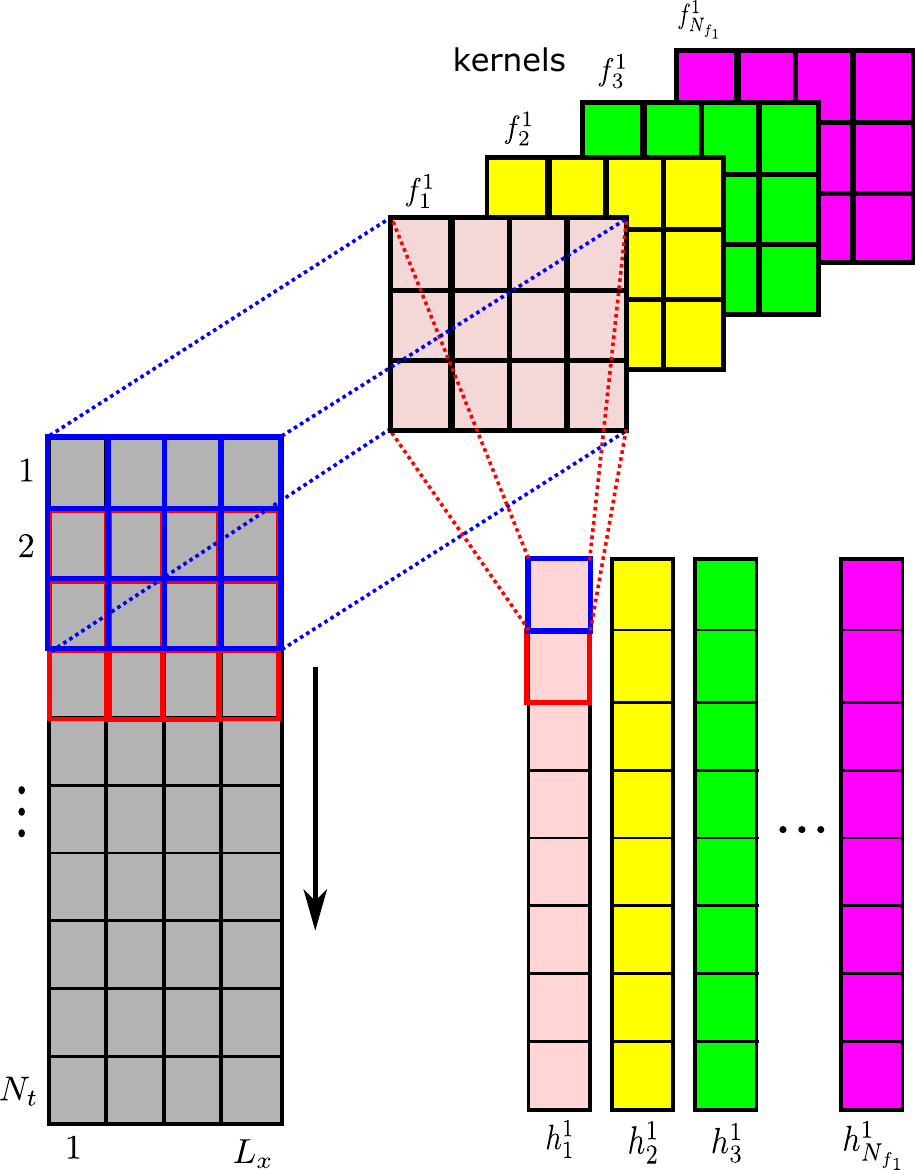}
         \caption{Convolution through the temporal dimension}
         \label{fig:1D_conv_multi_channel}
     \end{subfigure}  
 %++++++++++++++++++++++++++++++ 
   %++++++++++++
   \caption{Schematic representation of the 1D-convolutional operation along the spatial and temporal dimensions. For interpretation of the references to color in this figure, the reader is referred to the web version of this article.}
   \label{fig:1D_conv_one_multi_channel}
\end{figure}
%++++++++++++++++
%========================================

\subsubsection{Spatial compression}\label{space_comp}
The spatial compression represents the first compression level of the proposed technique which aims to reduce the dimension of the input dataset along the spatial dimension from $N_{x}$ to $L_{x}$ with $L_{x}\ll N_{x}$. Each snapshot matrix $\mathcal{U}_{s}\in\mathbb{R}^{N_{x}\times N_{t}}$, given by Eq.\eqref{<eq_loc_snap_mat>}, corresponding to the parameter value $\eta_{s}$ is reshaped by considering its transpose $\mathcal{U}_{s}^{T}\in\mathbb{R}^{N_{t}\times N_{x}}$ on which the space encoder $\mathcal{F}_{x_{enc}}$ is applied along the spatial dimension as follows:
\begin{linenomath*}
\begin{equation}\label{<x_enc>}
 \mathcal{V}_{x_{s}}=\mathcal{F}_{x_{enc}}\left(\mathcal{U}_{s}^{T}\right)\in\mathbb{R}^{N_{t}\times L_{x}} 
\end{equation}
\end{linenomath*}
where $\mathcal{V}_{x_{s}}\in\mathbb{R}^{N_{t}\times L_{x}}$ represents the encoded latent space corresponding to the parameters value $\eta_{s}$ with $s=1,\,\ldots,\,N_{s}$. It should be emphasized that for implementation purposes in the TensorFlow library \citep{abadi2016others}, the dimension of the input dataset has been extended to a 3d tensor of a shape $(N_{t},\, N_{x},\,1)$ through the use of a reshape operator: $\mathcal{R}:\mathbb{R}^{N_{t}\times N_{x}}\mapsto\mathbb{R}^{N_{t}\times N_{x}\times 1}$ \citep{halder2022non}, where the third dimension stands for the number of channels which is equal to one in this case. A schematic representation of the 1d convolution operation performed on the first convolution layer by the space encoder is depicted in Fig.\ref{fig:1D_conv_one_channel}, where $N_{f_{1}}$ 1d-filters, whose width corresponds to the number of channels, convolve local patch of the input dataset along the spatial dimension, represented by the vertical arrow, to generate the features of the convolved layer $h_{i}^{1},\,i=1,\,\ldots,\,N_{f_{1}}$. As shown in Fig.\ref{fig:1D_conv_one_channel}, each filter is distinguished by the same color as for its corresponding column features generated when it convolves along the spatial direction. The detailed structures of the space autoencoders adopted for different test cases are reported in tables \ref{tab:Tab_CAE_space_archit_Burg_Stok} and \ref{tab:Tab_CAE-space_archit_Miles_Iles}. 

\subsubsection{Temporal compression}\label{temp_comp}
Once the spatial dimension is encoded, a second compression level is further performed on the obtained latent space $\mathcal{V}_{x_{s}}\in\mathbb{R}^{N_{t}\times L_{x}}$ along the temporal dimension to reduce its size from $N_{t}$ to a given $L_{t}$ through a 1d time encoder $\mathcal{F}_{t_{enc}}$ that has the same structure as that shown in Fig.\ref{fig:CAE_structure}. The encoding process can be mathematically expressed as follows:
\begin{linenomath*}
\begin{equation}\label{<t_enc>}
\mathcal{V}_{t_{s}}\in\mathbb{R}^{L_{t}}=\mathcal{F}_{t_{enc}}\left(\mathcal{V}_{x_{s}}\right) 
\end{equation}
\end{linenomath*}
where $\mathcal{V}_{t_{s}}\in\mathbb{R}^{L_{t}}$ is the encoded latent vector corresponding to the single parameter value $\eta_{s}$. The convolution operation of the first layer 1d-time encoder, expressed by Eq.\eqref{<eq_conv1d>}, is presented in Fig.\ref{fig:1D_conv_multi_channel}, where $N_{f_{1}}$ filters convolve local patches of the input dataset $\mathcal{V}_{x_{s}}\in\mathbb{R}^{N_{t}\times L_{x}}$ along the temporal dimension ($N_{t}$). $L_{x}$ denotes the number of channels of the input dataset, which corresponds to the width of each filter. The detailed structures of the time autoencoders applied to different test cases addressed in this study are reported in tables \ref{tab:Tab_CAE_time_archit_Burg_Stok} and \ref{tab:Tab_CAE-time_archit_Miles_Iles} where information about the number of convolution layers, kernel sizes and activation functions can be found.  

\subsubsection{Regression based multilayer perceptron (MLP)}\label{MLP}
The third level of the proposed approach concerns the encoded latent vector $\mathcal{V}_{t_{s}}\in\mathbb{R}^{L_{t}}$, obtained through two compression levels of the input dataset $\mathcal{U}_{s}\in\mathbb{R}^{N_{x}\times N_{t}}$, which is mapped to the input parameter $\eta_{s}$ via a multilayer perceptron regression (MLP). This regression that links the input parameter to the final latent vector, composed of multiple stacked fully-connected layers, can be expressed as follows \citep{xu2020multi}:
\begin{linenomath*}
\begin{equation}\label{<mlp>}
\mathcal{V}_{t_{s}}=\Pi_{MLP}\left(\eta_{s}\right) 
\end{equation}
\end{linenomath*}
The detailed architecture of the MLP used in this work for the addressed test cases is presented in table \ref{tab:MLP_archit}.
The mathematical framework presented in subsections \ref{dataset} through \ref{MLP} represents the main steps of the offline stage of the proposed NIROM-CAEs technique.\\

It is worth mentioning that the proposed model was built and trained using the open-source deep learning library TensorFlow \citep{abadi2016others} using Adam optimizer with its default parameters. To accelerate the convergence and optimize the training processes, the data used during the training process of the offline stage are all normalized with the same min-max scaling \citep{xu2020multi, fu2021data}:
\begin{linenomath*}
\begin{equation}\label{<normaliz>}
\tilde{u}^{s}=\frac{u^{s}-min(\mathbf{u}_{j})}{max(\mathbf{u}_{j})-min(\mathbf{u}_{j})}-0.5\;;\quad j=1,\,\ldots,\,N_{t};\quad s=1,\,\ldots,\,N_{s}
\end{equation}
\end{linenomath*}
with $\tilde{u}^{s}\in\left[-0.5,\,0.5\right]$. The inverse transformation must also be applied to the predicted data matrix during the online stage to recover the original scaling.

\subsubsection{Online surrogate prediction}\label{on_pred}
Once the three models are trained during the offline stage described above, the online surrogate prediction becomes straightforward. A new set of the random parameter is generated following its probability distribution function $\boldsymbol{\widehat{\eta}}=\lbrace\widehat{\eta}_{1},\,\ldots,\,\widehat{\eta}_{N_{s'}}\rbrace$, and for each unseen value $\widehat{\eta}_{s}$, the encoded Spatio-temporal latent vector $\widehat{\mathcal{V}}_{t_{s}}$ is predicted through the built MLP regression model: $\widehat{\mathcal{V}}_{t_{s}}\in\mathbb{R}^{L_{t}}=\Pi_{MLP}\left(\widehat{\eta}_{s}\right)$. This predicted latent space is then decoded through the time decoder $\mathcal{F}_{t_{dec}}$ to approximate the spatial latent space: $\widehat{\mathcal{V}}_{x_{s}}\in\mathbb{R}^{N_{t}\times L_{x}}=\mathcal{F}_{t_{dec}}\left(\widehat{\mathcal{V}}_{t_{s}}\right)$. Finally, the obtained latent space is decoded using the space decoder $\mathcal{F}_{x_{dec}}$ to generate the predicted input dataset: $\widehat{\mathcal{U}}_{s}\in\mathbb{R}^{N_{t}\times N_{x}\times 1}=\mathcal{F}_{x_{dec}}\left(\widehat{\mathcal{V}}_{x_{s}}\right)$, which is transformed to its final shape through the inverse reshape operator $\mathcal{R}^{-1}:\mathbb{R}^{N_{t}\times N_{x}\times 1}\mapsto\mathbb{R}^{N_{t}\times N_{x}}$. Thus, the predicted snapshot matrix corresponding to the single value parameter $\widehat{\eta}_{s}$ through the online stage can be expressed as follows: 
\begin{linenomath*}
\begin{equation}\label{<normaliz>}
\widehat{\mathcal{U}}_{s}=\mathcal{F}_{x_{dec}}\left(\mathcal{F}_{t_{dec}}\left(\Pi_{MLP}\left(\widehat{\eta}_{s}\right)\right)\right)
\end{equation}
\end{linenomath*}

The statistical moments can be estimated through the constructed surrogate model of the stochastic output response $\widehat{u}$ as follows:

\begin{linenomath*}
\begin{equation} \label{<eq_pod_bsbem_mean>}
 \mu_{\widehat{u}}=\mathbb{E}\left[\widehat{u}\right]=\int_{\Xi}\widehat{u}\varrho(\widehat{\eta})d\widehat{\eta} 
\end{equation} 
\end{linenomath*}
and
\begin{linenomath*}
\begin{equation}\label{<eq_pod_bsbem_std>}
\sigma_{\widehat{u}}^{2}=\mathbb{E}\left[\widehat{u}^{2}\right]- \mu_{\widehat{u}}^{2} =\left[ \int_{\Xi}\left(\widehat{u}\right)^{2}\varrho(\widehat{\eta})d\widehat{\eta}\right]-\mu_{\widehat{u}}^{2}  
\end{equation}
\end{linenomath*}

\begin{figure}[ht!]
 \centering
\includegraphics[width=1\textwidth]{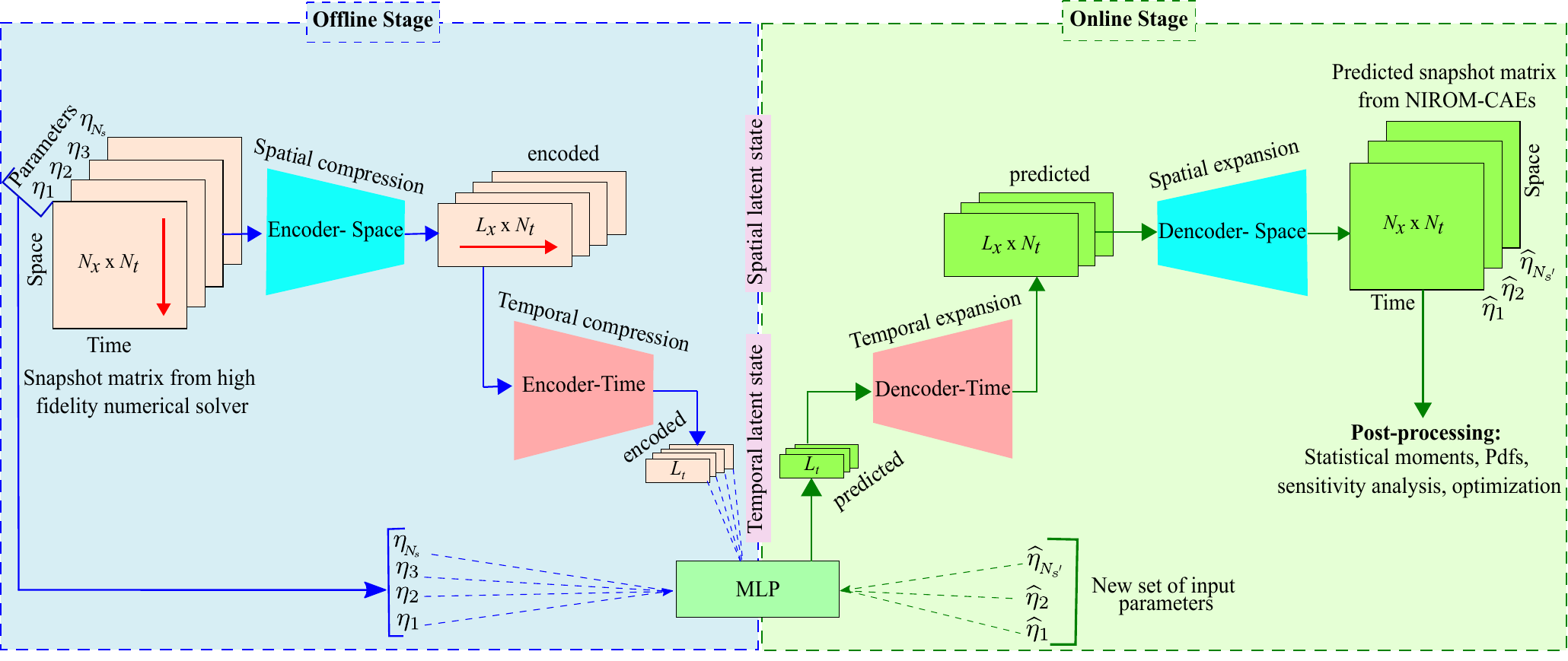}
 \caption{Flowchart illustrating the offline training and online predictive stages of the NIROM-CAEs model. The red arrow indicates the encoding direction (For interpretation of the references to color in this figure, the reader is referred to the web version of this article).}
   \label{fig:CAEs_Offline_and_Online_stages_diagram}
\end{figure}

The fundamental framework of the proposed NIROM-CAEs approach is summarized in the flowchart presented in Fig.\ref{fig:CAEs_Offline_and_Online_stages_diagram} where the most relevant steps of the offline training and online predictive stages are detailed.\\

%========================================
%+++++++++++++++++++++++++++++++
%+++++++++++++++++++++++++++++++
%========================================

The predictive abilities of the proposed NIROM-CAEs are assessed in comparison with the Latin Hypercube Sampling approach (LHS) whose results are considered reference solutions. In addition, another method based on a proper orthogonal decomposition concept combined with an artificial neural network (POD-ANN) is retained as a linear non-intrusive reduced order modeling technique for comparison purposes. Thus, a detailed presentation of the development and implementation steps of POD-ANN is beyond the scope of this paper, and the reader is encouraged to consult the corresponding references \citep{wang2019non, hesthaven2018non, JACQUIER2021109854} for a deeper insight into the modeling procedure.\\

\section{Results and discussion}\label{res}
In this section, The application of the proposed non-intrusive reduced order model to two well-known 1d benchmark test cases (Burgers' and Stokers' analytical solutions) and a hypothetical failure of a real dam is described. The obtained results are depicted to showcase the accuracy and efficiency of the proposed approach in approximating the statistical moments for large-scale and time-dependent problems.

%+=+=+=+=+=+=+=+=+=+=+=+=+=+=+=+=+=+=+=+=+=+=+=+=+=+=+=+
%++++++++++++++++++++++++++++++++++++++++
\subsection{One-dimensional Burgers equation test case} \label{Burg_sol}
The first test case concerns the one-dimensional viscous Burgers' equation for non-linear convection-diffusion PDE, expressed in its dimensionless form with the corresponding initial and Dirichlet boundary conditions, as follows \citep{burgers1948mathematical,guo2019data}: 

\begin{linenomath*}
\begin{equation}\label{<Burg_equ>}
\frac{\partial{u}}{\partial{t}}+u\frac{\partial{u}}{\partial{x}}=\frac{1}{Re}\frac{\partial^{2}{u}}{\partial{x^{2}}}\;,\qquad x\in\left[ 0,\;1\right],\qquad t\in\left[ 0,\;1\right]
\end{equation}
\end{linenomath*}
with: $u(x,0)=\frac{x}{1+exp\left( \frac{Re}{16}(4x^{2}-1)\right)}$ and $u(0,t)=u(1,t)=0$. An analytical solution for the velocity field of the convection-diffusion problem described above is given by \citep{dutta2022reduced}: 

\begin{linenomath*}
\begin{equation}\label{<Burg_analy>}
u(x,t)=\frac{\frac{x}{t+1}}{1+\sqrt{\frac{t+1}{t_{0}}}exp(Re\frac{x^{2}}{4t+4})}
\end{equation}
\end{linenomath*}  
with $t_{0}=exp(\frac{Re}{8})$. In this test case, the Reynolds number $Re$, which represents the diffusion coefficient, is considered as a random variable following uniform distribution within its variability range $Re_{\mu=1000,\,\sigma=200}\in\mathcal{U}\left[654,\;1\,346\right]$. The sample size used in this test case is about $N_{s}=200$ randomly divided into two sets: one for the training ($80\,\%$) and another for validation ($20\,\%$). For each selected value in the generated random parametric sets, the analytical solution, given by Eq.\eqref{<Burg_analy>}, is evaluated over a discretized spatial domain with $N_{x}=1\,000$ nodes for all the $N_{t}=104$ time-steps to construct the training and validation high-fidelity snapshot matrices that are used to train the 1D spatial-CAE ($\mathcal{F}_{x}$) of the NIROM-CAEs approach, as described above. \\

The detailed architectures of the spatial encoder ($\mathcal{F}_{x_{enc}}$) and spatial decoder ($\mathcal{F}_{x_{dec}}$), which consists of $1\,033\,773$ trainable parameters, are reported in table \ref{tab:Tab_CAE_space_archit_Burg_Stok}. The spatial encoder allows the compression of the spatial dimension from $N_{x}=1\,000$ nodes to a reduced-order latent space of dimension of $L_{x}=50$ through a succession of 1D convolutional and max-pooling layers with associated non-linear activation functions. The generated spatial latent space of dimension of $N_{t}=104\times L_{x}=50$ serves as input to the 1D-time autoencoder ($\mathcal{F}_{t}$), which reduces the temporal dimension from $N_{t}=104$ to $L_{t}=10$. The detailed structure of both encoder ($\mathcal{F}_{t_{enc}}$) and decoder ($\mathcal{F}_{t_{dec}}$), with a number of trainable parameters of $177\,494$, is shown in table \ref{tab:Tab_CAE_time_archit_Burg_Stok}. The obtained Spatio-temporal latent space is then mapped with the input parameters using a multi-layer perceptron whose structure is reported in table \ref{tab:MLP_archit} and it consists of $34\,570$ trainable parameters. The CAE-space, CAE-time and MLP was trained for $500$, $1\,000$ and $3\,000$ epochs, respectively, and the decay of the training and validation losses during the training phase are depicted in Fig.\ref{fig:Conv_hist_Burg}.\\

\begin{figure}[ht!]
  \centering
    \begin{subfigure}[b]{0.49\textwidth}
      \centering
        \includegraphics[width=\textwidth]{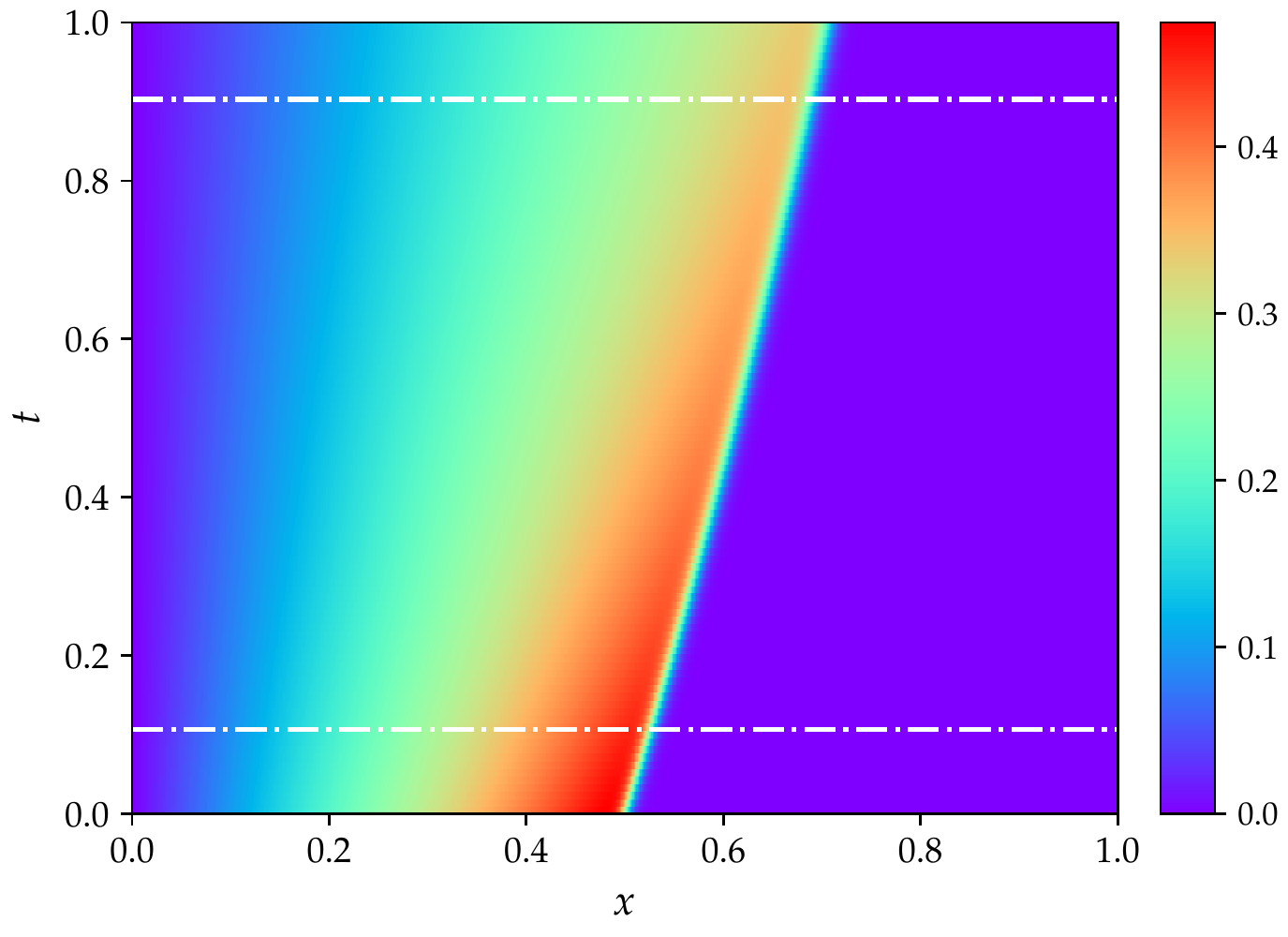}
         \caption{Mean}
         \label{fig:Burg_mean_Contours_Lhs_ns_5000}
    \end{subfigure}  
  \hfill
    \begin{subfigure}[b]{0.49\textwidth}
      \centering
        \includegraphics[width=\textwidth]{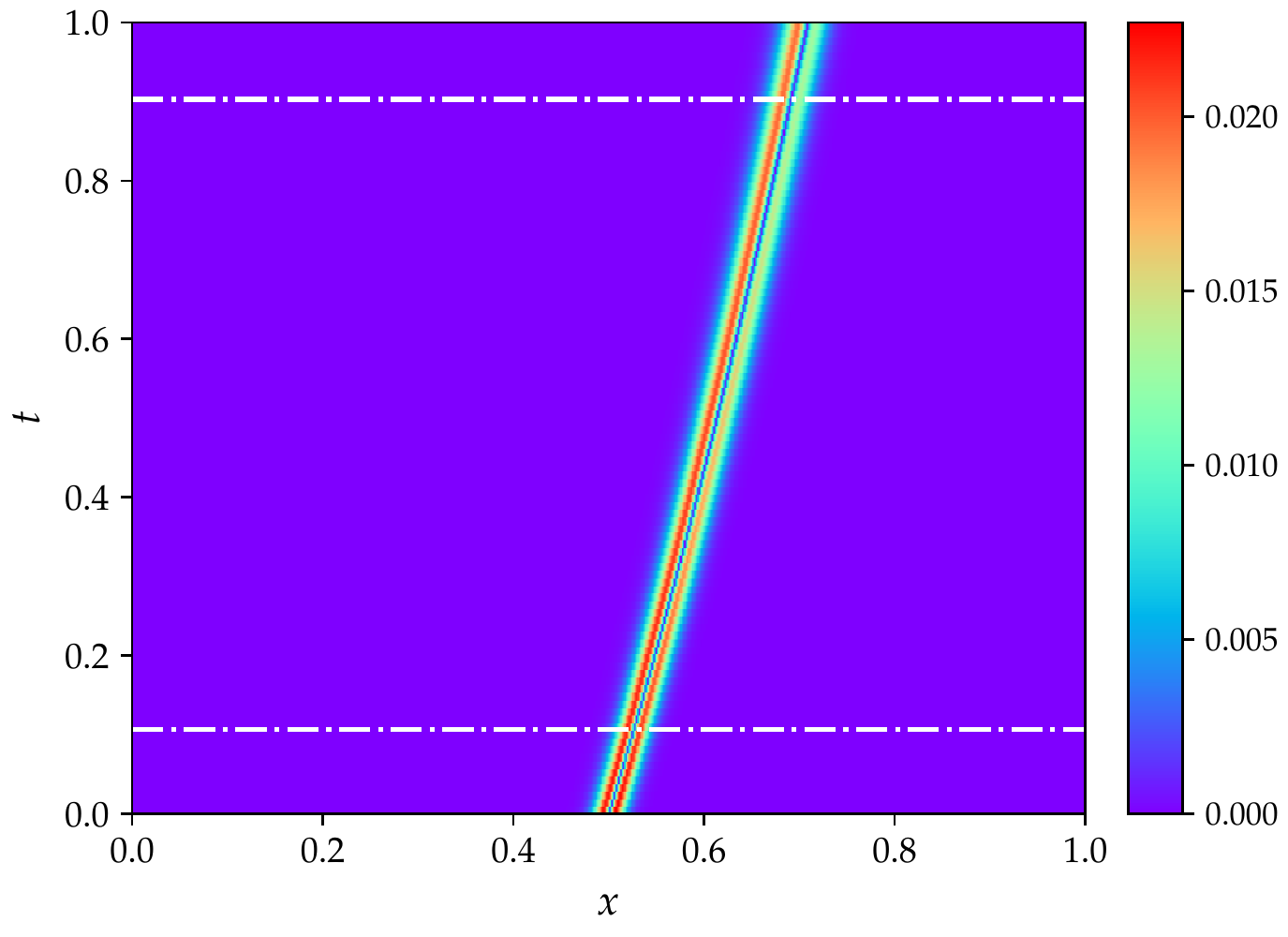}
         \caption{Std}
         \label{fig:Burg_Std_Contours_Lhs_ns_5000}
     \end{subfigure} 
   
 %++++++++++++++++++++++++++++++ 
   %++++++++++++
   \caption{Contour plots of the mean and standard deviation of the Burgers' analytical solution obtained from $5\,000$ LHS realizations with uniform distribution $Re_{\mu=1000,\,\sigma=200}\in\mathcal{U}\left[654,\;1\,346\right]$. The two horizontal dashed white lines indicate the times for which the mean and standard deviation of the POD-ANN and NIROM-CAEs are compared to those of the LHS reference solutions (For interpretation of the references to color in this figure, the reader is referred to the web version of this article).}
   \label{fig:Burg_mean_std_cont_lhs_ns_5000}
\end{figure}

The most relevant results for this test case obtained by the proposed NIROM-CAEs model are presented in terms of the statistical moments' and relative $L^{2}$-error profiles and compared with those from the POD-ANN and the LHS solutions. In Fig.\ref{fig:Burg_mean_std_cont_lhs_ns_5000}, the 2d contours of the mean and standard deviation from the LHS method with $5\,000$ realizations, considered as a reference solution, are presented in the space-time plane, where the longitudinal and transversal directions represent space and time, respectively. These plots show clearly the advecting discontinuity in the velocity field (Fig.\ref{fig:Burg_mean_Contours_Lhs_ns_5000}) and the variability surrounding its propagation with time over the longitudinal positive direction (Fig.\ref{fig:Burg_Std_Contours_Lhs_ns_5000}), which may present a challenging test case for the proposed reduced-order modeling approach. The two horizontal dashed lines represent the time location for which the mean and the standard deviation profiles are displayed along with the spatial domain.\\
%========================================
%++++++++++++++++
%++++++++++++++++
%========================================

\begin{figure}[ht!]
  \centering
    \begin{subfigure}[b]{0.49\textwidth}
      \centering
        \includegraphics[width=\textwidth]{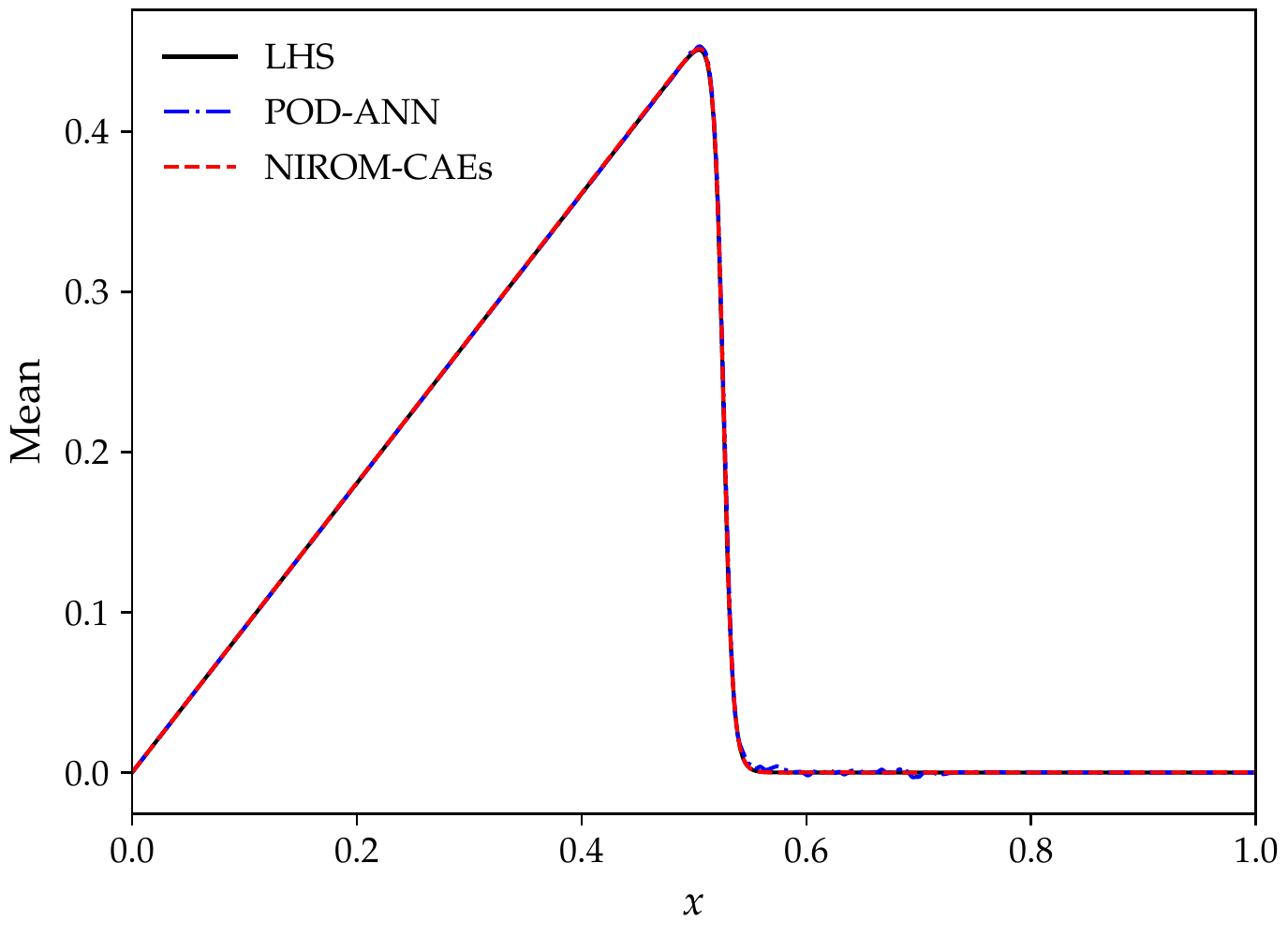}
         \caption{$t\approx 0.1$}
         \label{fig:Mean_Burg_U_compari_t_0_1}
    \end{subfigure}  
  \hfill
    \begin{subfigure}[b]{0.49\textwidth}
      \centering
        \includegraphics[width=\textwidth]{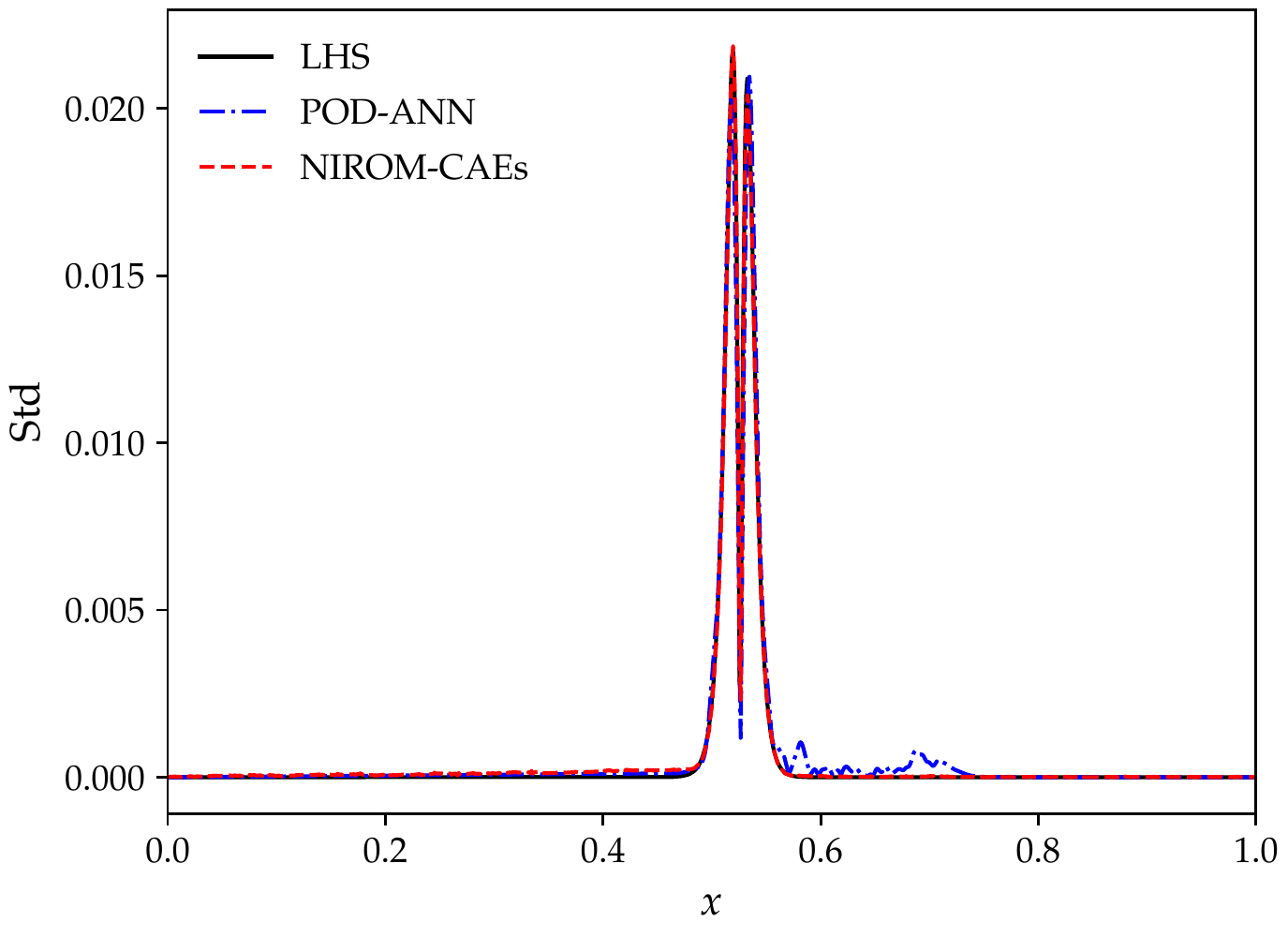}
         \caption{$t\approx 0.1 $}
         \label{fig:Std_Burg_U_compari_t_0_1}
     \end{subfigure} 
     \begin{subfigure}[b]{0.49\textwidth}
      \centering
        \includegraphics[width=\textwidth]{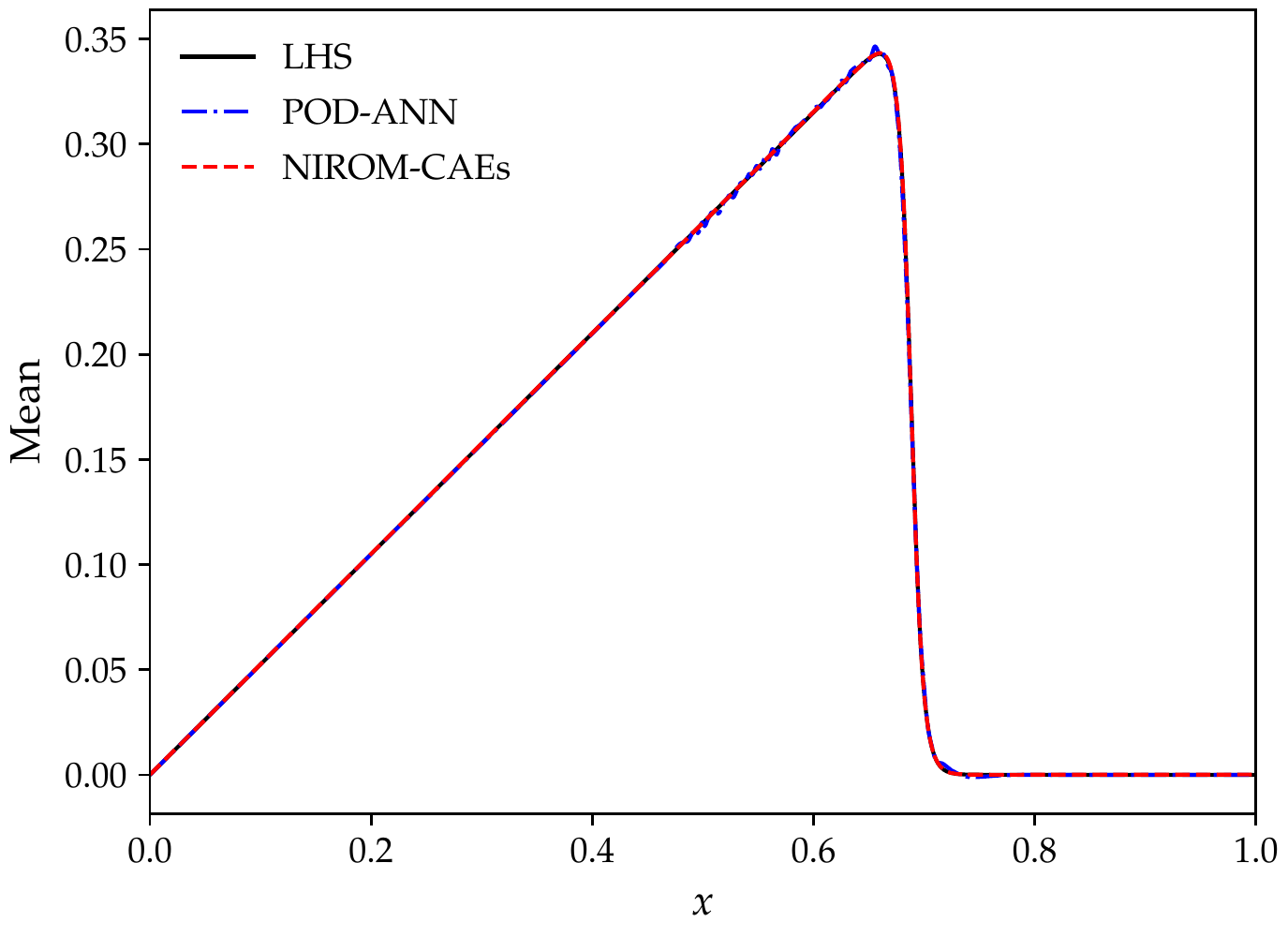}
         \caption{$t\approx 0.9 $}
         \label{fig:Mean_Burg_U_compari_t_0_9}
    \end{subfigure}  
  \hfill
    \begin{subfigure}[b]{0.49\textwidth}
      \centering
        \includegraphics[width=\textwidth]{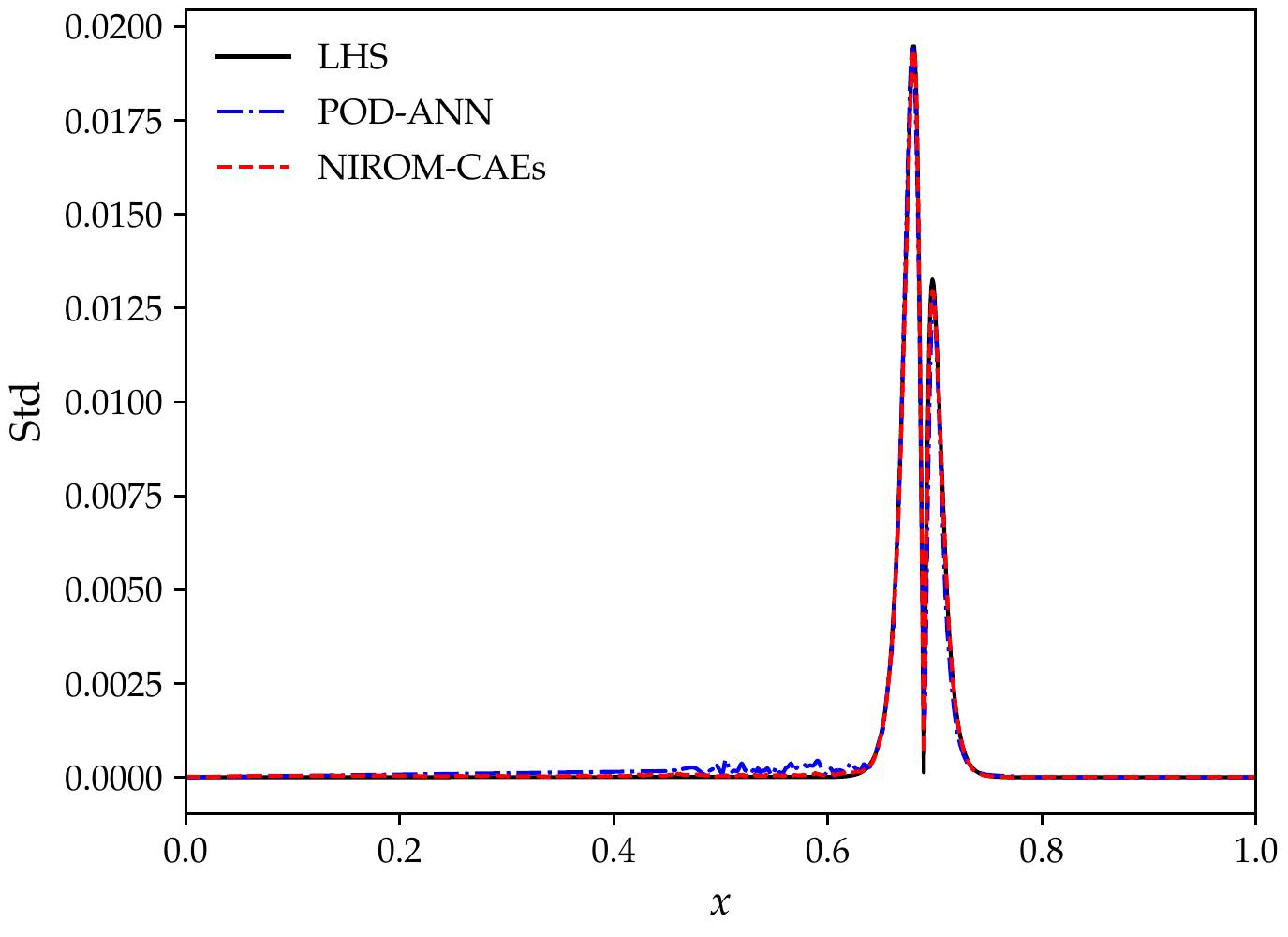}
         \caption{$t\approx 0.9 $}
         \label{fig:Std_Burg_U_compari_t_0_9}
     \end{subfigure}
   
 %++++++++++++++++++++++++++++++ 
   %++++++++++++
   \caption{Comparison of the mean (left) and standard deviation (right) of the Burger's solution obtained with the POD-ANN and NIROM-CAEs approaches with that of the LHS reference solution ($5\,000$ realizations) at different times: $t\approx0.1$ (a and b) and $t\approx0.9$ (c and d).}
   \label{fig:Burg_mean_std_compa_U_t_0_1_and_0_9}
\end{figure}

Fig.\ref{fig:Burg_mean_std_compa_U_t_0_1_and_0_9} reports the comparison of the mean and standard deviation profiles as a function of the $x$-coordinate at different time-steps ($t\approx0.1$ and $0.9$) obtained with the NIROM-CAEs (with $L_{x}=50$ and $L_{t}=10$) and POD-ANN (with $\epsilon_{s}=\epsilon_{t}=10^{-10}$ which generates a latent space of dimension of $L_{POD}=77$ modes) with those from the reference LHS solutions (with $N_{s}=5\,000$ realizations). It can be observed from these results that the proposed non-intrusive surrogate model (NIROM-CAEs) shows predicted profiles of the mean (left column) and standard deviation (right column) in close agreement with those from the LHS reference solution. Conversely, an oscillatory behavior can be showcased in the statistical moments' curves obtained with the POD-ANN technique with pronounced deviations in comparison with those from the LHS solution. It should be emphasized that the predicted statistical moments' from the NIRON-CAEs approach during the predictive online phase are obtained with new values of the random input parameter different from those used in the offline training phase. As evidenced by the presented results, the surrogate NIROM-CAEs technique allows a better approximation of the output quantities of interest due to nonlinearities that effectively capture the dynamics of the viscous advection shock in contrast with the linear-based POD approach.

%========================================
%++++++++++++++++
%++++++++++++++++
%========================================

\begin{figure}[ht!]
  \centering
    \begin{subfigure}[b]{0.49\textwidth}
      \centering
        \includegraphics[width=\textwidth]{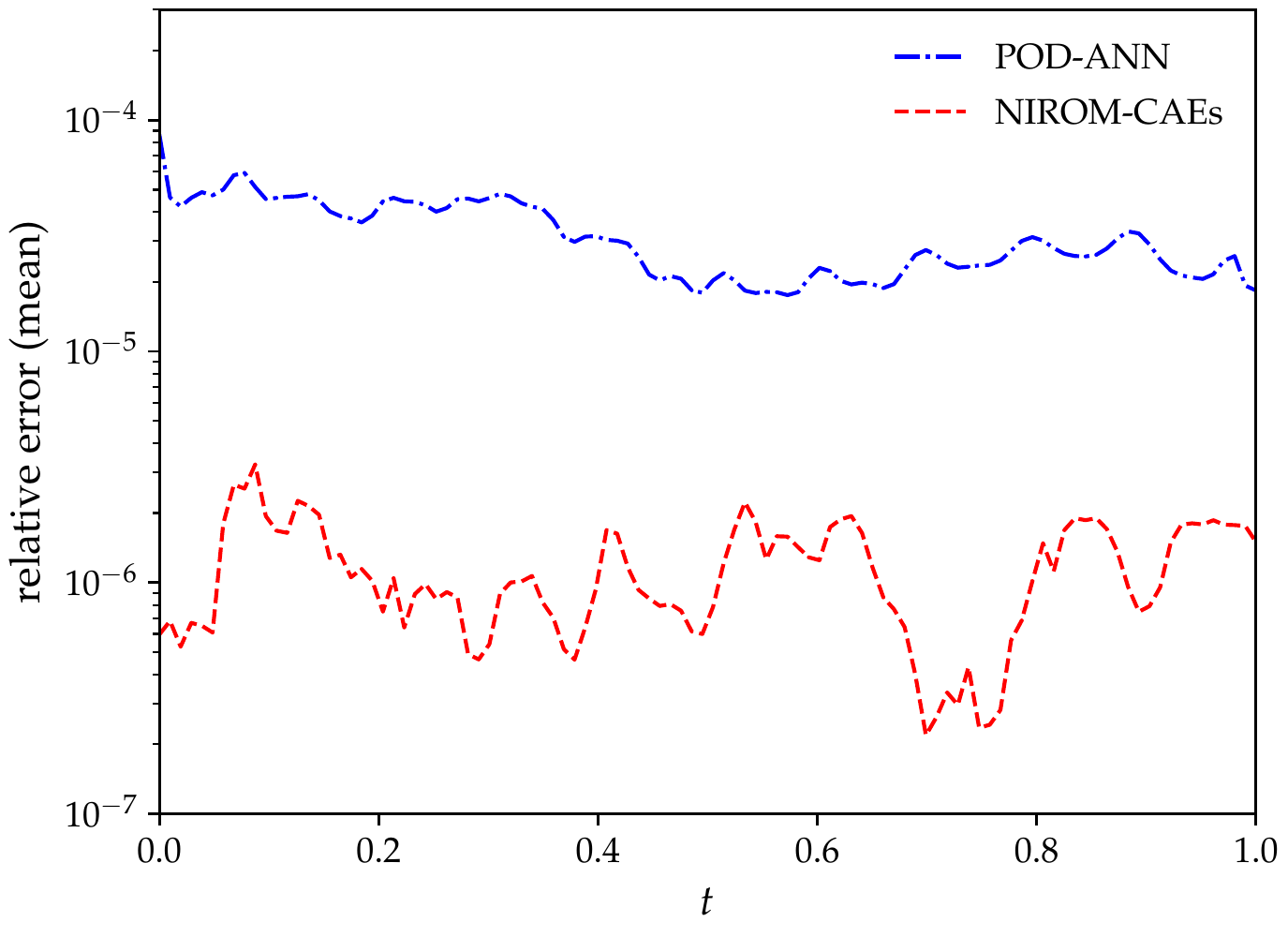}
         \caption{Mean}
         \label{fig:Error_Mean_Burg_U_compari}
    \end{subfigure}  
  \hfill
    \begin{subfigure}[b]{0.49\textwidth}
      \centering
        \includegraphics[width=\textwidth]{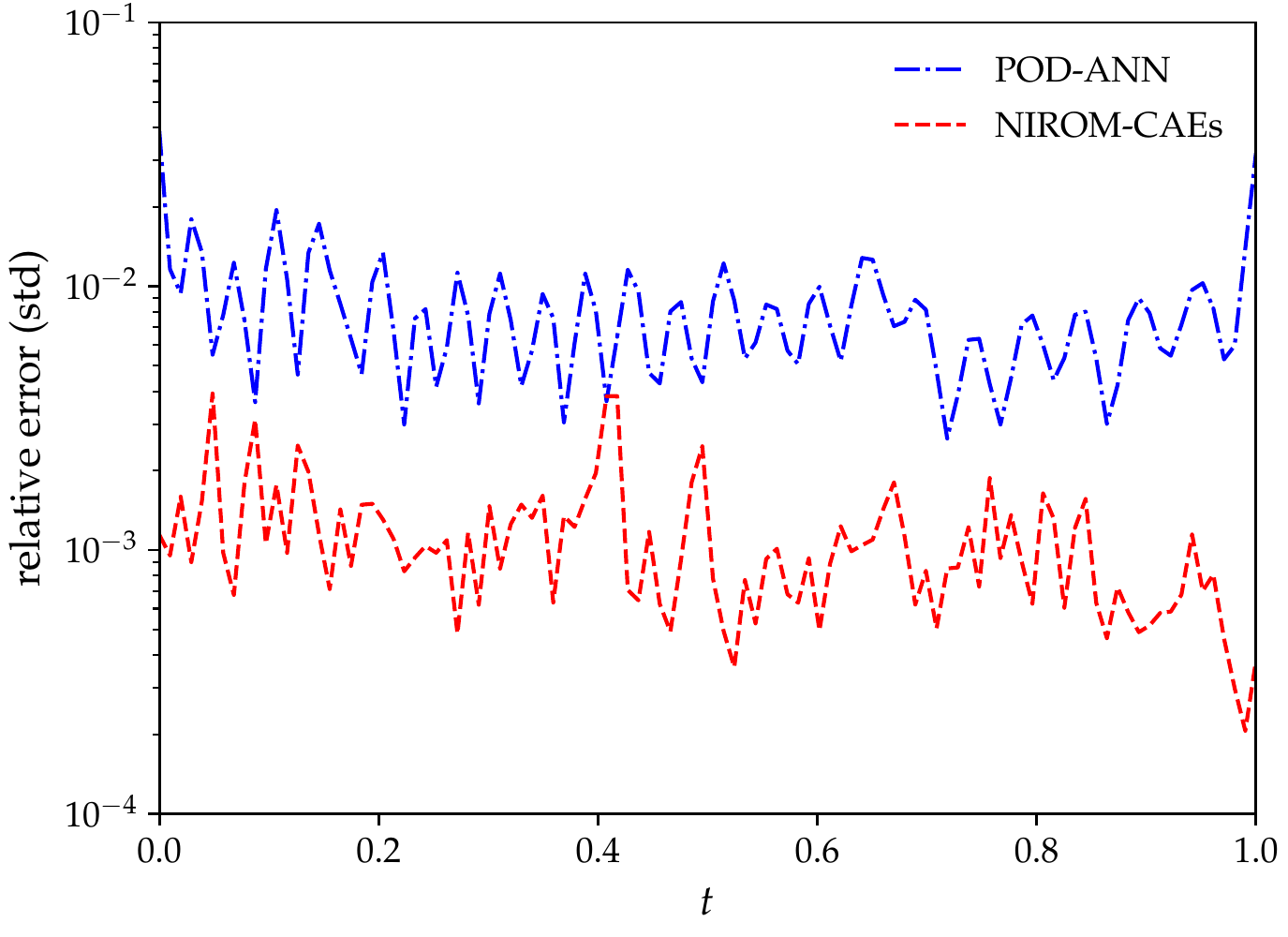}
         \caption{Std}
         \label{fig:Error_Std_Burg_U_compari}
     \end{subfigure}  
 %++++++++++++++++++++++++++++++ 
   %++++++++++++
   \caption{Variation of the $L^{2}$-relative error of the mean (left) and standard deviation (right) as a function of time obtained with the POD-ANN and NIROM-CAEs. Errors are evaluated with the LHS reference solution obtained with $5\,000$ realizations.}
   \label{fig:Error_mean_Std_Burg_U_compari}
\end{figure}

To further investigate the performance of the proposed approach, the time trajectory of the spatial relative $L^{2}$-errors of the mean and standard deviation of the predicted solutions obtained by NIROM-CAEs are compared with those from the POD-ANN model. The relative errors are computed for the LHS reference solutions with $5\,000$ realizations over the whole computational nodes and for each time step. Its mathematical expression is defined as follows:

\begin{linenomath*}
\begin{equation}\label{<L2_max>}
Err_{L^{2},\Phi}^{Surr}(t)= \sqrt{\frac{\sum_{i=1}^{N_{e}}\left(\Phi_{i,Surr}(t)-\Phi_{i,LHS}(t)\right)^2}{\sum_{i=1}^{N_{e}}(\Phi_{i,LHS}(t))^2}}
\end{equation}
\end{linenomath*}
with $\Phi$ denotes either the mean or the standard deviation, and $Surr$ stands for NIROM-CAEs or POD-ANN. The comparison of the mean and standard deviation error profiles highlights the predictive abilities of the NIROM-CAEs model, which presents lower values of the errors compared to those from the POD-ANN approach as shown in Fig.\ref{fig:Error_mean_Std_Burg_U_compari}. These plots confirm indeed the capacity of the proposed non-linear reduced-order model to accurately estimate the statistics of the outputs even for challenging time-dependent physical problems as for the advective viscous shock Burgers equation.

%========================================
%++++++++++++++++
%++++++++++++++++
%========================================

\subsection{One dimensional dam break test case} \label{Stok_sol}
\begin{figure}[ht!]
 \centering
\includegraphics[width=0.85\textwidth]{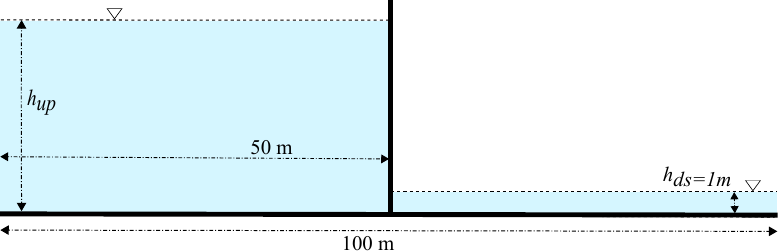}
 \caption{Schematic representation of the initial condition of the Stokers analytical solution of the 1D dam break over a wet flat bottom.}
   \label{fig:Stoker_solution_physical_domain}
\end{figure}

The second test case concerns Stoker's analytical solution \citep{stoker1957water} which describes the propagation and rarefaction wave resulting from a one-dimensional dam break over a wet flat frictionless bottom. The dynamic is initiated by unequal water levels of both the upstream and downstream sides located in the middle of the studied domain of $100\;m$ length as shown in Fig.\ref{fig:Stoker_solution_physical_domain}. The mathematical expression for both water level and velocity of the parametrized Stoker's analytical solution fields is given as follows \citep{delestre2013swashes, seyedashraf2017two}:
\begin{linenomath*}
\begin{equation} \label{<eq_stocker>}
h(x,t) = 
  \begin{cases} 
    h_{up}                                         \\[8pt]                             
    \frac{4}{9g}(\sqrt{gh_{up}}-\frac{x}{2t})^{2}  \\[8pt] 
    \frac{c_{m}^{2}}{g}                            \\[8pt]              
    h_{ds}                              
  \end{cases},\quad
 u(x,t) = 
  \begin{cases} 
    0\quad m/s                                    &\text{if } x \leq x_{A}(t) \\[8pt]
    \frac{2}{3}(\frac{x}{t}+\sqrt{gh_{up}})   & \text{if } x_{A}(t)\leq x \leq x_{B}      
                                                      (t) \\[8pt]
    2(\sqrt{gh_{up}}-c_{m})                  & \text{if } x_{B}(t)\leq x \leq x_{C}      
                                                      (t) \\[8pt]
    0\quad m/s                               &\text{if } x_{C}(t) \leq x
  \end{cases}
\end{equation}
\end{linenomath*}
where $x$ denotes the axial position, $x_{A}(t)=x_{0}-t\sqrt{gh_{up}}$, $x_{B}(t)=x_{0}+t(\sqrt{gh_{up}}-3c_{m})$ and $x_{C}=x_{0}+t\frac{2c_{m}^{2}(\sqrt{gh_{up}}-c_{m})}{c_{m}^{2}-gh_{ds}}$, and where $c_{m}=\sqrt{gh_{m}}$ denotes the selected solution of $-8gh_{ds}c_{m}^{2}(gh_{up}-c_{m}^{2})^{2}+(c_{m}^{2}+gh_{ds})(c_{m}^{2}-gh_{ds})^{2}=0$.\\

%========================================
%+++++++++++++++++++++++++++++++
%+++++++++++++++++++++++++++++++
%========================================
The upstream water level ($h_{up}$) is considered as an input random variable whose values are uniformly sampled within its plausible variability range $h_{up}\in\mathcal{U}\left[8,\;11\right]$, whereas the downstream water depth is kept constant at a deterministic value $h_{ds}=1\:m$. For each selected value in the generated sample set of the upstream water level, the analytical solution, given by Eq\eqref{<eq_stocker>}, is evaluated over the $N_{x}=1\,000$ nodes that contain the computational domain $x\in\left[0,\;100\right] m$ for all the $N_{t}=104$ time-steps of the temporal domain $t\in\left[1,\;4\right] s$. The obtained solution vectors are then concatenated to construct the so-called high fidelity snapshot matrix to train the proposed non-intrusive reduced order model (NIROM-CAEs) during the offline training phase.\\

\begin{figure}[ht!]
  \centering
    \begin{subfigure}[b]{0.49\textwidth}
      \centering
        \includegraphics[width=\textwidth]{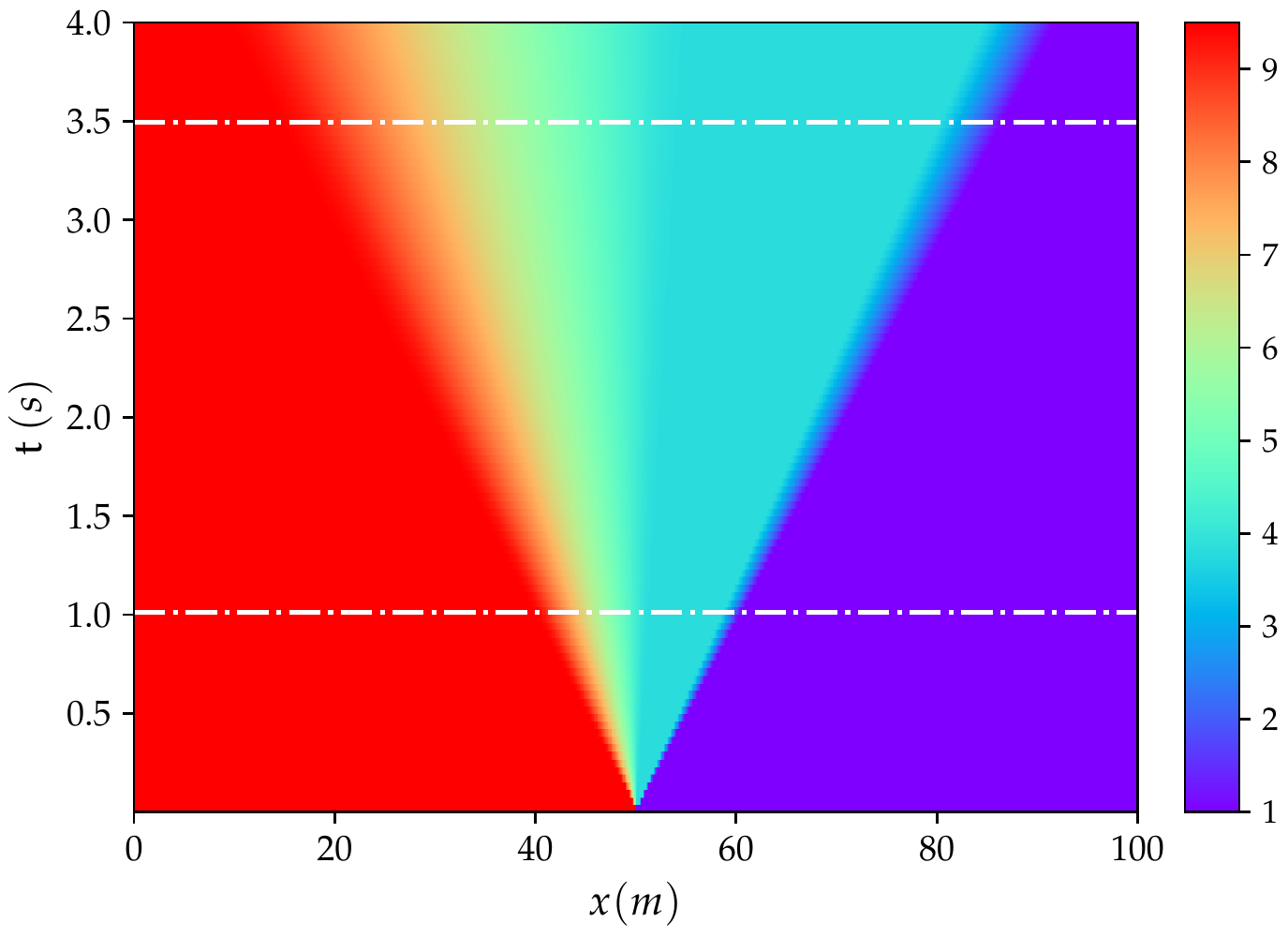}
         \caption{Mean (water level)}
         \label{fig:Stok_H_contour_mean_lhs_ns_5000}
    \end{subfigure}  
  \hfill
    \begin{subfigure}[b]{0.49\textwidth}
      \centering
        \includegraphics[width=\textwidth]{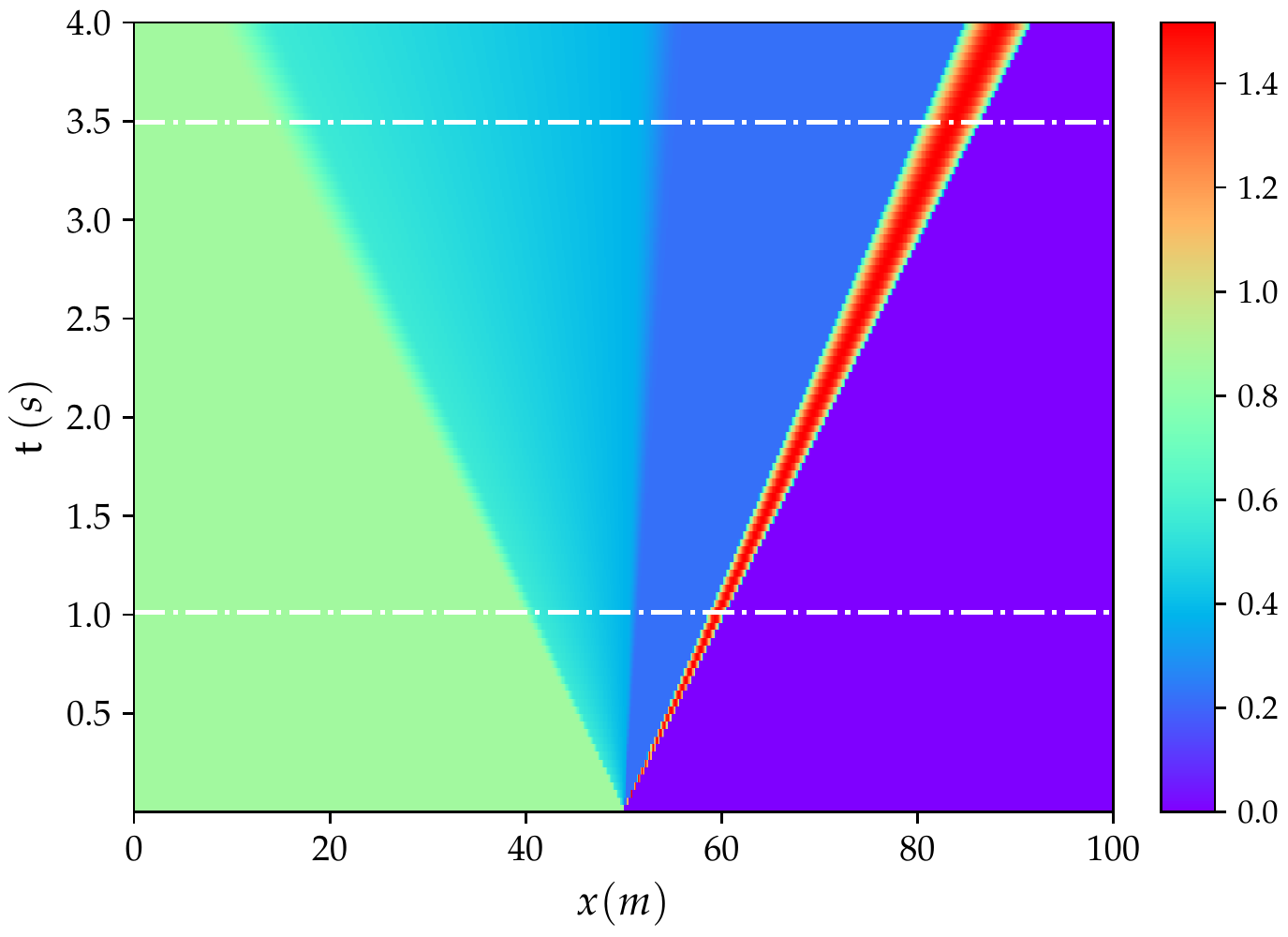}
         \caption{Std (water level)}
         \label{fig:Stok_H_contour_std_lhs_ns_5000}
     \end{subfigure} 
     \begin{subfigure}[b]{0.49\textwidth}
      \centering
        \includegraphics[width=\textwidth]{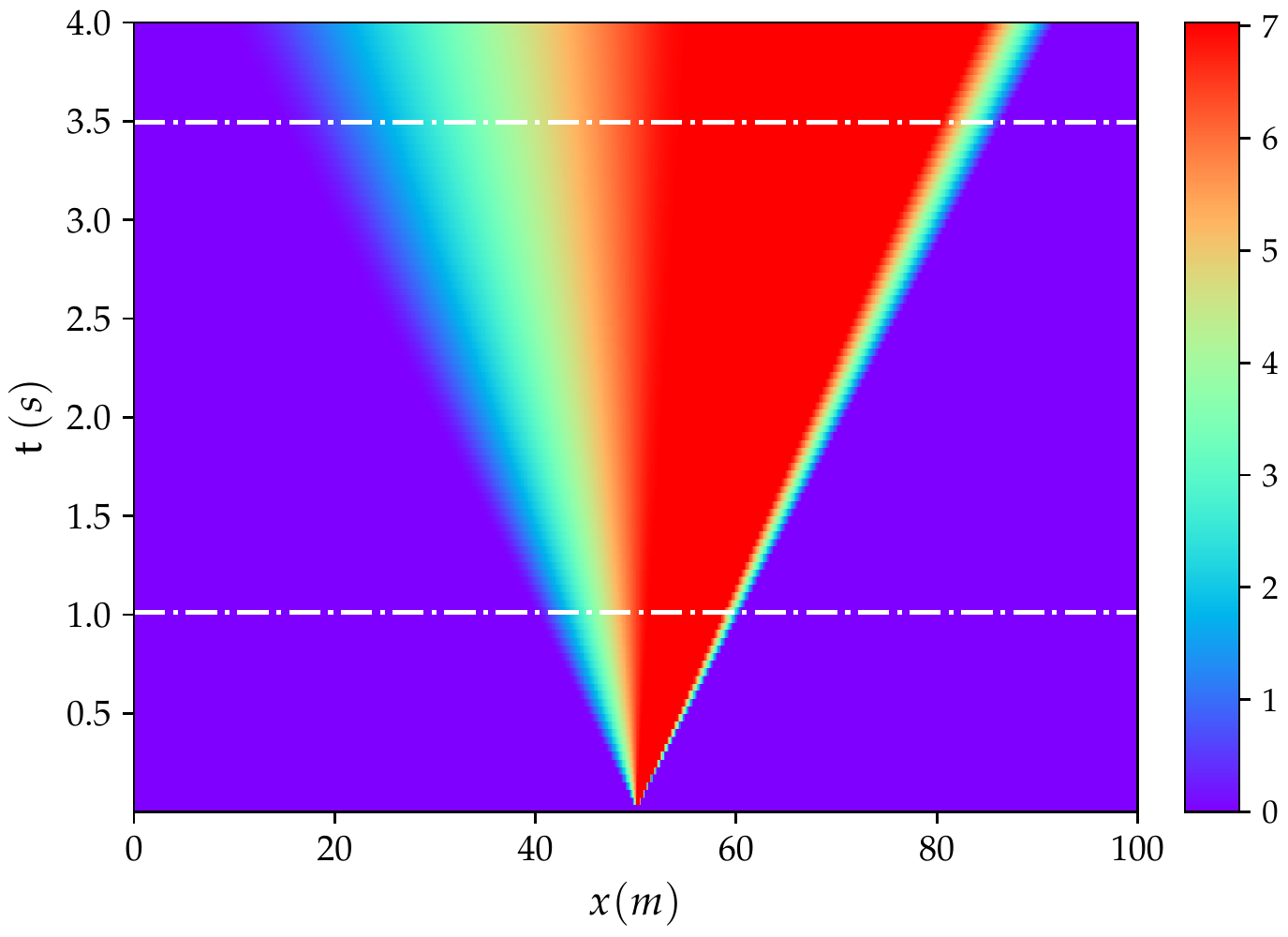}
         \caption{Mean (velocity)}
         \label{fig:Stok_U_contour_mean_lhs_ns_5000}
    \end{subfigure}  
  \hfill
    \begin{subfigure}[b]{0.49\textwidth}
      \centering
        \includegraphics[width=\textwidth]{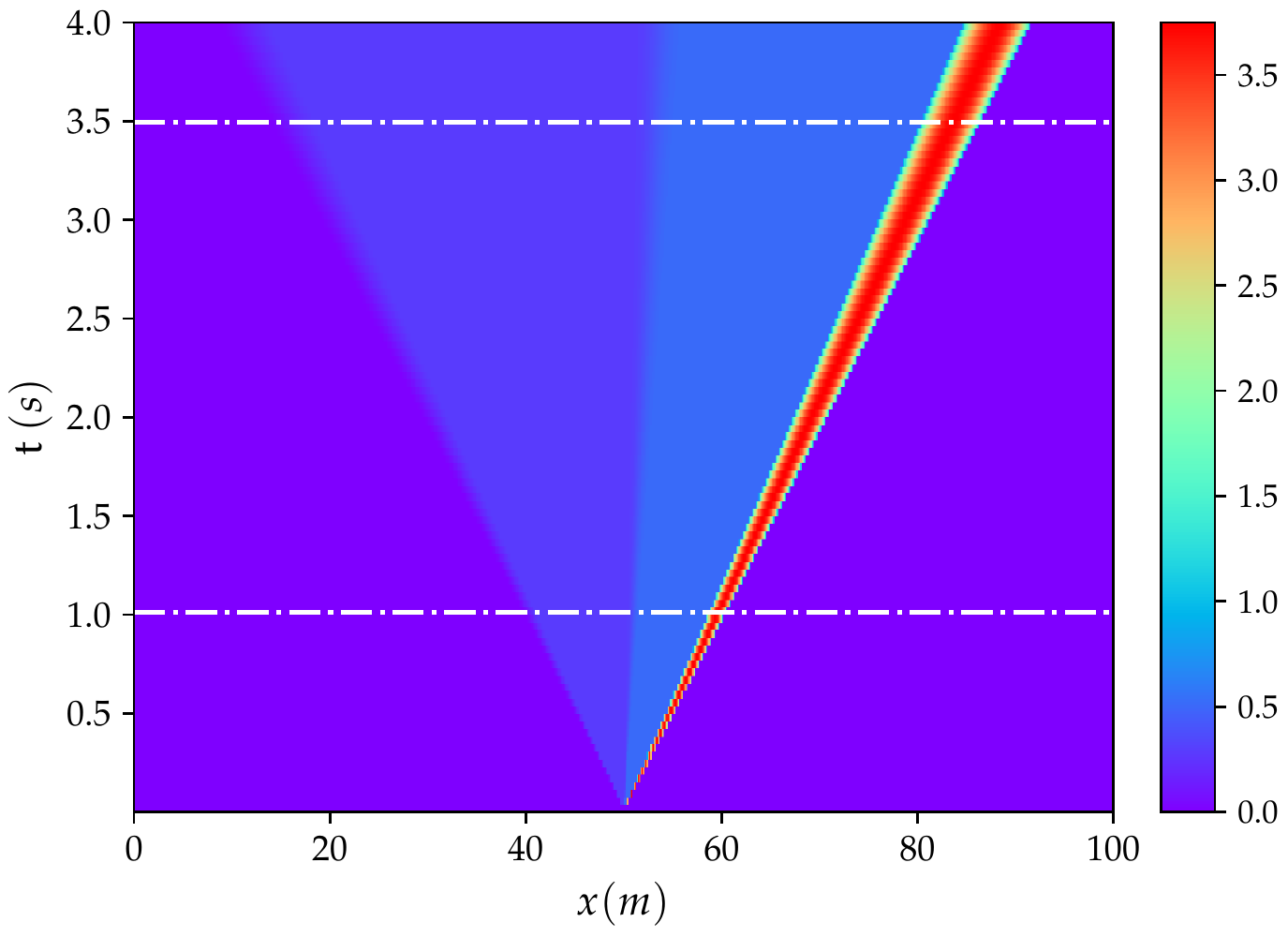}
         \caption{Std (velocity)}
         \label{fig:Stok_U_contour_std_lhs_ns_5000}
     \end{subfigure}
   
 %++++++++++++++++++++++++++++++ 
   %++++++++++++
   \caption{Contour plots of the mean (left column) and standard deviation (right column) of the Stoker's solution of the water level (upper row) and the velocity (bottom row) obtained from the LHS solution with $5\,000$ realizations. The two horizontal dashed lines represent the time locations for which results from POD-ANN and NIROM-CAEs are compared with those from the LHS solution.}
   \label{fig:Stok_mean_std_cont_H_U_lhs_ns_5000}
\end{figure}

Stoker's problem is considered among the most challenging benchmark test cases due to its strong hyperbolic behavior and the discontinuity accompanying the propagation of the front wave resulting from the initial breaking. The performance of the proposed NIROM-CAEs is assessed over this test case by comparing the obtained statistical moments' with those from the LHS reference solution whose contours for the water level and velocity fields are depicted in the Spatio-temporal plane as shown in Fig.\ref{fig:Stok_mean_std_cont_H_U_lhs_ns_5000}, where the longitudinal and transversal axes represent the spatial and temporal dimensions, respectively. The two horizontal lines are presented to show the time locations, $t\approx1$ and $t\approx3.5$ s, respectively, for which the evolution of the mean and standard deviation profiles over the channel length is compared with the LHS and POD-ANN results.\\

%========================================
%++++++++++++++++
%++++++++++++++++
%========================================
Similar to the former numerical test case, the spatial and temporal autoencoders structures are composed of three 1d-convolutional layers with $32$, $64$ and $128$ channels, thus reducing the spatial dimension from $N_{x}=1\,000$ to $L_{x}=50$ and then to $L_{t}=10$, representing the reduced latent space. The obtained spatio-temporal latent space is then linked with the input parameter vector through an MLP mapping. The detailed architectures of the proposed NIROM-CAEs model with its CAE-space, CAE-time, and MLP are summarized in table \ref{tab:Tab_CAE_space_archit_Burg_Stok}, table \ref{tab:Tab_CAE_time_archit_Burg_Stok} and table \ref{tab:MLP_archit}, respectively. The convergence history of the spatial, temporal CAEs and the MLP during the training phase are depicted in Fig.\ref{fig:Conv_hist_Stoker} with the number of epochs of $500$, $1\,000$, and $3\,000$, respectively.\\

%========================================
%++++++++++++++++
\begin{figure}[ht!]
  \centering
    \begin{subfigure}[b]{0.49\textwidth}
      \centering
        \includegraphics[width=\textwidth]{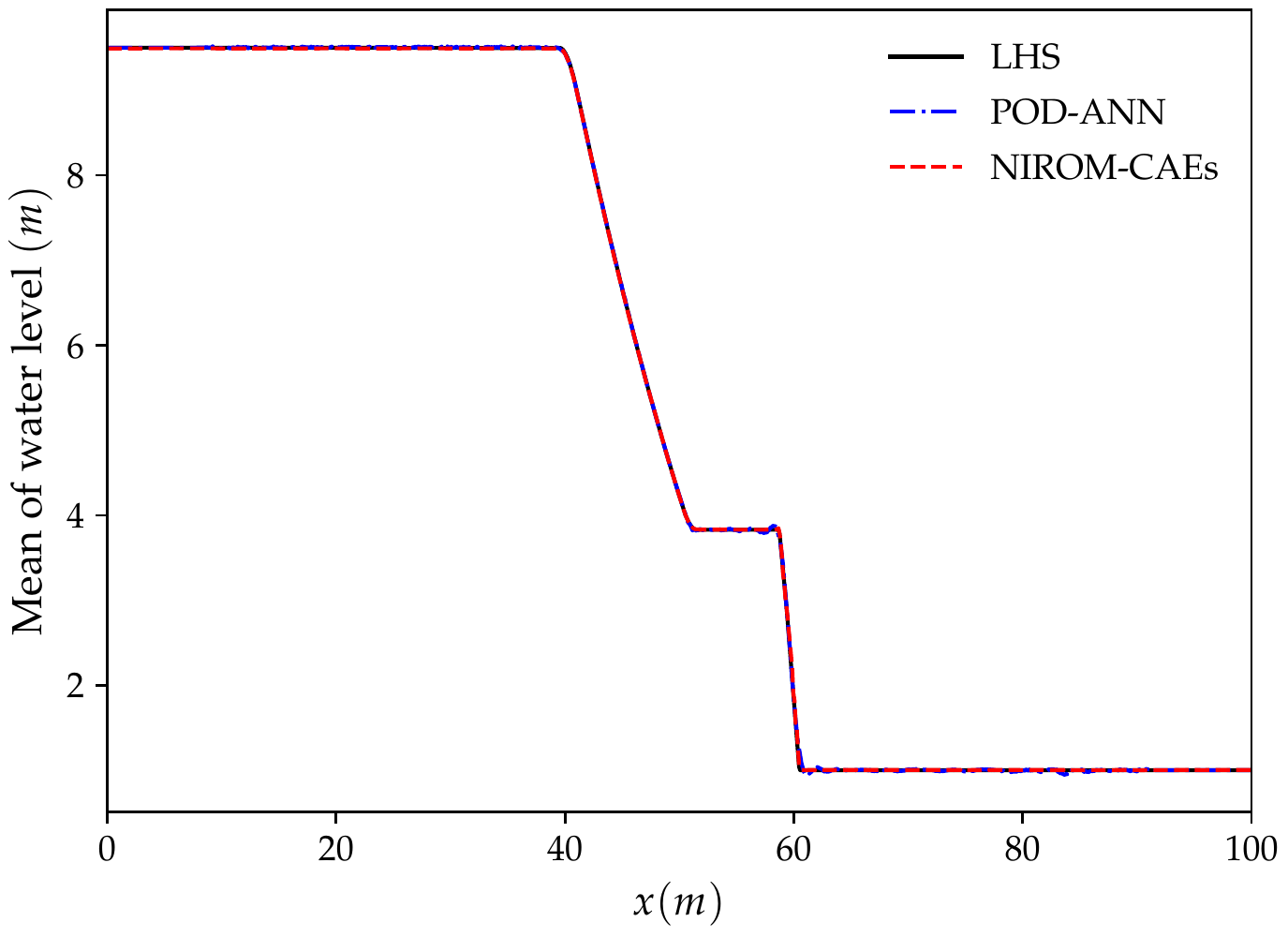}
         \caption{$t\approx 1\,(s)$}
         \label{fig:Mean_Stoker_H_compari_t_1}
    \end{subfigure}  
  \hfill
    \begin{subfigure}[b]{0.49\textwidth}
      \centering
        \includegraphics[width=\textwidth]{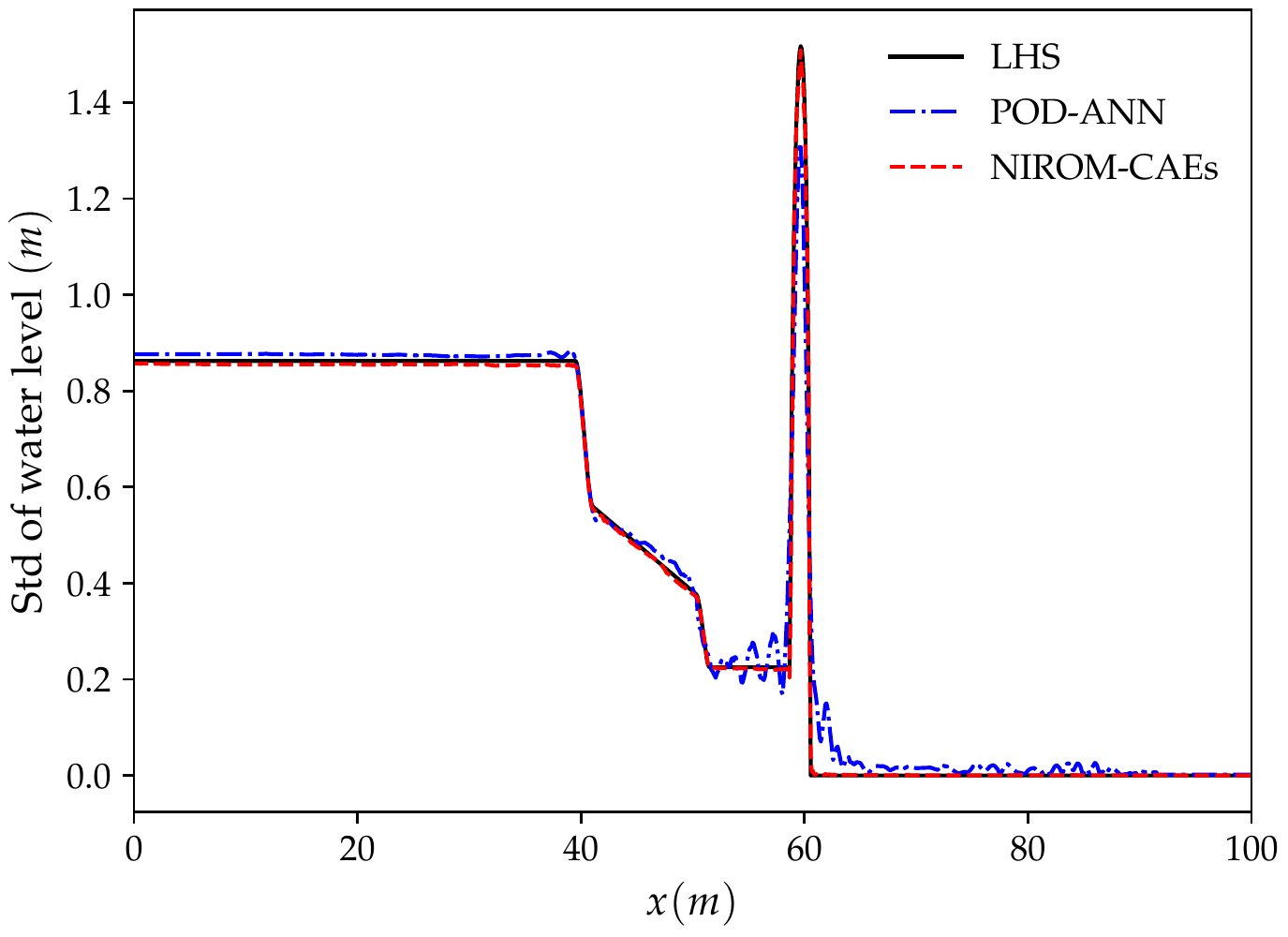}
         \caption{$t\approx 1\,(s)$}
         \label{fig:Std_Stoker_H_compari_t_1}
     \end{subfigure} 
     \begin{subfigure}[b]{0.49\textwidth}
      \centering
        \includegraphics[width=\textwidth]{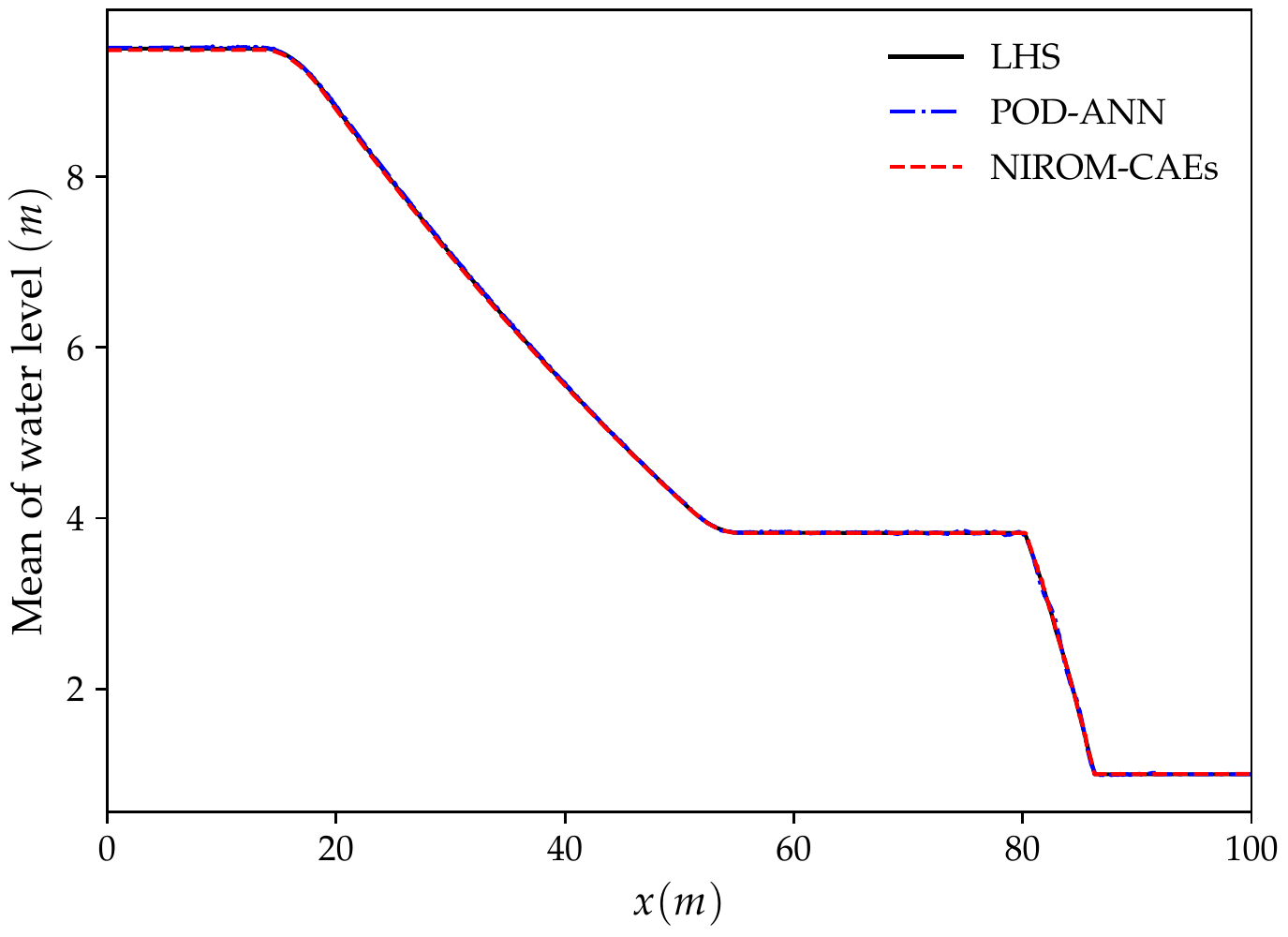}
         \caption{$t\approx 3.5\,(s)$}
         \label{fig:Mean_Stoker_H_compari_3_5}
    \end{subfigure}  
  \hfill
    \begin{subfigure}[b]{0.49\textwidth}
      \centering
        \includegraphics[width=\textwidth]{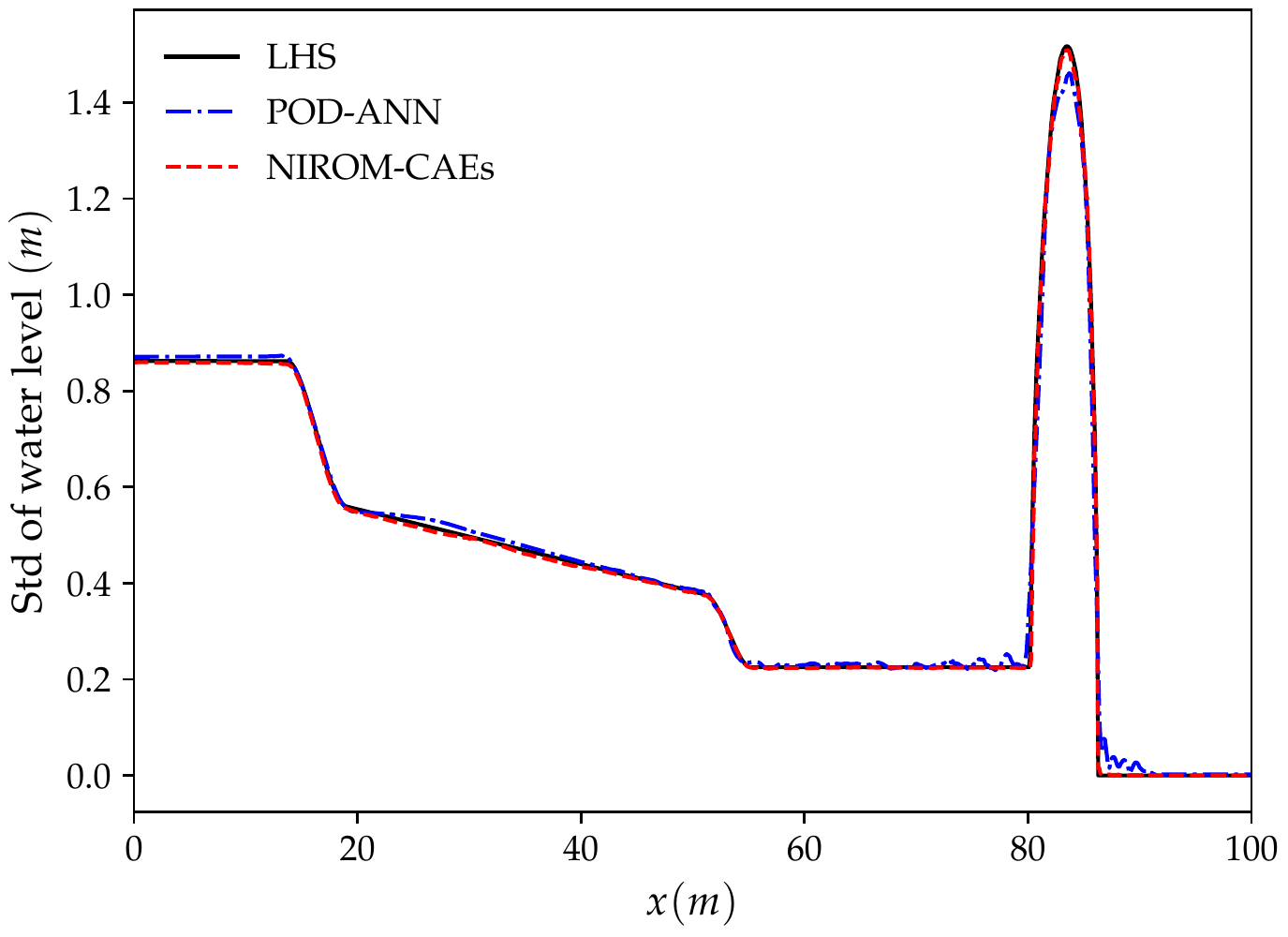}
         \caption{$t\approx 3.5\,(s)$}
         \label{fig:Std_Stoker_H_compari_t_3_5}
     \end{subfigure}
   
 %++++++++++++++++++++++++++++++ 
   %++++++++++++
   \caption{Distribution of the mean (left) and standard deviation (right) of the water level along the channel length at different time steps ($t\approx1$ and $t\approx3.5\; s$). The results obtained from the POD-ANN and NIROM-CAEs techniques are compared with those from the LHS reference solution.}
   \label{fig:Stok_mean_std_compa_H_t_1_and_3_5}
\end{figure}
%++++++++++++++++
%========================================

\begin{figure}[ht!]
  \centering
    \begin{subfigure}[b]{0.49\textwidth}
      \centering
        \includegraphics[width=\textwidth]{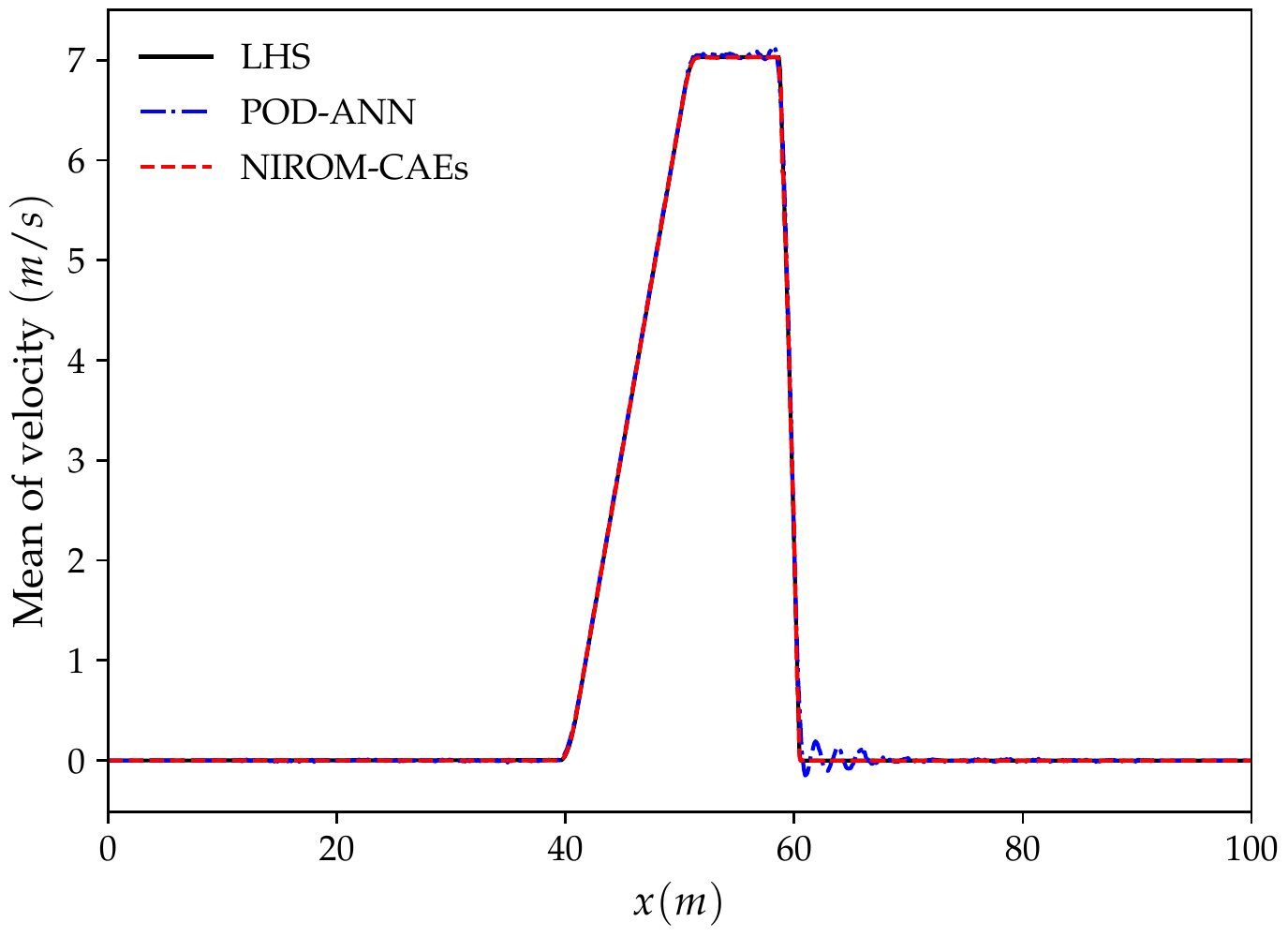}
         \caption{$t\approx 1\,(s)$}
         \label{fig:Mean_Stoker_U_compari_t_1}
    \end{subfigure}  
  \hfill
    \begin{subfigure}[b]{0.49\textwidth}
      \centering
        \includegraphics[width=\textwidth]{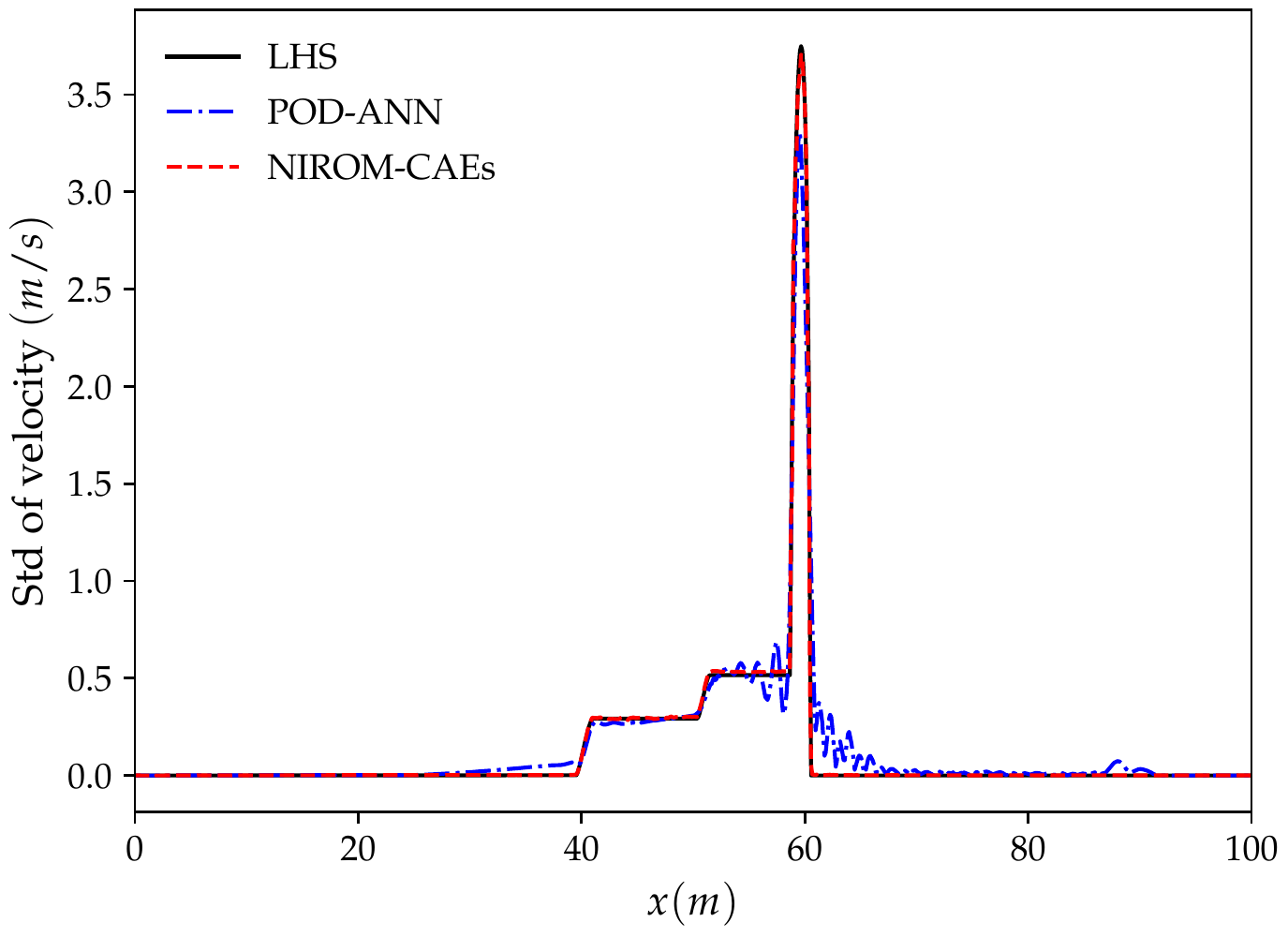}
         \caption{$t\approx 1\,(s)$}
         \label{fig:Std_Stoker_U_compari_t_1}
     \end{subfigure} 
     \begin{subfigure}[b]{0.49\textwidth}
      \centering
        \includegraphics[width=\textwidth]{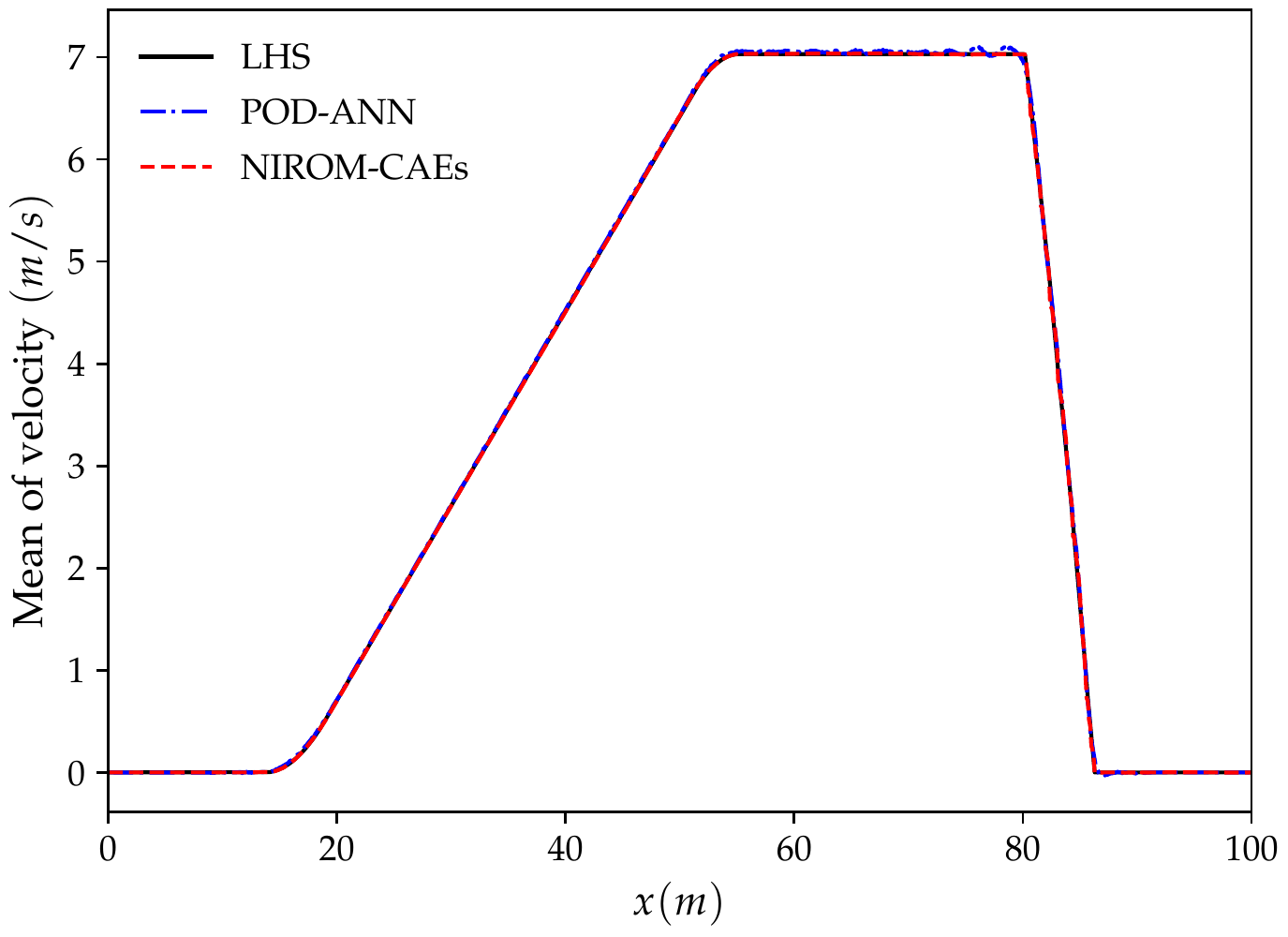}
         \caption{$t\approx 3.5\,(s)$}
         \label{fig:Mean_Stoker_U_compari_3_5}
    \end{subfigure}  
  \hfill
    \begin{subfigure}[b]{0.49\textwidth}
      \centering
        \includegraphics[width=\textwidth]{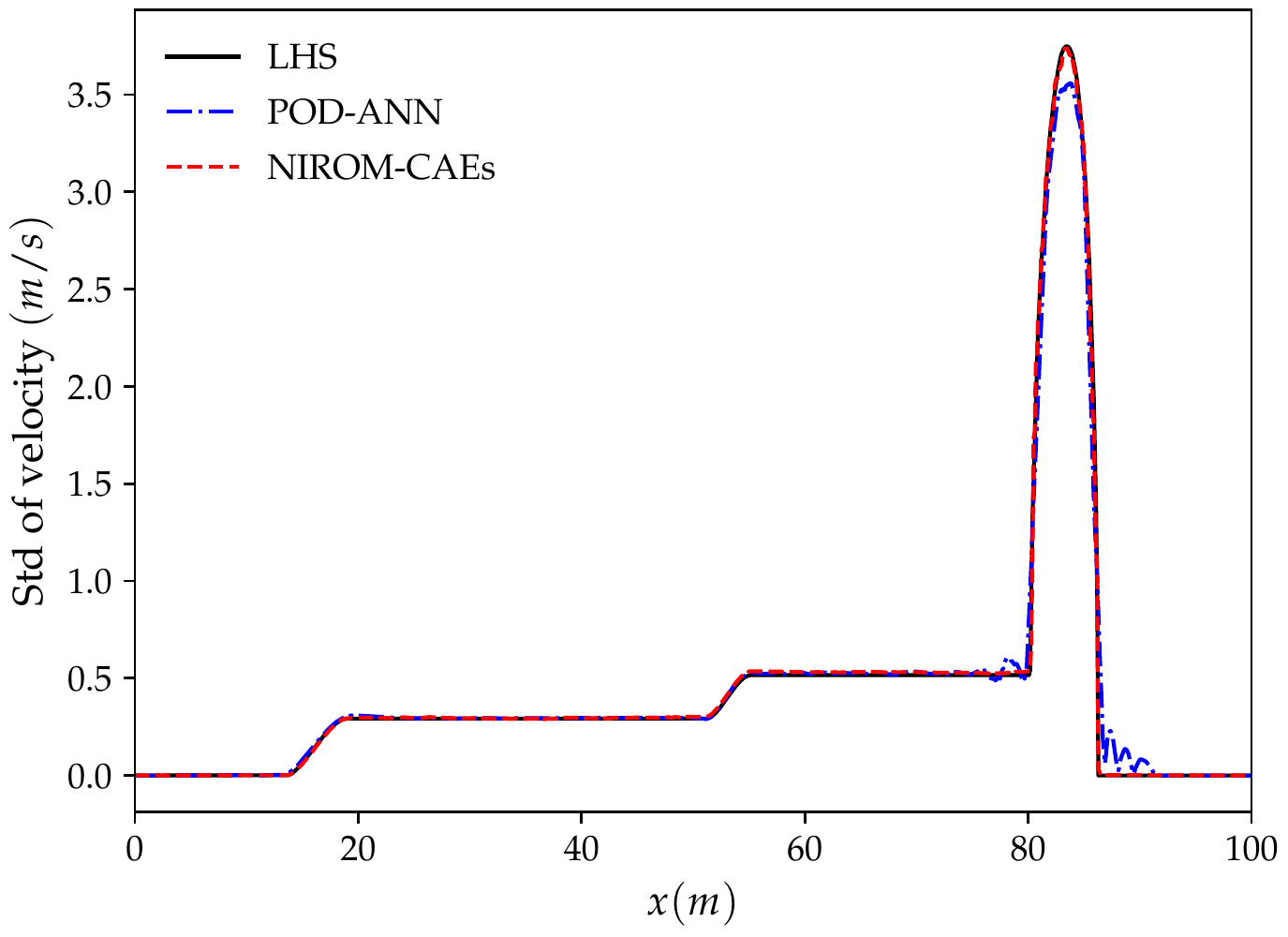}
         \caption{$t\approx 3.5\,(s)$}
         \label{fig:Std_Stoker_U_compari_t_3_5}
     \end{subfigure}
   
 %++++++++++++++++++++++++++++++ 
   %++++++++++++
   \caption{Distribution of the mean (left) and standard deviation (right) of the velocity along the channel length at different time steps ($t\approx1$ and $t\approx3.5\; s$). The results obtained from the POD-ANN and NIROM-CAEs techniques are compared with those from the LHS reference solution.}
   \label{fig:Stok_mean_std_compa_U_t_1_and_3_5}
\end{figure}

The results obtained by the proposed NIROM-CAEs model (presented in terms of the variation of the mean and standard deviation profiles as a function of the $x$-coordinate at different times ($t\approx1,\, 3.5\;s$) are compared with those from the LHS reference solution (which is obtained with $5\,000$ realizations),  as shown in Fig.\ref{fig:Stok_mean_std_compa_H_t_1_and_3_5} and Fig.\ref{fig:Stok_mean_std_compa_U_t_1_and_3_5}. In addition to the LHS solution, the POD-ANN model, known as a linear reduced-order model, is considered another solution with which the predictive ability of the non-linear NIROM-CAEs is assessed. In addition to the LHS solution, the POD-ANN model, known as a linear reduced-order The structure of the neural network introduced in the POD-ANN is constituted by three hidden layers, each of which contains $50$ neurons. A sample set of $N_{s}=300$ values of the upstream water level is randomly generated in its plausible variability range to build the snapshots matrix for both models by collecting the corresponding high-fidelity solutions whose $80\,\%$ is used for training, and the remaining $20\,\%$ for validation.\\

%========================================
%++++++++++++++++
%++++++++++++++++
%========================================
The comparison of the mean and standard deviation profiles of the water level presented in Fig.\ref{fig:Stok_mean_std_compa_H_t_1_and_3_5} shows that both POD-ANN and NIROM-CAEs models predict accurately the distribution of the mean over the whole channel for the two simulation times where a good agreement with the reference LHS profiles can be observed. However, the POD-ANN predictions for the standard deviation exhibit spurious oscillations in the area surrounding the front shock wave, where significant deviations appear in comparison with the profiles from the LHS solution. In contrast, the predicted profiles from the proposed NIROM-CAEs show an excellent agreement with those from the high fidelity LHS solutions, even in the area where the discontinuity occurs. The same trend can be observed in the velocity statistical moments' profiles, as shown in Fig.\ref{fig:Stok_mean_std_compa_U_t_1_and_3_5}, conversely to the approximated outputs from the POD-ANN which are characterized by an oscillatory behavior. It can be concluded that, for this test case,  the use of convolutional neural network autoencoders with non-linear activation functions provides a powerful non-linear compression model with high predictive abilities in capturing the dynamic of the outputs in contrast with the linear compression model based on the traditional POD approach.\\

\begin{figure}[ht!]
  \centering
    \begin{subfigure}[b]{0.49\textwidth}
      \centering
        \includegraphics[width=\textwidth]{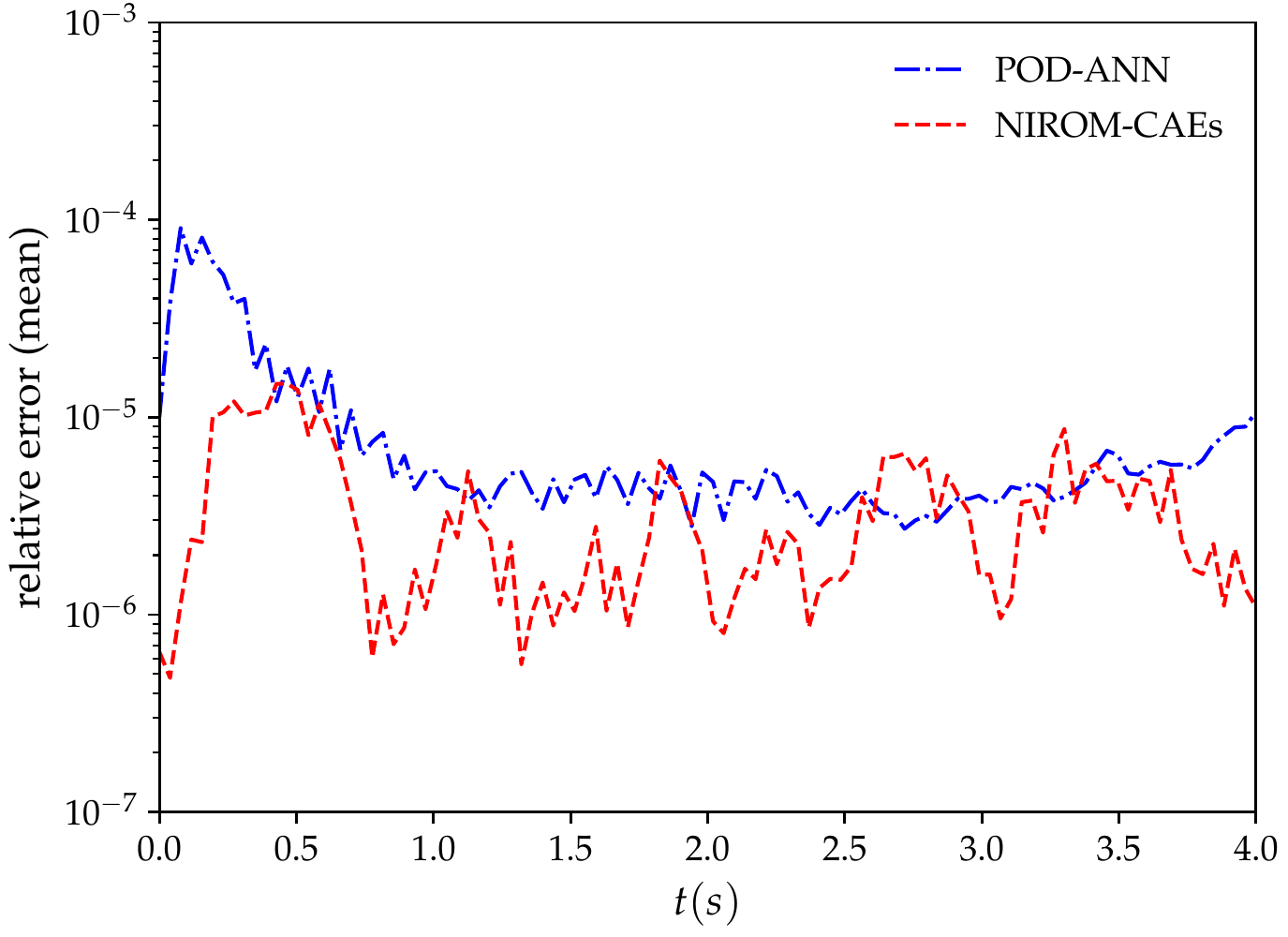}
         \caption{Water level (mean)}
         \label{fig:Error_Mean_Stoker_H_compari}
    \end{subfigure}  
  \hfill
    \begin{subfigure}[b]{0.49\textwidth}
      \centering
        \includegraphics[width=\textwidth]{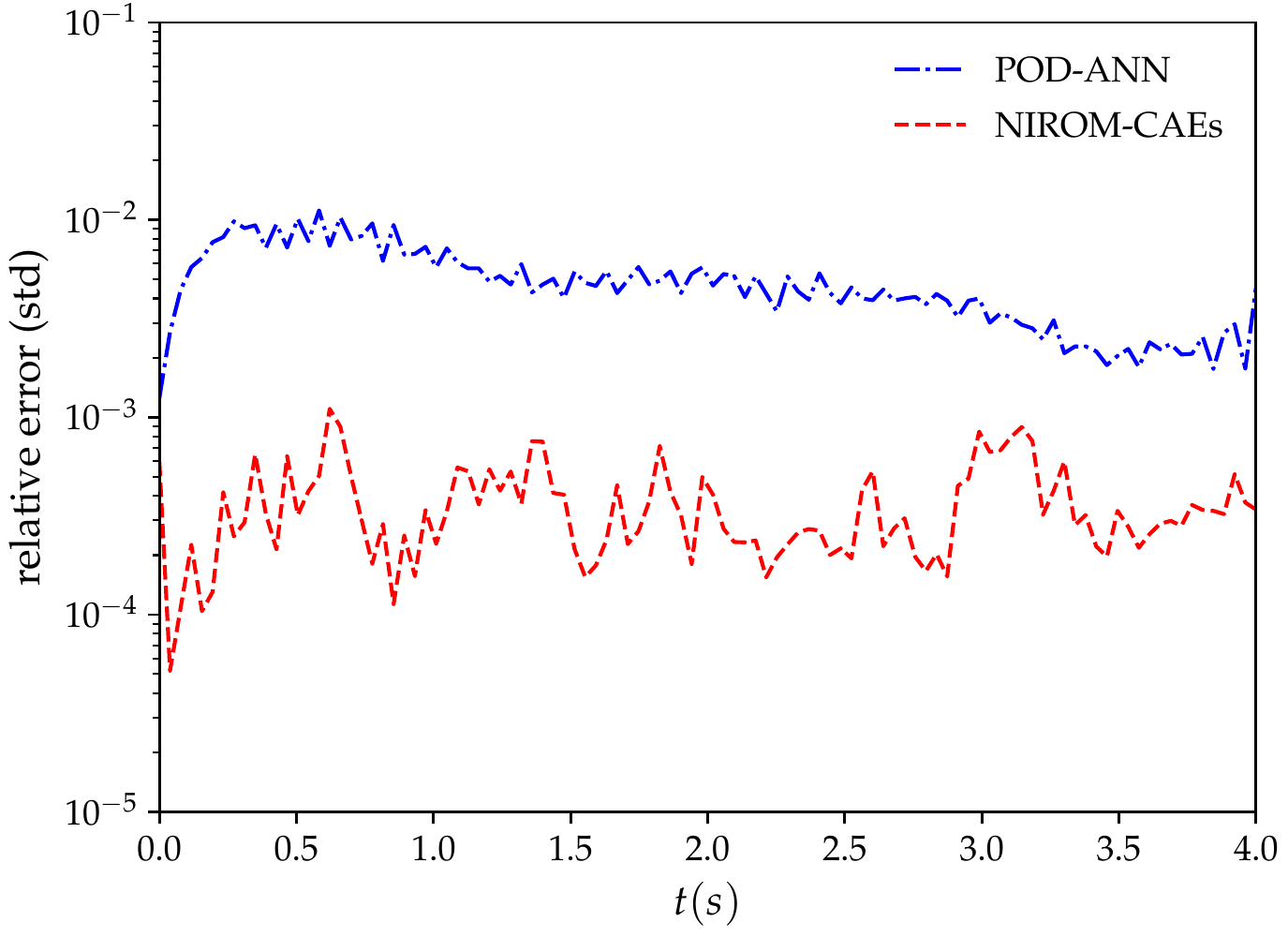}
         \caption{Water level (std)}
         \label{fig:Error_Std_Stoker_H_compari}
     \end{subfigure} 
     \begin{subfigure}[b]{0.49\textwidth}
      \centering
        \includegraphics[width=\textwidth]{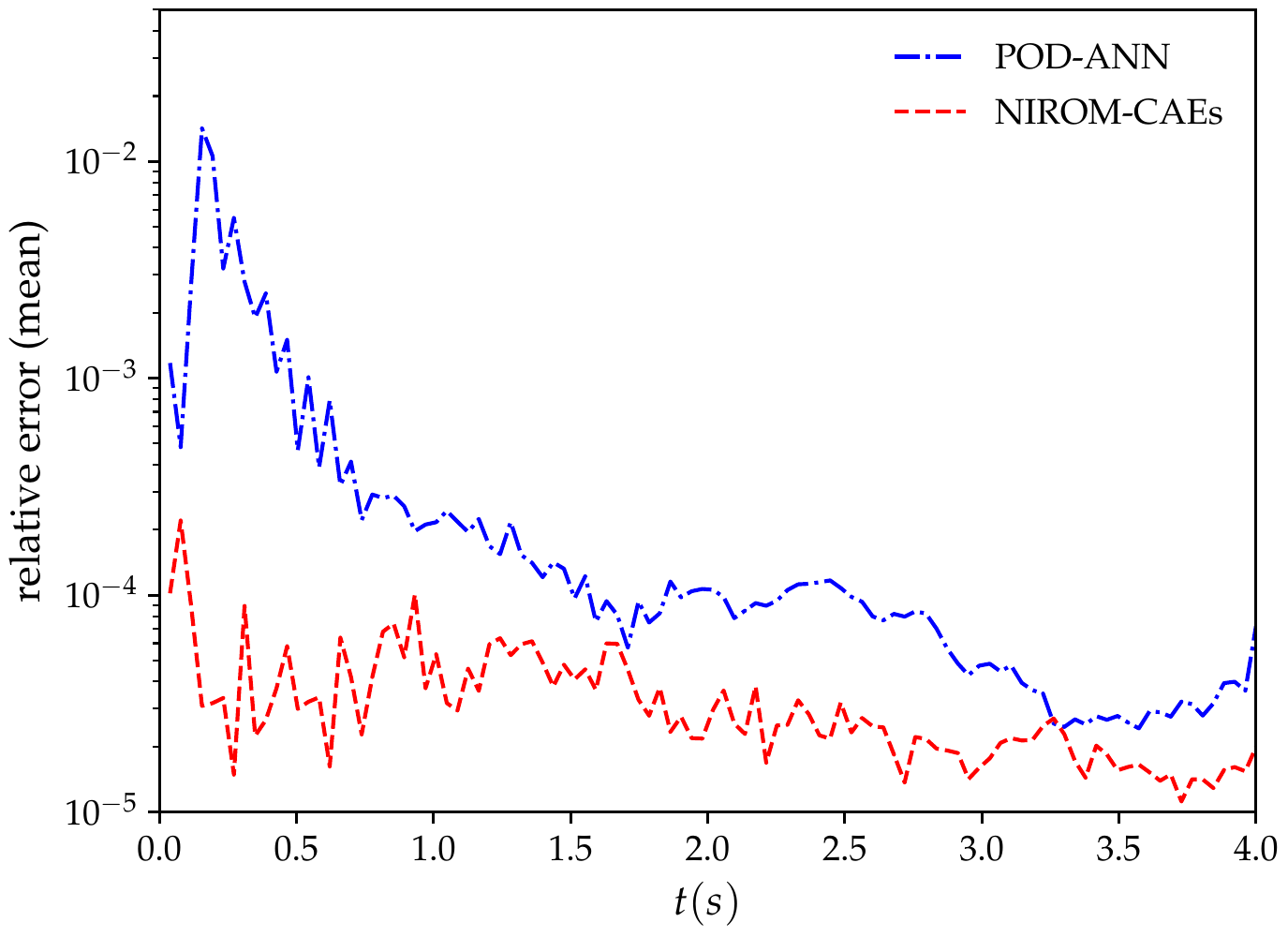}
         \caption{Velocity (mean)}
         \label{fig:Error_Mean_Stoker_U_compari}
    \end{subfigure}  
  \hfill
    \begin{subfigure}[b]{0.49\textwidth}
      \centering
        \includegraphics[width=\textwidth]{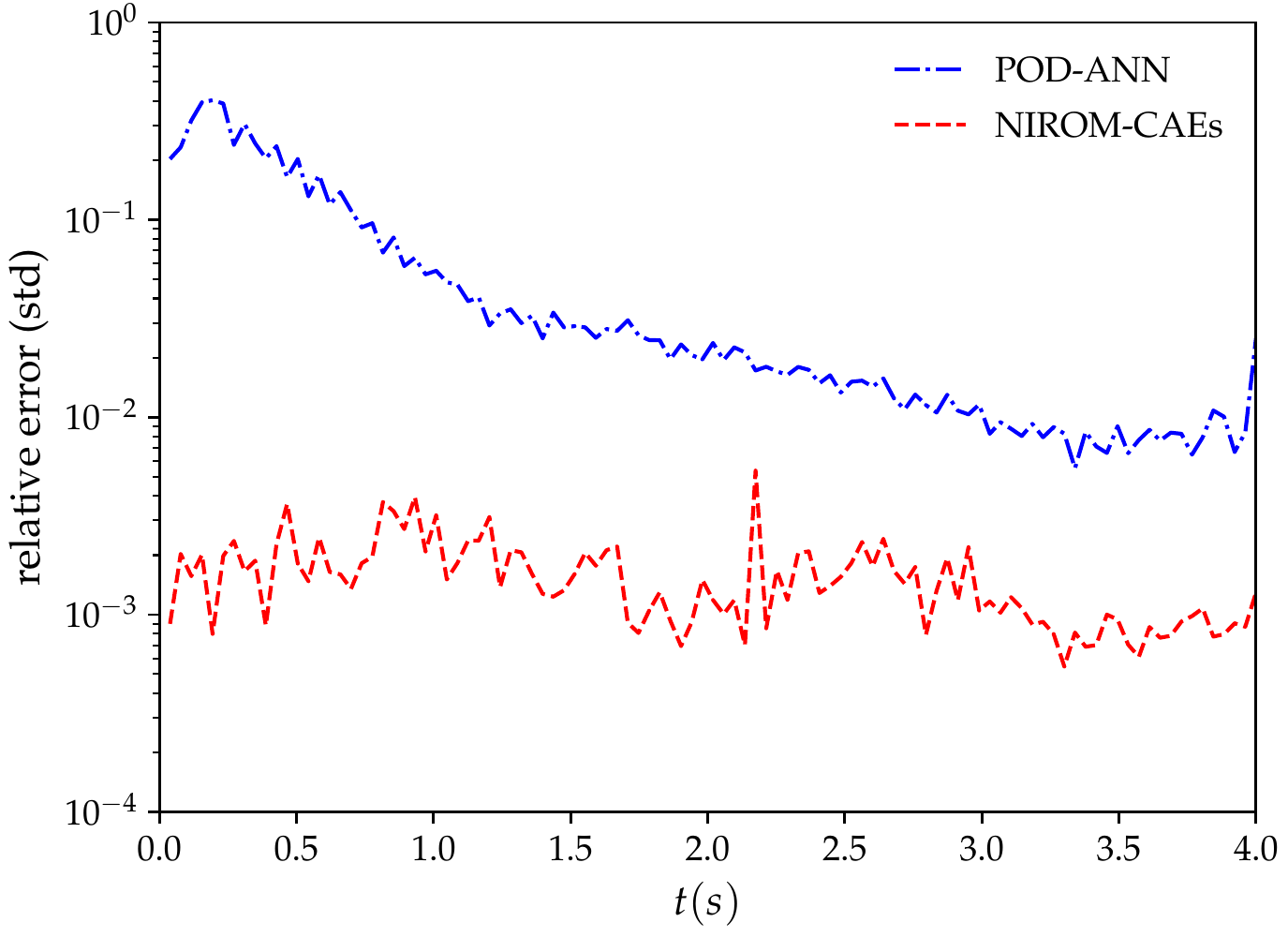}
         \caption{Velocity (std)}
         \label{fig:Error_Std_Stoker_U_compari}
     \end{subfigure}
   
 %++++++++++++++++++++++++++++++ 
   %++++++++++++
   \caption{Comparison of the relative $L^{2}$-error profiles of the mean (left) and standard deviation (right) of the water level (first row) and velocity (second row) as a function of time, obtained with the POD-ANN and NIROM-CAEs. Errors are evaluated with respect to the $5\,000$ realizations LHS reference solution.}
   \label{fig:Error_mean_Std_Stoker_H_U_compari}
\end{figure}

To better illustrate the potential of nonlinear model reduction approaches to efficiently predict the statistical moments of the output responses of high-complexity problems, an evaluation of the relative $L^{2}$ error is performed for both the water level and velocity fields as shown in Fig.\ref{fig:Error_mean_Std_Stoker_H_U_compari}. The mean and the standard deviation errors are calculated for each time by evaluating the difference between the predicted statistical moments with those from the LHS reference ones over the entire calculation domain as formulated by Eq.\eqref{<L2_max>}. The evolution of the relative errors as a function of time shows that the NIROM-CAEs approach presents a better approximation of the output responses with errors of the order of  $10^{-7}$ for the mean and $10^{-4}$ for the standard deviation, unlike the linear POD-ANN technique which presents higher error values that can reach up to two orders of magnitude in the case of the standard deviation for both velocity and water level. This quantitative analysis has shown that the introduction of non-linearities contributes to a better approximation of the statistical moments, thus allowing the construction of efficient spatio-temporal compression models for time-dependent complex problems.    

%========================================
%++++++++++++++++
%++++++++++++++++
%========================================

%+++++++++++++++++++++++++++++++++++++++++++++++++++++++++++
\subsection{Application to a hypothetical dam-break in a river} \label{dam_break}

\begin{figure}[ht!]
  \centering
    \begin{subfigure}[b]{0.49\textwidth}
      \centering
        \includegraphics[width=\textwidth]{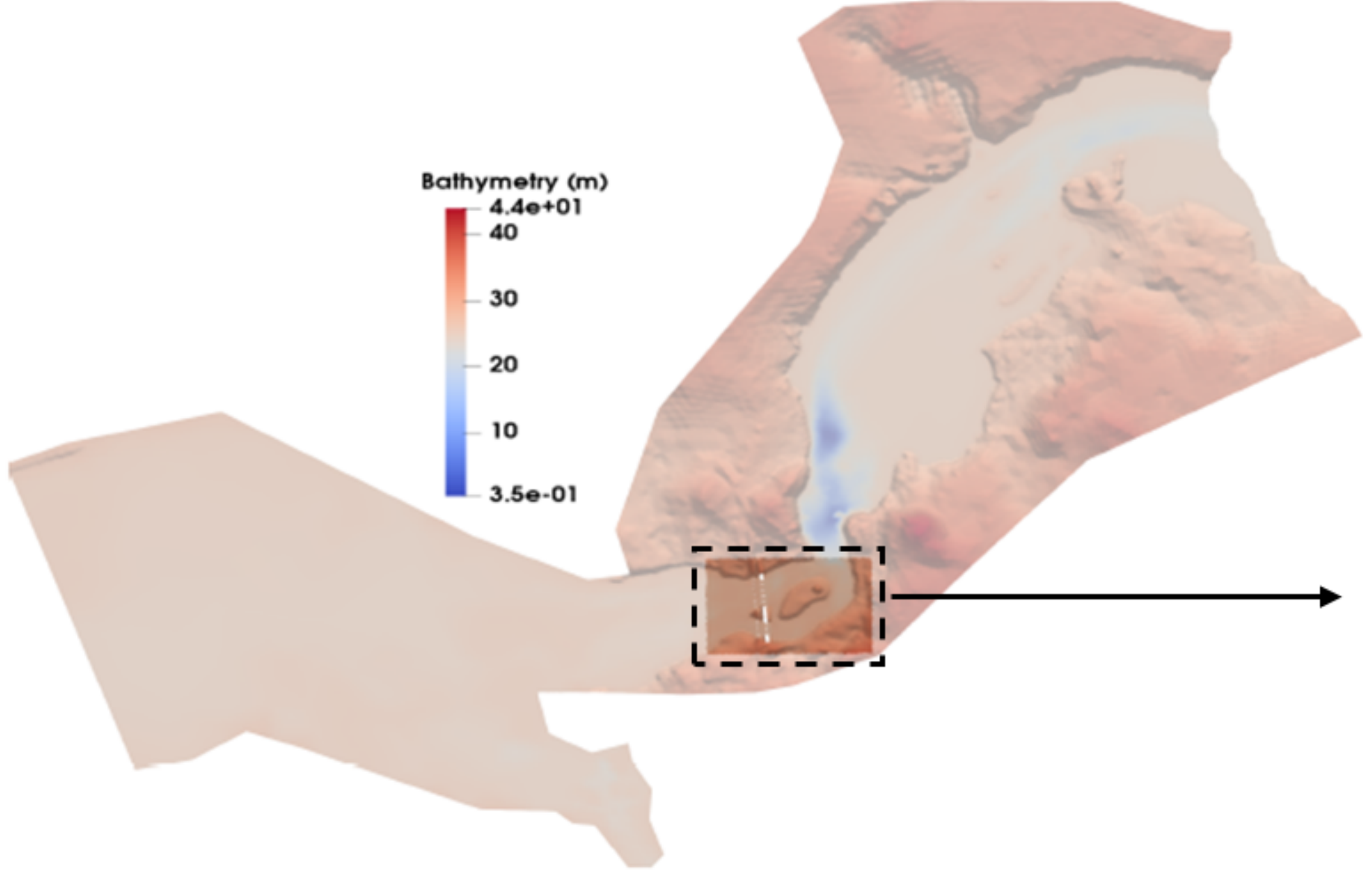}
         \caption{Reach of the Mille-Iles river}
         \label{fig:M_Iles_reach_bathy}
    \end{subfigure}  
  \hfill
    \begin{subfigure}[b]{0.49\textwidth}
      \centering
        \includegraphics[width=\textwidth]{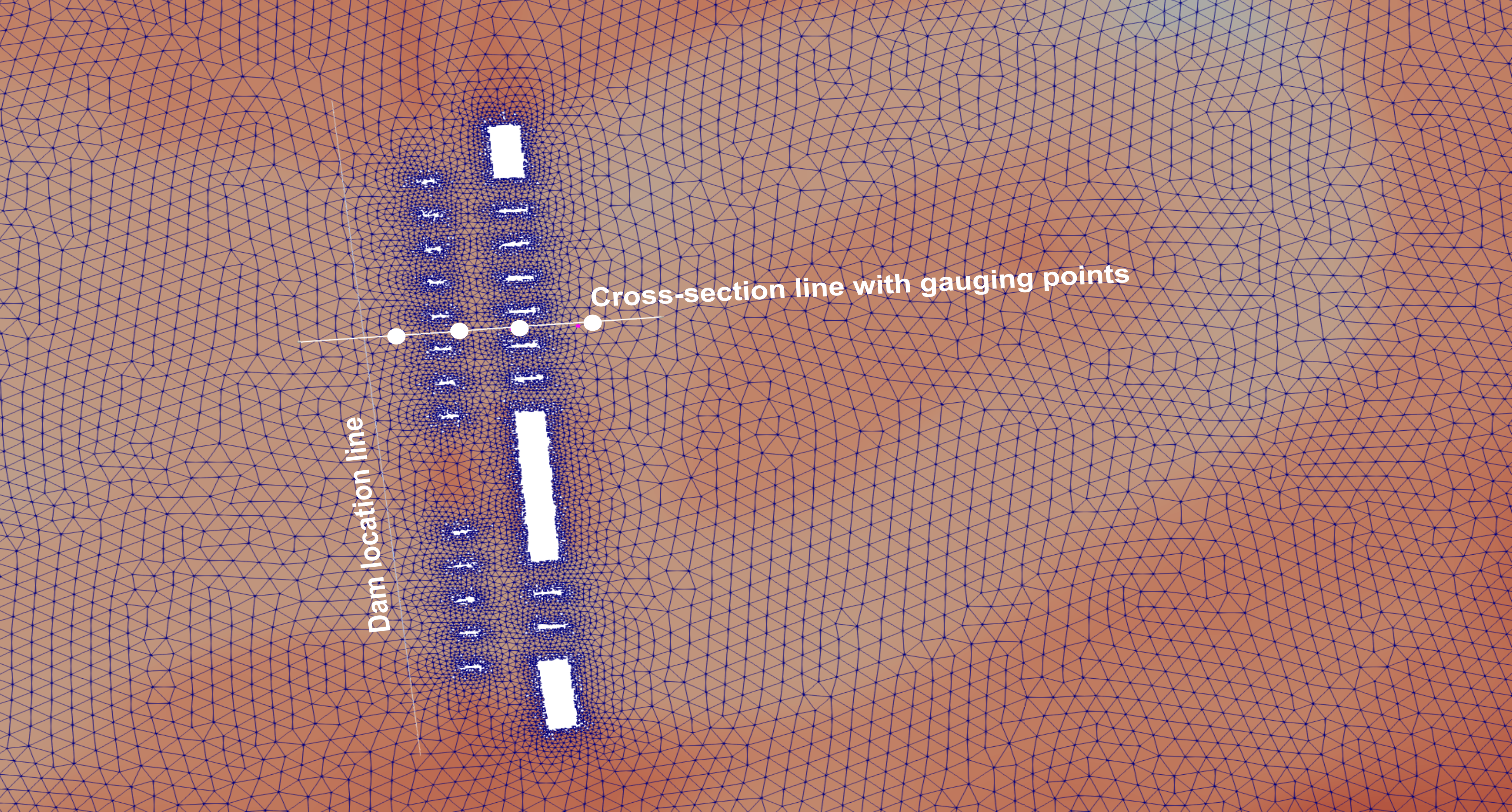}
         \caption{Zoom of the studied sub-domain}
         \label{fig:M_Iles_bathy_zoom}
     \end{subfigure} 
 %++++++++++++++++++++++++++++++ 
   \caption{Sketch of the reach of the Mille Iles river with a close-up view of the studied zone. The cross-section line and the gauging points represent locations where results are represented as a function of the longitudinal direction and time, respectively.}
   \label{fig:M_Iles_domain}
\end{figure}

The proposed approach is applied to a third test case to evaluate its ability to perform an efficient uncertainty propagation analysis on a spatial domain with complex bathymetry. The test concerns a reach of the Milles-Iles River (province of Québec, Canada) including a dam as shown in Fig.\ref{fig:M_Iles_domain}. The data relating to the bathymetry and the roughness coefficient are provided by the Communeauté Métropolitaine de Montréal (CMM) from measurements and observations. The sub-domain of study is composed of an unstructured triangular mesh with $16\,763$ elements and $N_{e}=10\,200$ nodes over which high fidelity solutions of the quantities of interest are collected from an in-house multi-GPU finite volume solver for shallow water equations \citep{delmas2022multi}. A detailed description of the physical domain of this test case was addressed in \citep{abdedou2021non}.

%========================================
%++++++++++++++++++++++++++++++++
%========================================

A fictitious breaching process was initialized from unequal water levels on both parts of a hypotetical dam  located is indicated by a line in Fig.\ref{fig:M_Iles_bathy_zoom}. The downstream part of the dam is considered as dry whereas the free surface of the upstream part is considered as a random input parameter whose values are uniformly generated within its plausible variability range $\eta_{up}\in\mathcal{U}\left[29,\;32\right]\,m$. The snapshot matrix is obtained by running the numerical solver for each value of the upstream free surface, selected randomly from the generated sample set, for the whole $N_{t}=100$ simulation time steps that constitute the temporal domain ($t\in\left[ 0,\, 50\right]\,s$). For each parameter-time combination, a so-called high-fidelity solution is stored in a vector of dimension of $N_{x}=10\,200$ representing the free surface values at each node of the computational domain. The concatenation of these solution vectors thus allows the construction of the snapshot matrix.\\

The spatial compression is performed by the proposed NIROM-CAEs using a space-autoencoder constituted of two 1D-convolutional layers with 32 and 64 channels followed by a max-pooling and dense layers reducing the initial spatial dimension from $N_{x}=10\,200$ nodes to $L_{x}=50$ representing the latent dimension. A second time-autoencoder is then applied to the generated spatial latent space to reduce the temporal dimension from $N_{t}=100$, representing the number of time-steps, to $L_{t}=10$. The time-autoencoder is constituted of a succession of three 1D-convolutional-max pooling layers followed by a dense layer that links the channels to the temporal latent space. The obtained latent space is then linked to the input parameter through an MLP of three layers each one with 128 neurons width to map the values of the input parameter to the final latent space. A detailed description of the architectures of the space, time autoencoders, and MLP are presented in tables \ref{tab:Tab_CAE-space_archit_Miles_Iles}, \ref{tab:Tab_CAE-time_archit_Miles_Iles}, and \ref{tab:MLP_archit}. The numbers of trainable parameters are $6\,915\,197$, $153\,230$ and $34\,570$, respectively. The convergence history of the space, time CAEs and the MLP during the training phase, performed on a Tesla P100 GPU with 32 GB of memory, are depicted in Fig.\ref{fig:Conv_hist_Mille_Iles} with the number of epochs of $200$, $1\,000$ and $2\,000$, respectively.\\

\begin{figure}[ht!]
  \centering
    \begin{subfigure}[b]{0.49\textwidth}
      \centering
        \includegraphics[width=\textwidth]{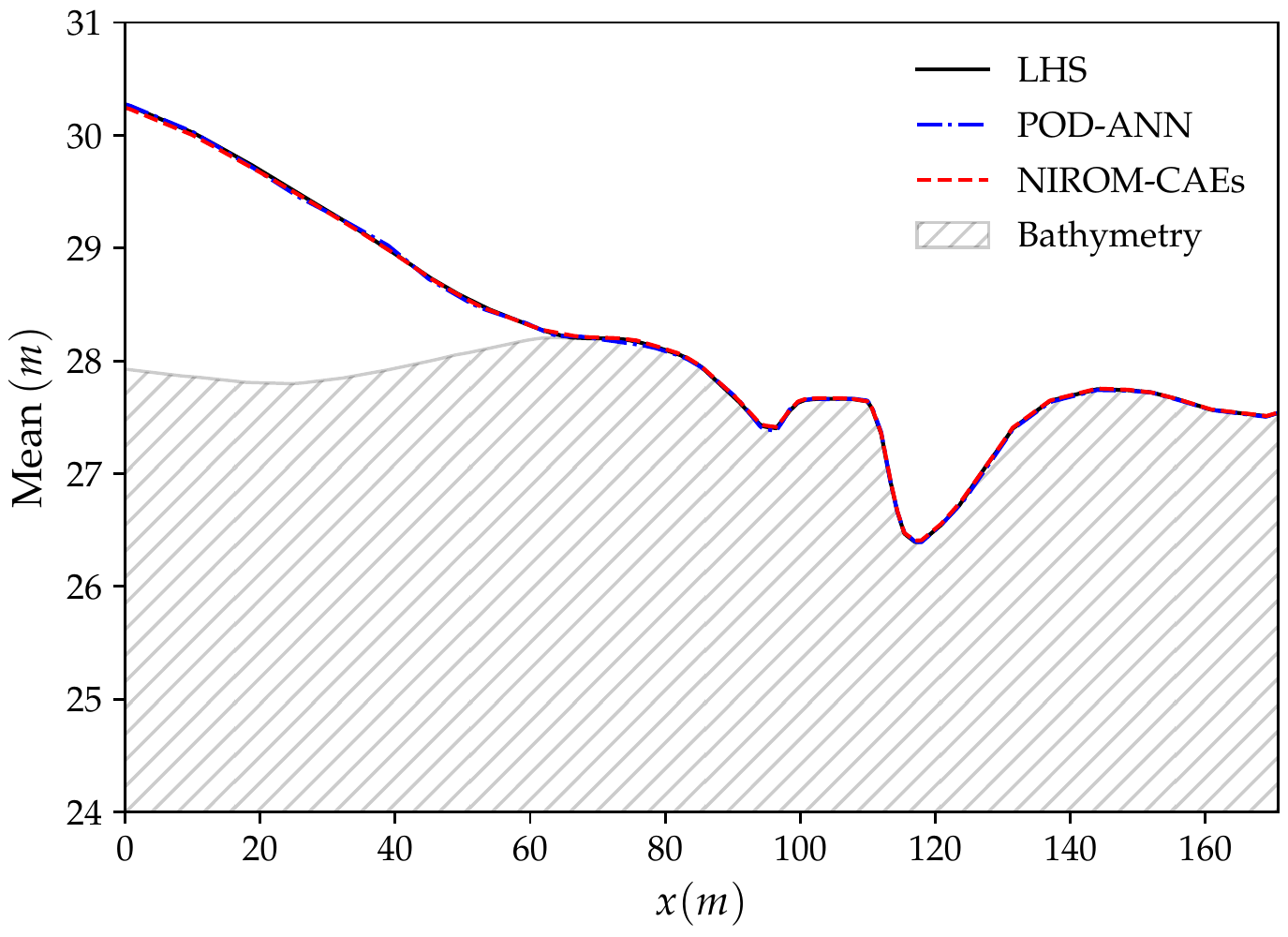}
         \caption{$t\approx\,5\,(s)$}
         \label{fig:Mean_Milles_Iles_compa_t_10}
    \end{subfigure}  
  \hfill
    \begin{subfigure}[b]{0.49\textwidth}
      \centering
        \includegraphics[width=\textwidth]{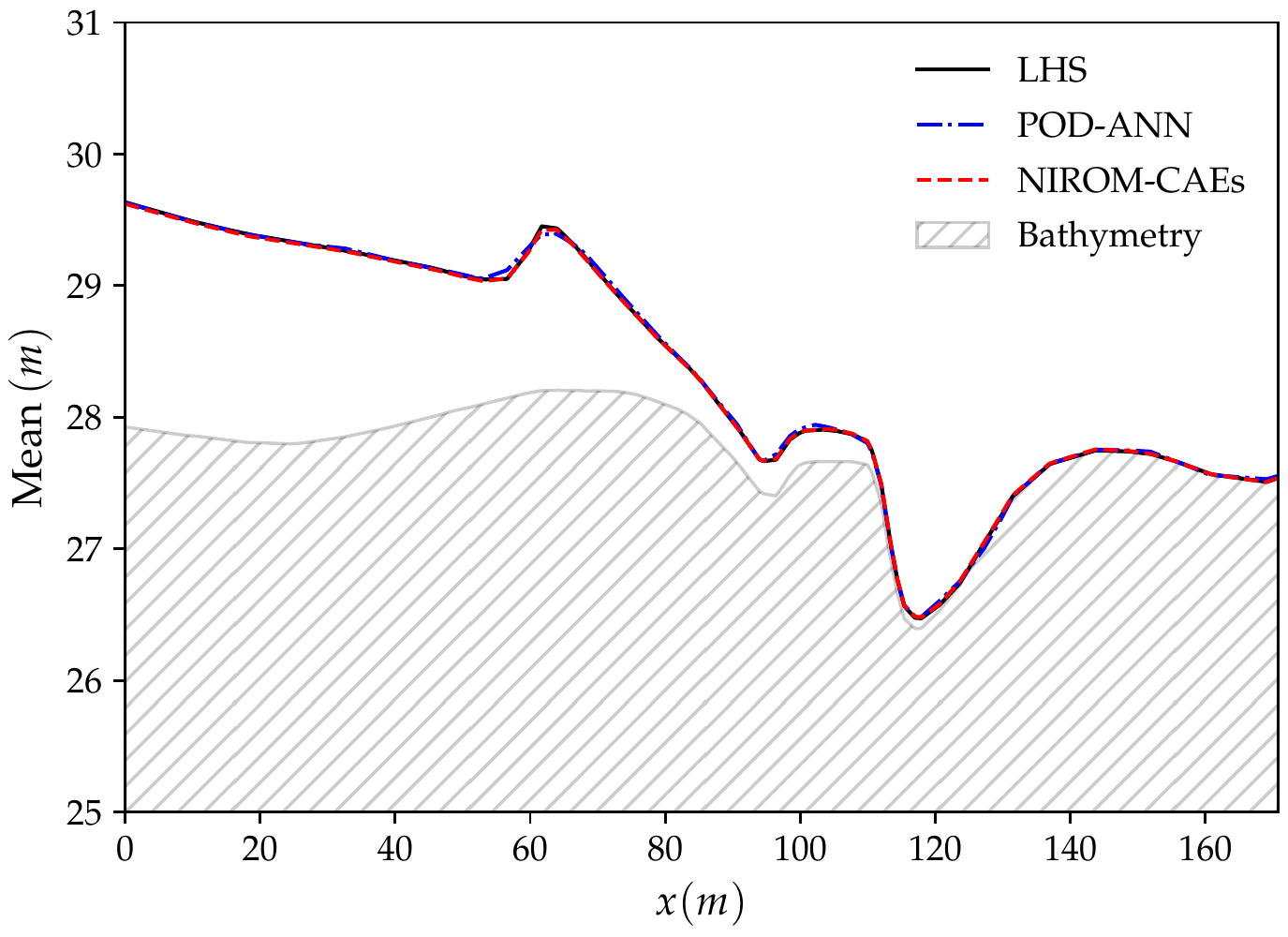}
         \caption{$t\approx\,15\,(s)$}
         \label{fig:Mean_Milles_Iles_compa_t_20}
     \end{subfigure} 
     \begin{subfigure}[b]{0.49\textwidth}
      \centering
        \includegraphics[width=\textwidth]{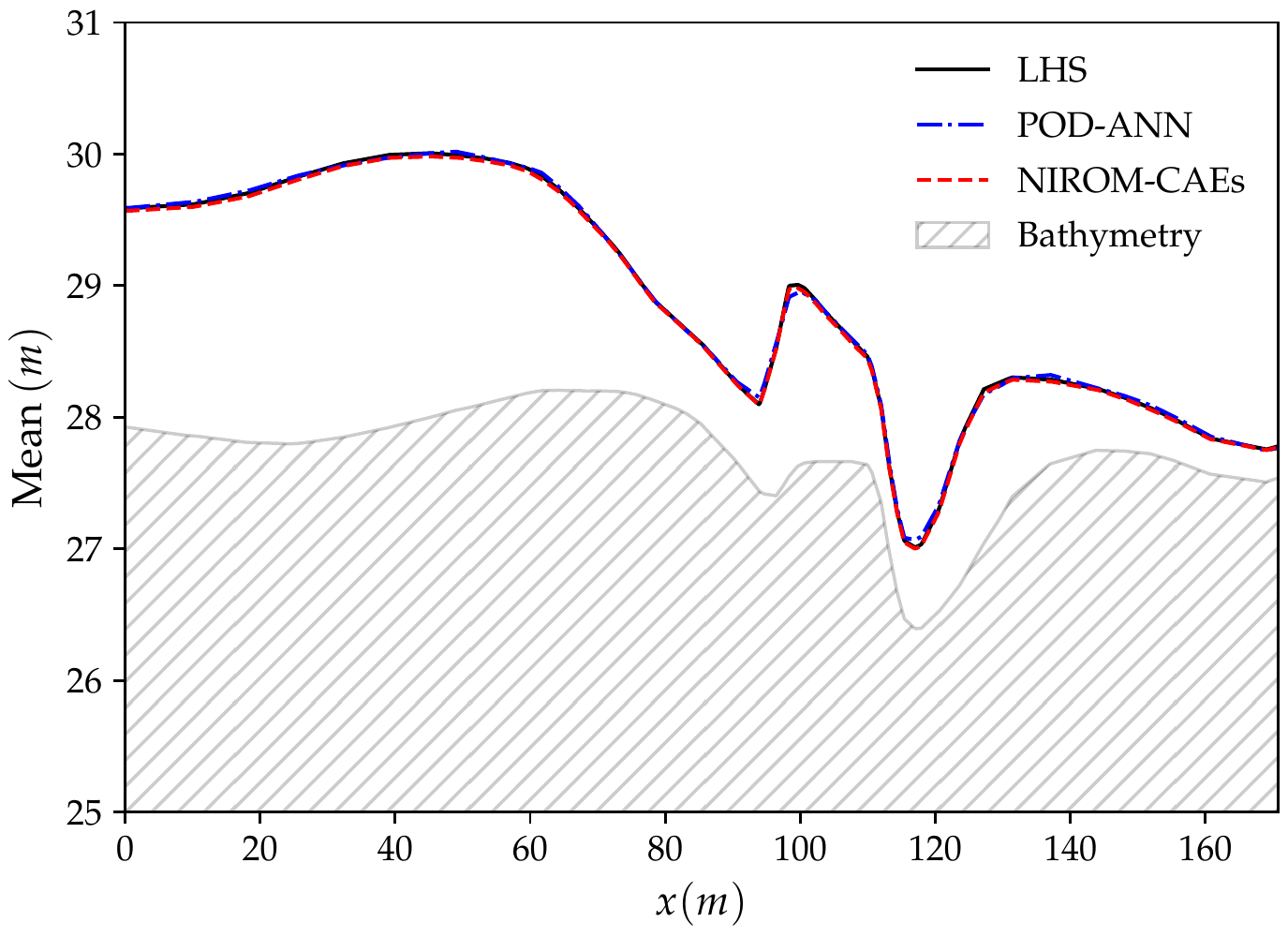}
         \caption{$t\approx\,30\,(s)$}
         \label{fig:Mean_Milles_Iles_compa_t_30}
    \end{subfigure}  
  \hfill
    \begin{subfigure}[b]{0.49\textwidth}
      \centering
        \includegraphics[width=\textwidth]{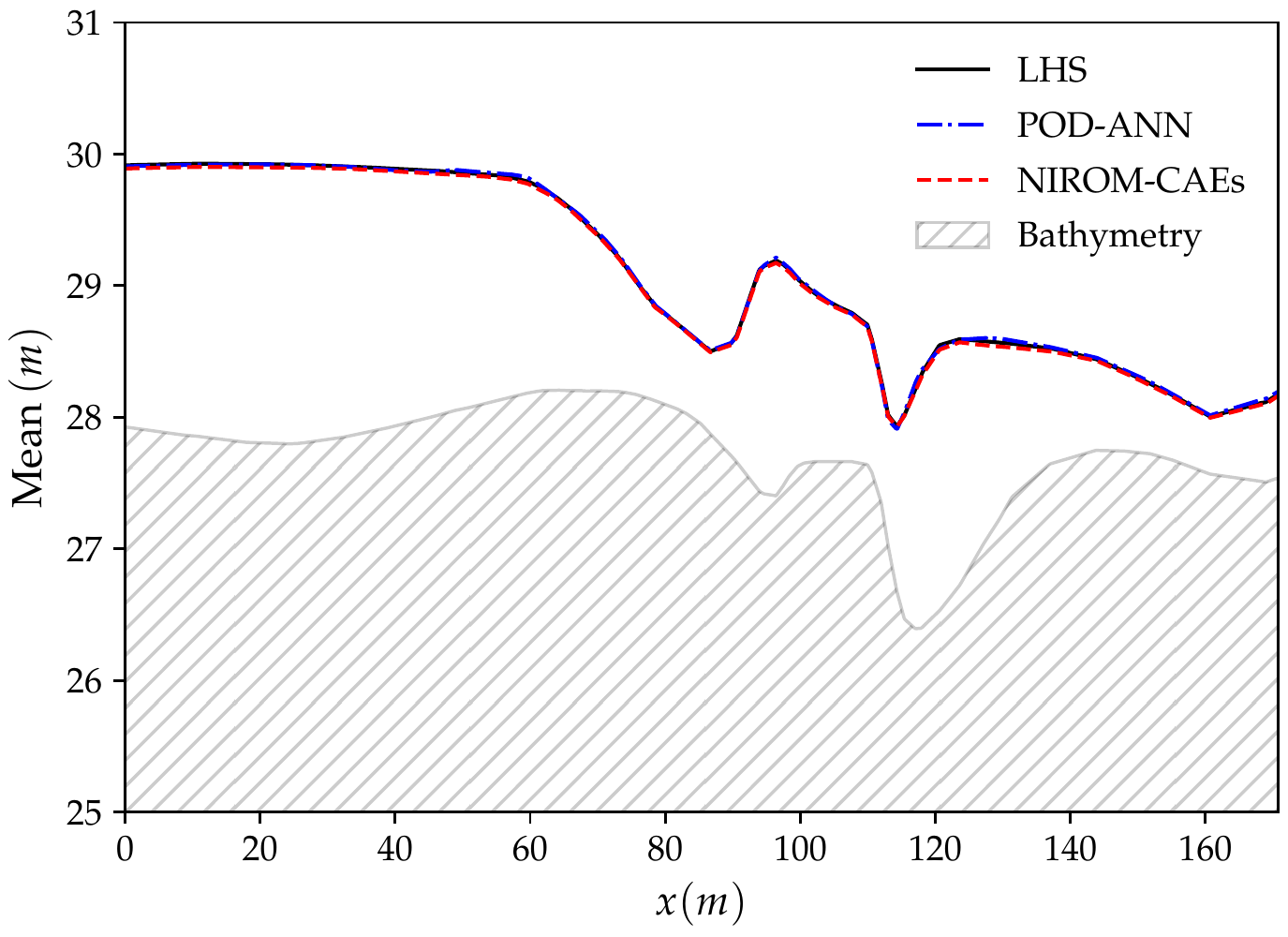}
         \caption{$t\approx\,45\,(s)$}
         \label{fig:Mean_Milles_Iles_compa_t_45}
     \end{subfigure}
   
 %++++++++++++++++++++++++++++++ 
   %++++++++++++
   \caption{Comparison of the mean profiles of the water level over the cross-section line at various time-steps, obtained with the POD-ANN and NIROM-CAEs with those from the LHS reference solution (with $2\,000$ realizations). (a): $t\approx 5\,s$, (b): $t\approx 15\,s$, (c): $t\approx 30\,s$ and (d): $t\approx 45\,s$.}
   \label{fig:Mean_Milles_Iles_compa_diff_times}
\end{figure}

\begin{figure}[ht!]
  \centering
    \begin{subfigure}[b]{0.49\textwidth}
      \centering
        \includegraphics[width=\textwidth]{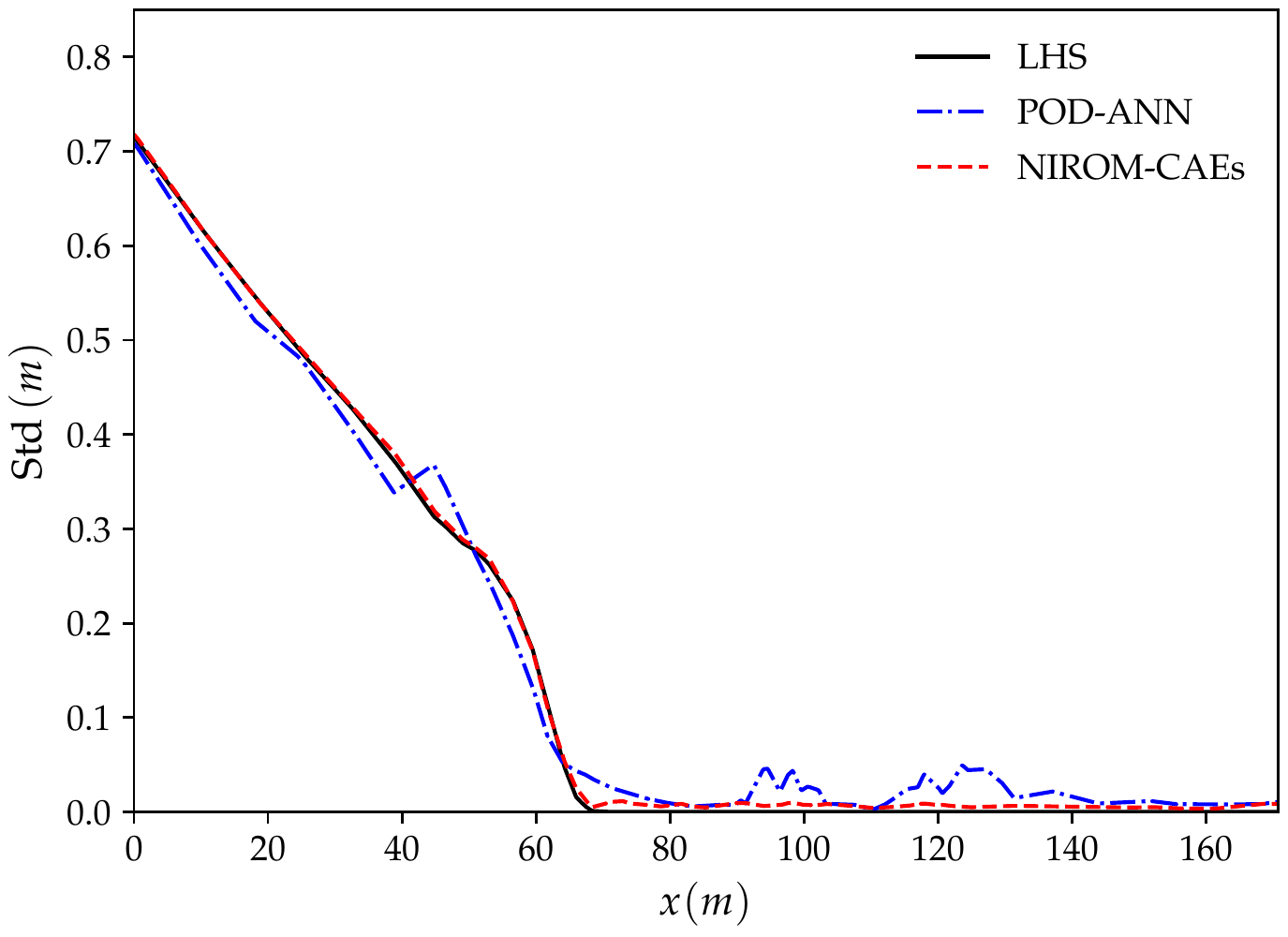}
         \caption{$t\approx\,5\,(s)$}
         \label{fig:Std_Milles_Iles_compa_t_10}
    \end{subfigure}  
  \hfill
    \begin{subfigure}[b]{0.49\textwidth}
      \centering
        \includegraphics[width=\textwidth]{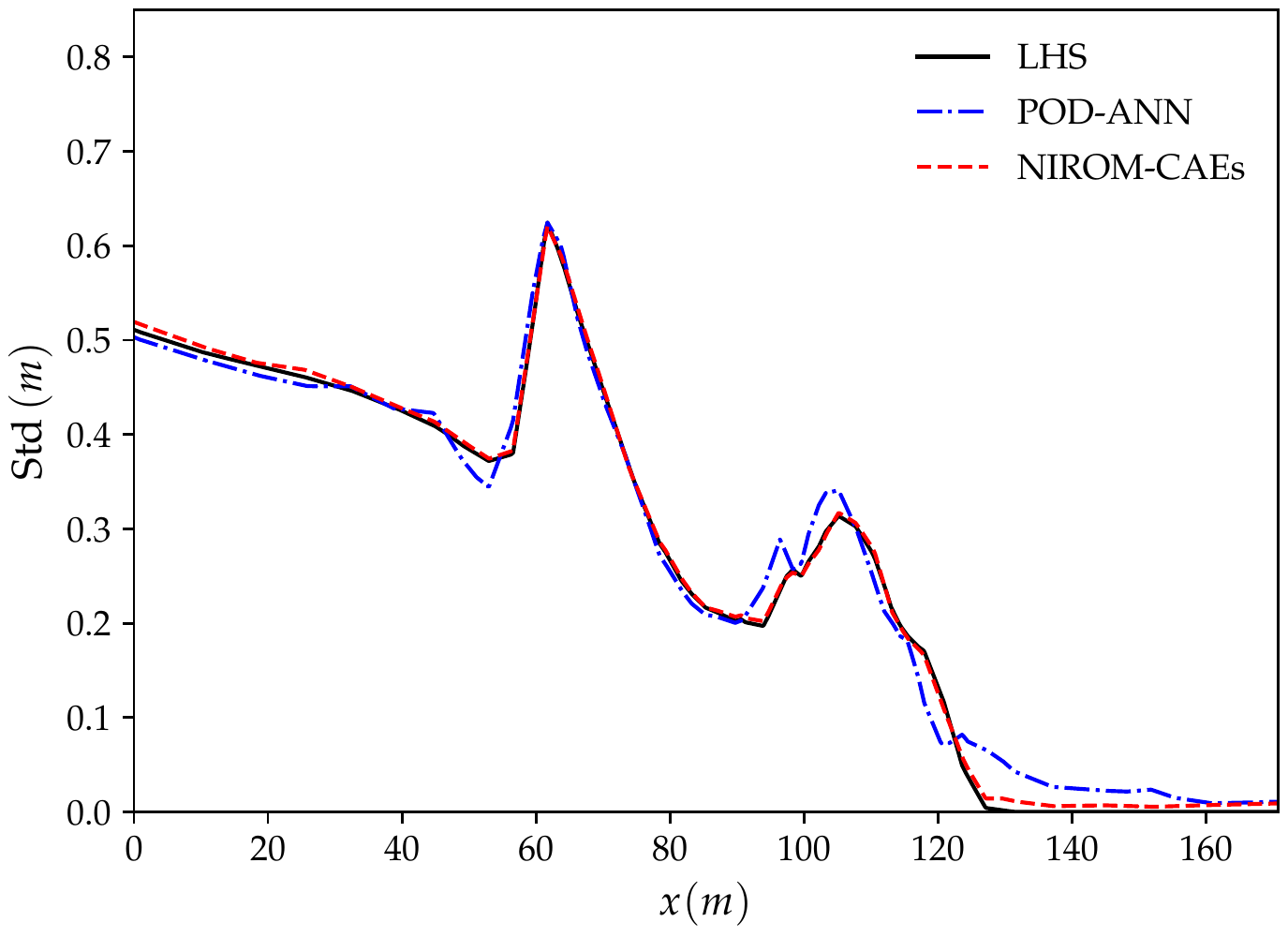}
         \caption{$t\approx\,15\,(s)$}
         \label{fig:Std_Milles_Iles_compa_t_20}
     \end{subfigure} 
     \begin{subfigure}[b]{0.49\textwidth}
      \centering
        \includegraphics[width=\textwidth]{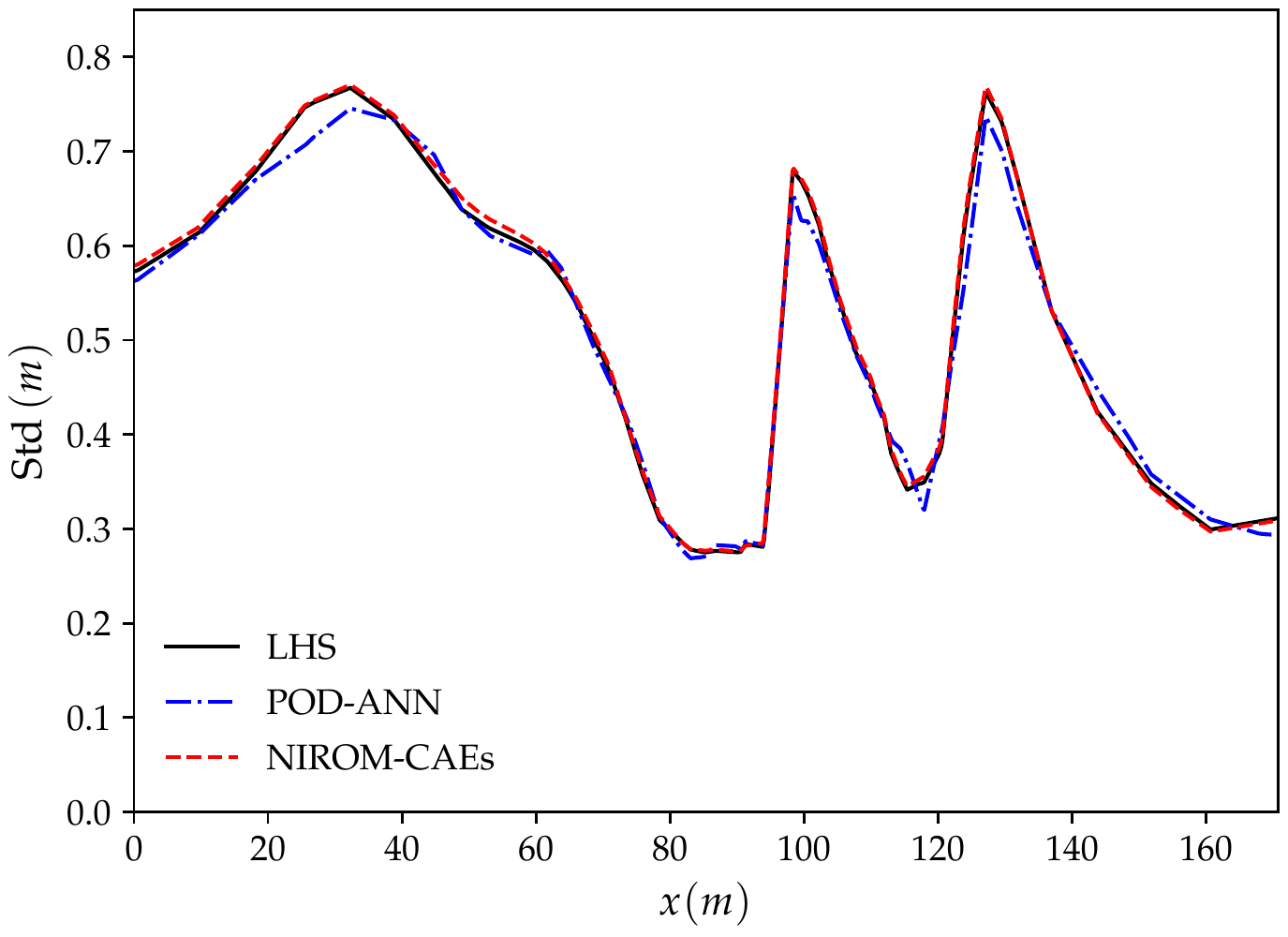}
         \caption{$t\approx\,30\,(s)$}
         \label{fig:Std_Milles_Iles_compa_t_30}
    \end{subfigure}  
  \hfill
    \begin{subfigure}[b]{0.49\textwidth}
      \centering
        \includegraphics[width=\textwidth]{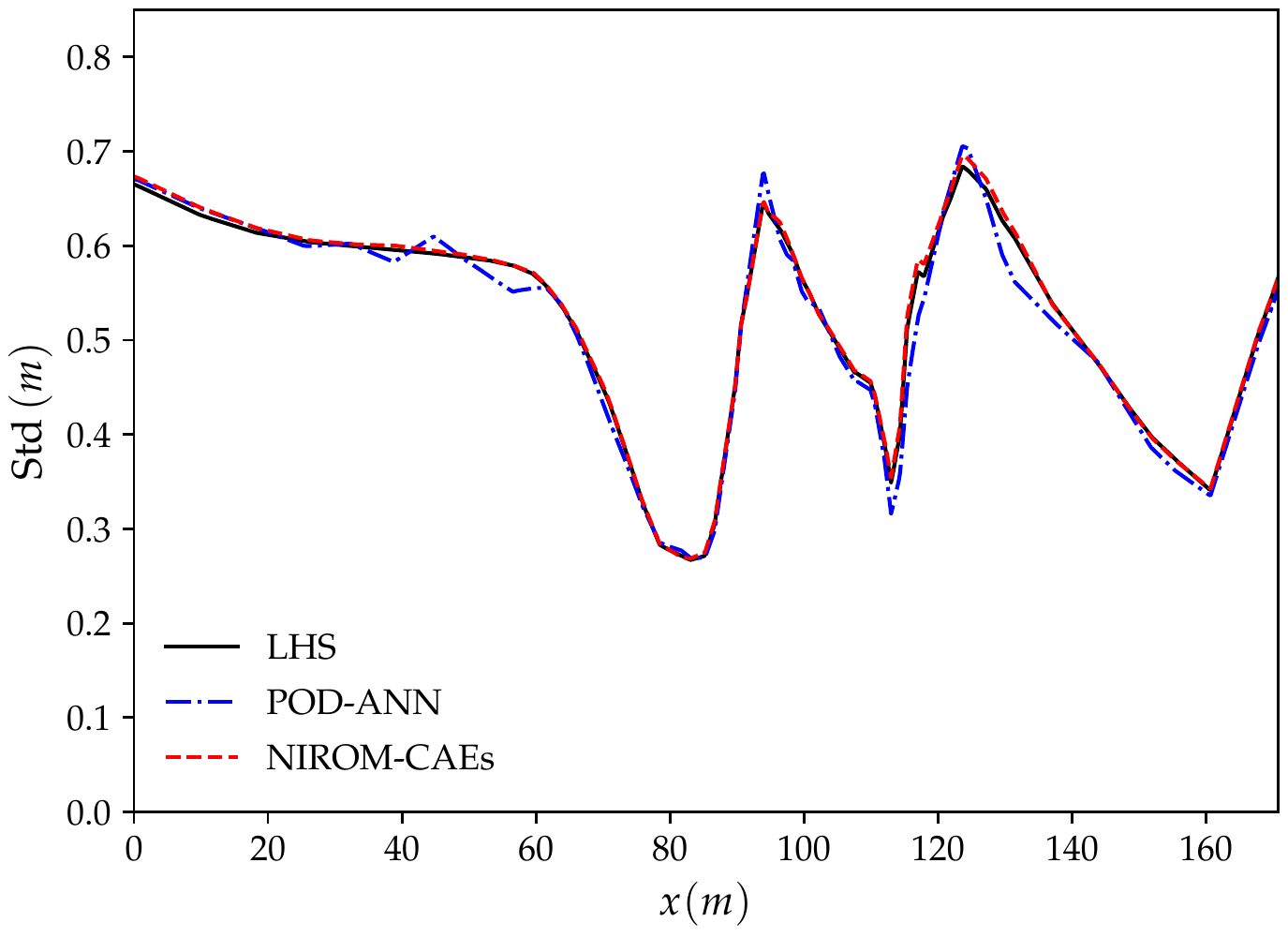}
         \caption{$t\approx\,45\,(s)$}
         \label{fig:Std_Milles_Iles_compa_t_45}
     \end{subfigure}
   
 %++++++++++++++++++++++++++++++ 
   %++++++++++++
   \caption{Comparison of the standard deviation profiles of the water level over the cross-section line at various time-steps, obtained with the POD-ANN and NIROM-CAEs with those from the LHS reference solution (with $2\,000$ realizations). (a): $t\approx 5\,s$, (b): $t\approx 15\,s$, (c): $t\approx 30\,s$ and (d): $t\approx 45\,s$.}
   \label{fig:Std_Milles_Iles_compa_diff_times}
\end{figure}

The results stemming from this test case are presented mainly in terms of the variation of the mean and standard deviation profiles of the water surface level over the computational sub-domain of study. The results obtained with the proposed NIROM-CAEs are compared with those from the LHS solutions. These are obtained by running the deterministic numerical solver using a sample set of $N_{s}=2\,000$ points of the uncertain initial upstream water level. The predictive model of the POD-ANN approach is based on a neural network composed of three hidden layers with $50$ neurons in each that map the reduced basis coefficient to the input parameter. Fig.\ref{fig:Mean_Milles_Iles_compa_diff_times} and \ref{fig:Std_Milles_Iles_compa_diff_times} present the variation of the mean and the standard deviation profiles, respectively, as a function of the longitudinal coordinate along the cross-section line, as shown in Fig.\ref{fig:M_Iles_bathy_zoom}. The comparison of the predicted mean profile of the water level at different simulation times shows that the NIROM-CAEs predictions are in good concordance with the profiles from the LHS reference solutions. Also, the POD-ANN model presents satisfactory predictions for the mean. One can also see  the shock wave propagation as the water level over the cross-section line increases with  time. In Fig.\ref{fig:Mean_Milles_Iles_compa_t_10} through Fig.\ref{fig:Mean_Milles_Iles_compa_t_45} is shown the hatched area which represents the bathymetry of the terrain. Despite accurate results obtained by the POD-ANN approach for the mean profiles of the water level, the standard deviation  predictions over the cross-section line reveals spurious oscillations, as shown in Fig.\ref{fig:Std_Milles_Iles_compa_t_10} through Fig.\ref{fig:Std_Milles_Iles_compa_t_45}. In contrast, the predictions from the NIROM-CAEs approach show good agreement with reference solution for all the presented simulation times. This further illustrates the good abilities of the proposed nonlinear reduced-order approach in approximating  the outputs of high-dimensional time-dependent problems.\\

%========================================
%++++++++++++++++
%++++++++++++++++
%========================================

%========================================
%++++++++++++++++
%++++++++++++++++
%========================================

\begin{figure}[ht!]
  \centering
    \begin{subfigure}[b]{0.49\textwidth}
      \centering
        \includegraphics[width=\textwidth]{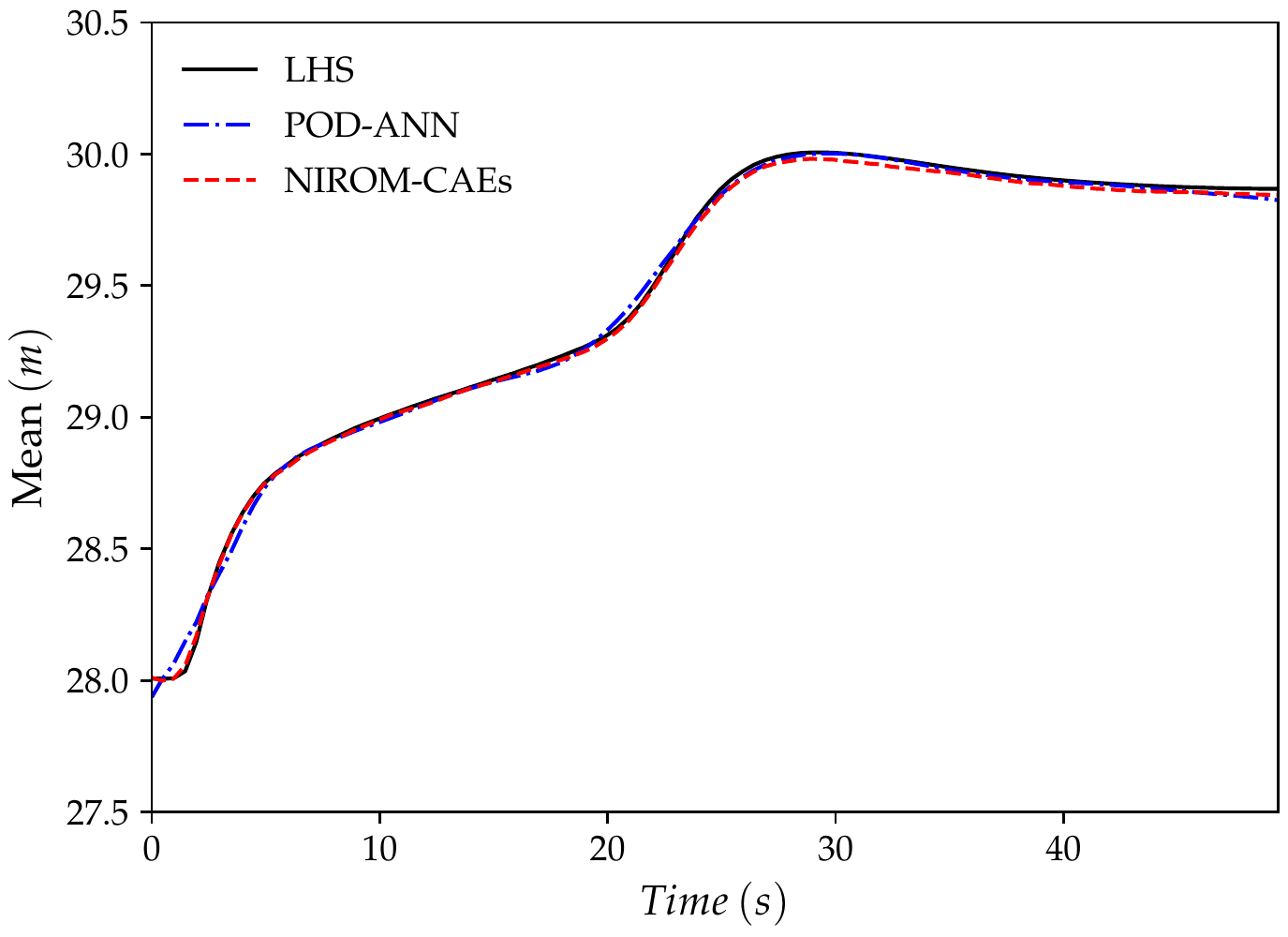}
         \caption{Point 1}
         \label{fig:Mean_Milles_Iles_compa_x_2066}
    \end{subfigure}  
  \hfill
    \begin{subfigure}[b]{0.49\textwidth}
      \centering
        \includegraphics[width=\textwidth]{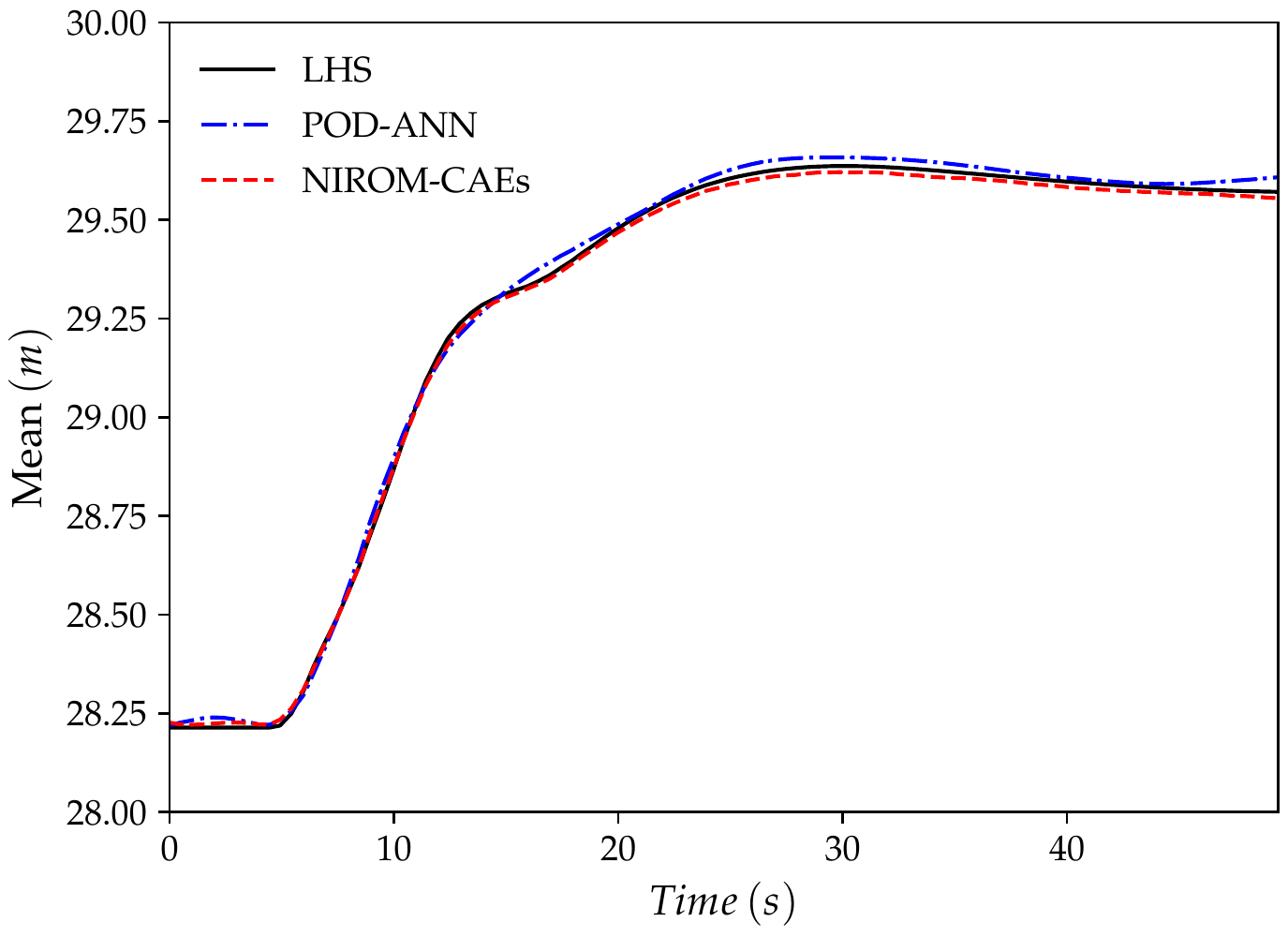}
         \caption{Point 2}
         \label{fig:Mean_Milles_Iles_compa_x_2837}
     \end{subfigure} 
     \begin{subfigure}[b]{0.49\textwidth}
      \centering
        \includegraphics[width=\textwidth]{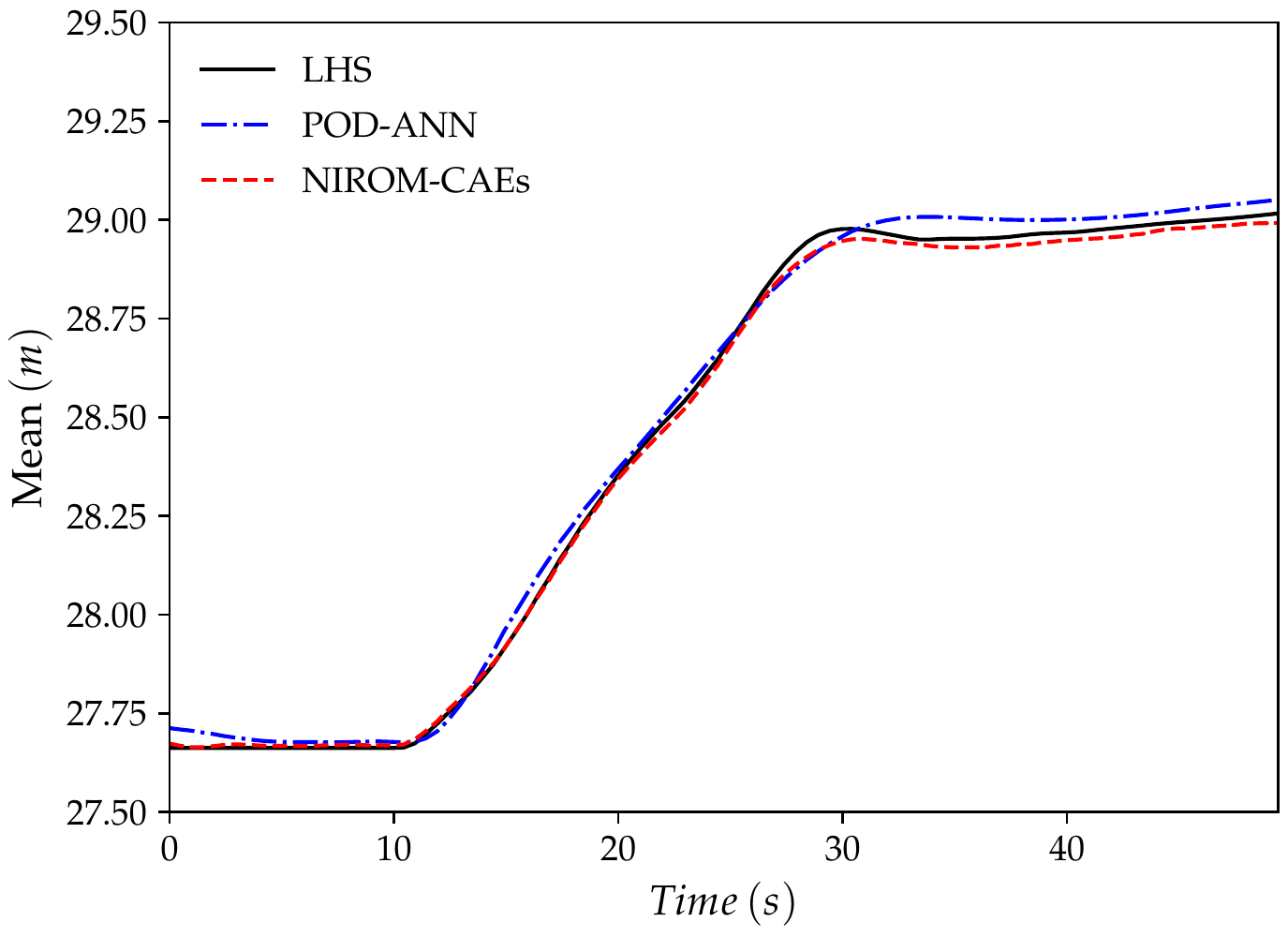}
         \caption{Point 3}
         \label{fig:Mean_Milles_Iles_compa_x_3834}
    \end{subfigure}  
  \hfill
    \begin{subfigure}[b]{0.49\textwidth}
      \centering
        \includegraphics[width=\textwidth]{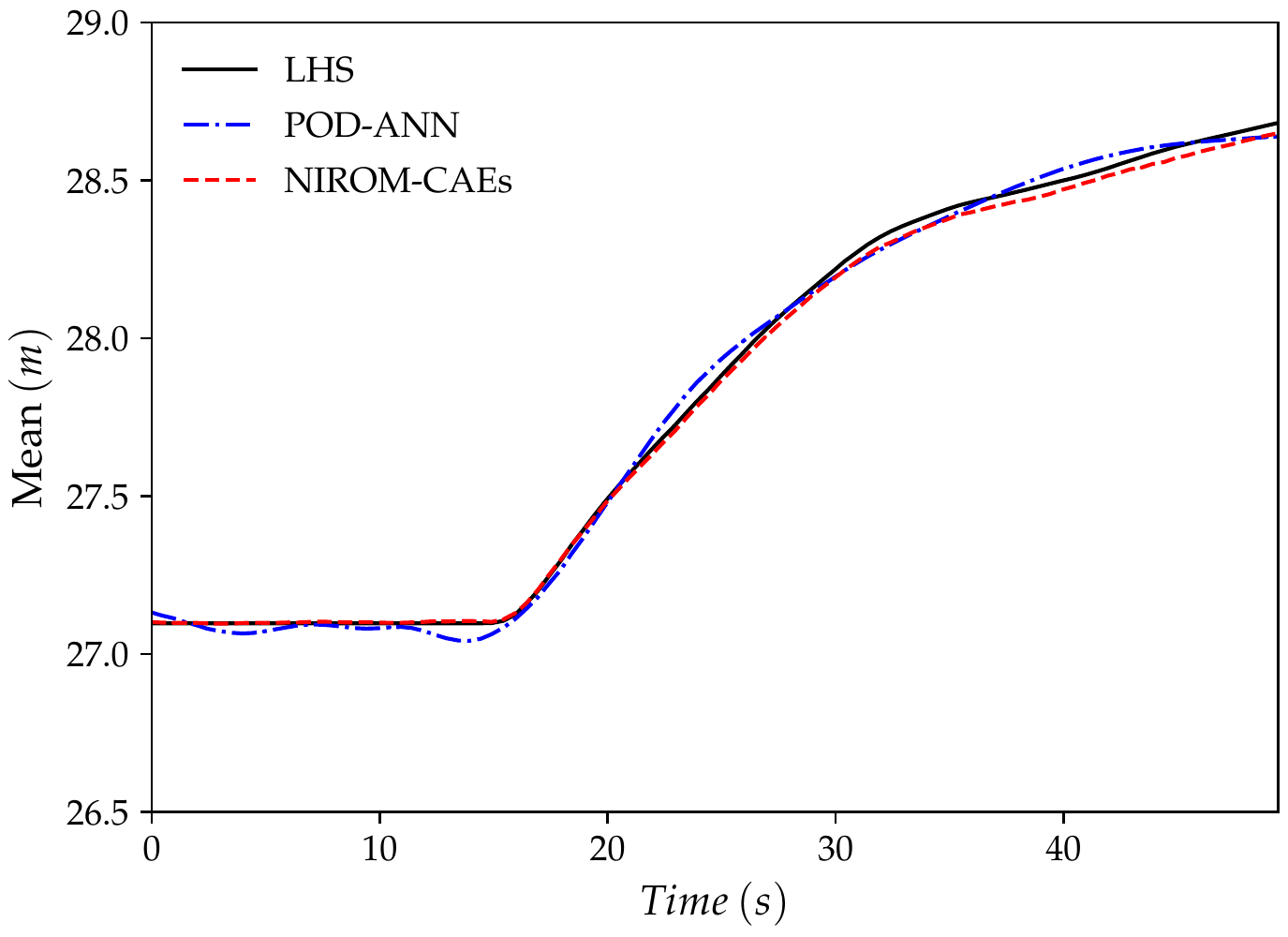}
         \caption{Point 4}
         \label{fig:Mean_Milles_Iles_compa_x_4590}
     \end{subfigure}
   
 %++++++++++++++++++++++++++++++ 
   %++++++++++++
   \caption{Evolution of the mean profiles of the water level as a function of time at four gauging points. The profiles obtained with the NIROM-CAEs are compared to those from the POD-ANN technique and the LHS reference solution (with $2\,000$ realizations). (a): Point 1, (b): Point 2, (c): Point 3 and (d): Point 4.}
   \label{fig:Mean_Milles_Iles_compa_t_diff_nodes}
\end{figure}

\begin{figure}[ht!]
  \centering
    \begin{subfigure}[b]{0.49\textwidth}
      \centering
        \includegraphics[width=\textwidth]{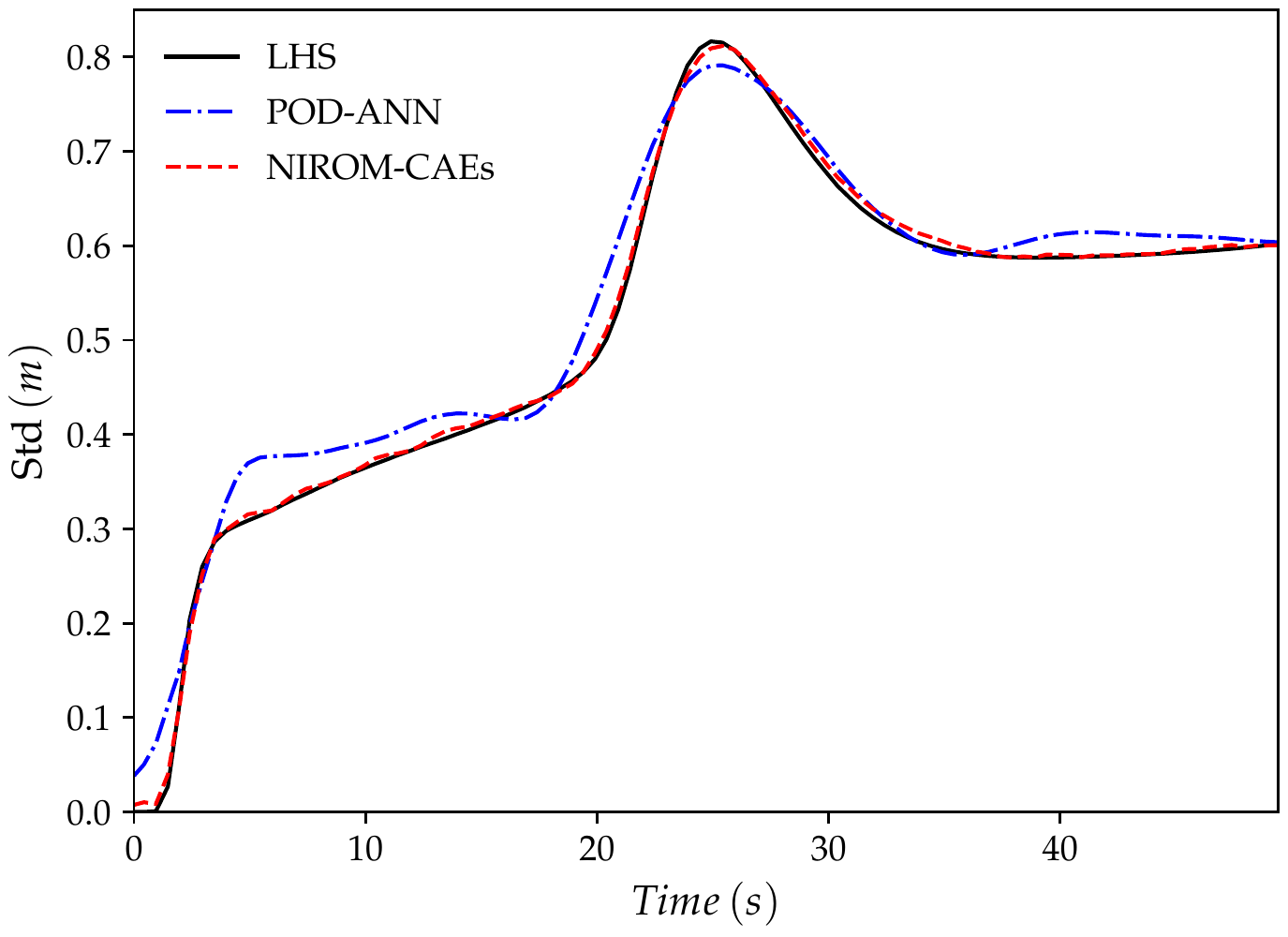}
         \caption{Point 1}
         \label{fig:Std_Milles_Iles_compa_x_2066}
    \end{subfigure}  
  \hfill
    \begin{subfigure}[b]{0.49\textwidth}
      \centering
        \includegraphics[width=\textwidth]{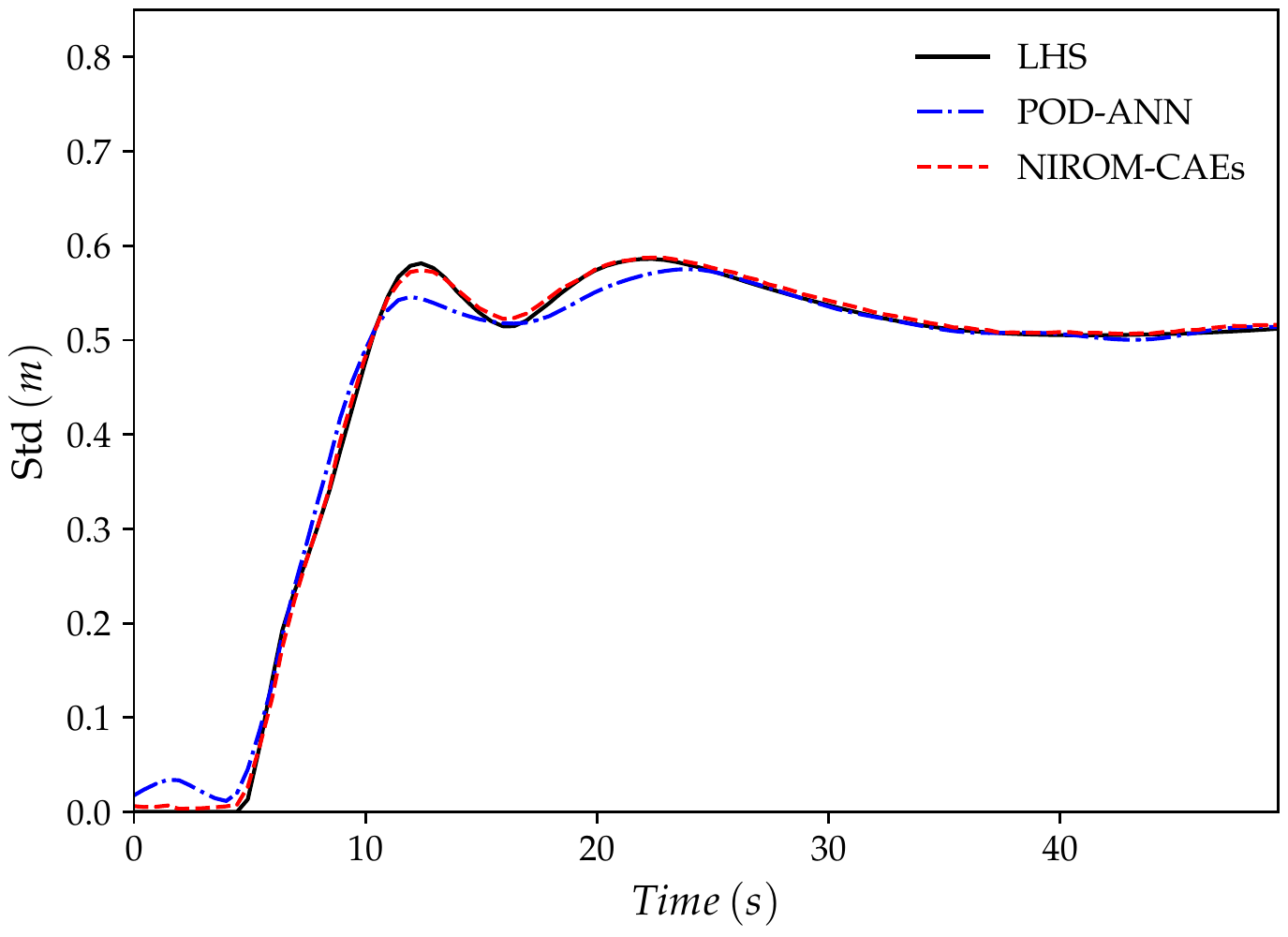}
         \caption{Point 2}
         \label{fig:Std_Milles_Iles_compa_x_2837}
     \end{subfigure} 
     \begin{subfigure}[b]{0.49\textwidth}
      \centering
        \includegraphics[width=\textwidth]{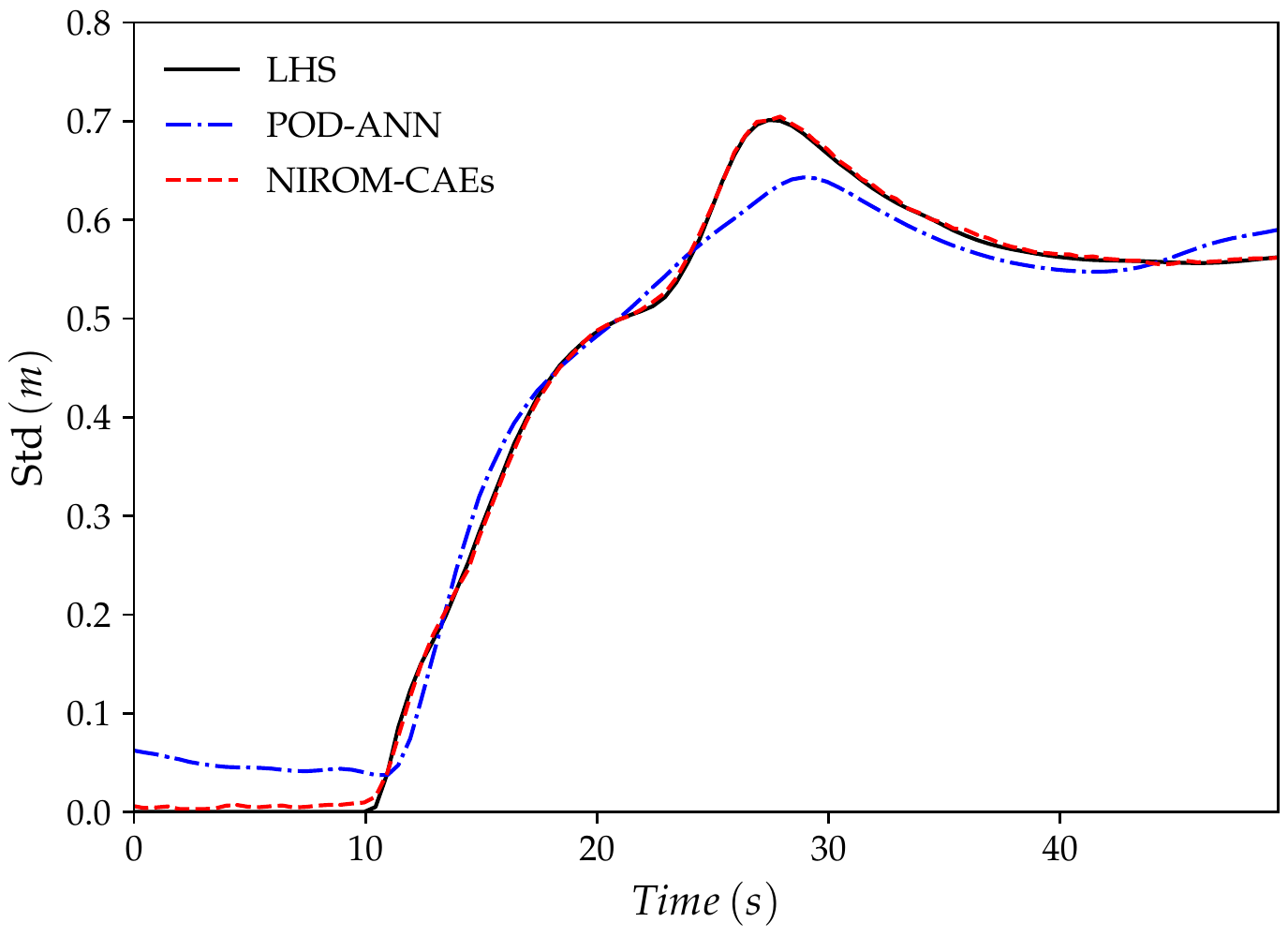}
         \caption{Point 3}
         \label{fig:Std_Milles_Iles_compa_x_3834}
    \end{subfigure}  
  \hfill
    \begin{subfigure}[b]{0.49\textwidth}
      \centering
        \includegraphics[width=\textwidth]{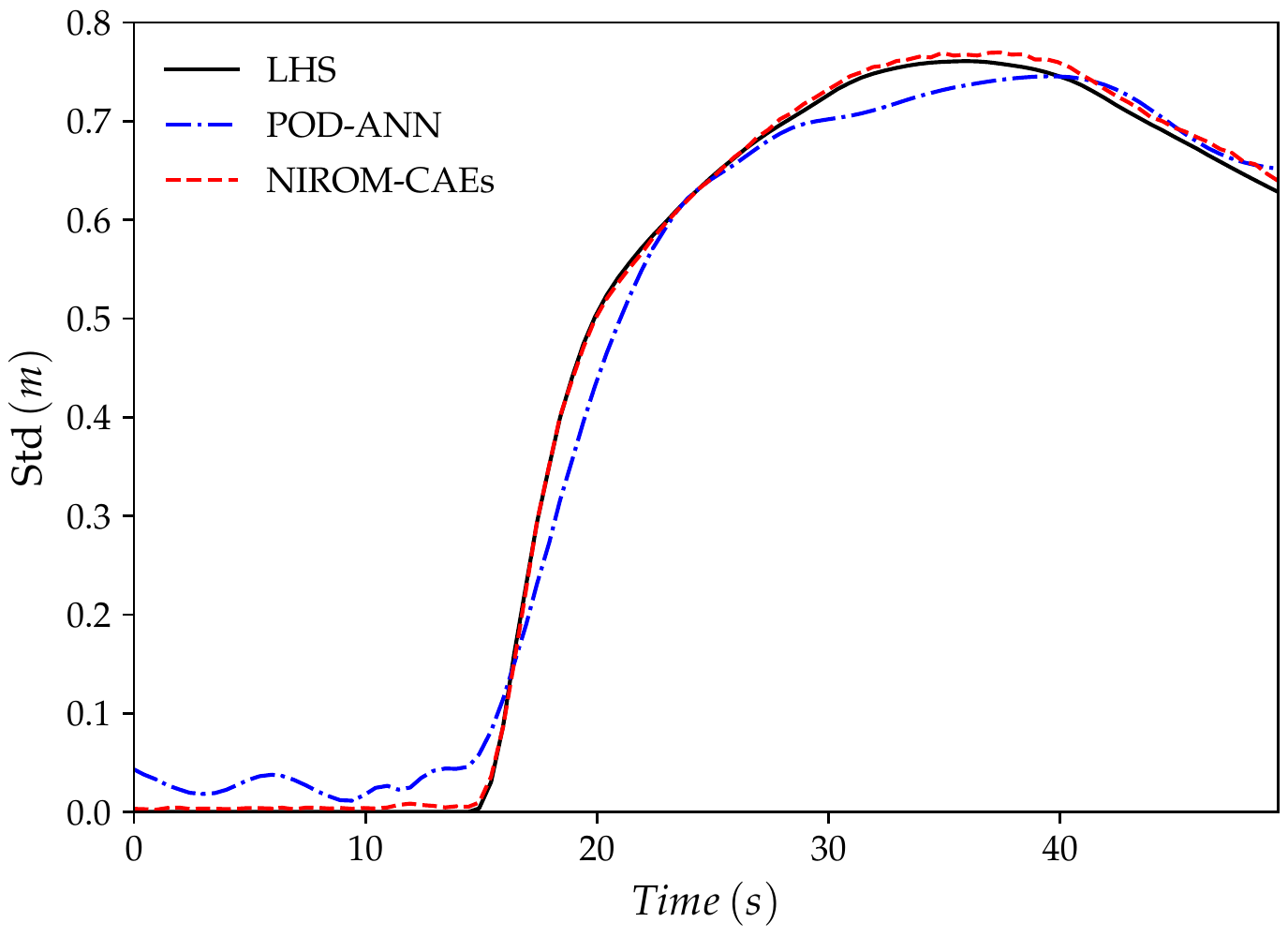}
         \caption{Point 4}
         \label{fig:Std_Milles_Iles_compa_x_4590}
     \end{subfigure}
   
 %++++++++++++++++++++++++++++++ 
   %++++++++++++
   \caption{Evolution of the standard deviation profiles of the water level as a function of time at four gauging points. The profiles obtained with the NIROM-CAEs are compared to those from the POD-ANN technique and the LHS reference solution (with $2\,000$ realizations). (a): Point 1, (b): Point 2, (c): Point 3 and (d): Point 4.}
   \label{fig:Std_Milles_Iles_compa_diff_x}
\end{figure}

In addition to the results presented above, four gauging points are chosen to depict the evolution of statistical moments as a function of time. These gauging points, whose approximate positions on the cross-section line are shown in Fig.\ref{fig:M_Iles_bathy_zoom}, are selected to show the ability of the proposed non-linear technique to capture the temporal dynamics. In Fig.\ref{fig:Mean_Milles_Iles_compa_t_diff_nodes} and Fig.\ref{fig:Std_Milles_Iles_compa_diff_x}, the time variations of the mean and standard deviation are presented at the four selected gauging points. Comparisons of the results from NIROM-CAEs, POD-ANN,  and those from the LHS solution are made. One can see from these plots the increase in the water level from point 1 (close to the dam location) to point 4 (far from the dam), which allows an estimation of the arrival time of the flooding wave. Slight deviations can be reported in the mean profiles from the POD-ANN at almost all the gauging points, while the profiles from NIROM-CAEs present a good match with the reference solution as shown in Fig.\ref{fig:Mean_Milles_Iles_compa_x_2066} through Fig.\ref{fig:Mean_Milles_Iles_compa_x_4590}. This tendency is further confirmed by the standard deviation profiles where excellent superimposition can be observed between the profiles from the NIROM-CAEs and those from the LHS solution at all the gauging points. Conversely, the predictions from the linear POD-ANN technique present remarkable deviations with poor accuracy in the approximations of the standard deviation profiles of the water level at all gauging points as depicted in Fig.\ref{fig:Std_Milles_Iles_compa_x_2066} through Fig.\ref{fig:Std_Milles_Iles_compa_x_4590}.\\   

%========================================
%++++++++++++++++
%++++++++++++++++
%========================================

%========================================
%++++++++++++++++
%++++++++++++++++
%========================================

\begin{figure}[ht!]
  \centering
    \begin{subfigure}[b]{0.49\textwidth}
      \centering
        \includegraphics[width=\textwidth]{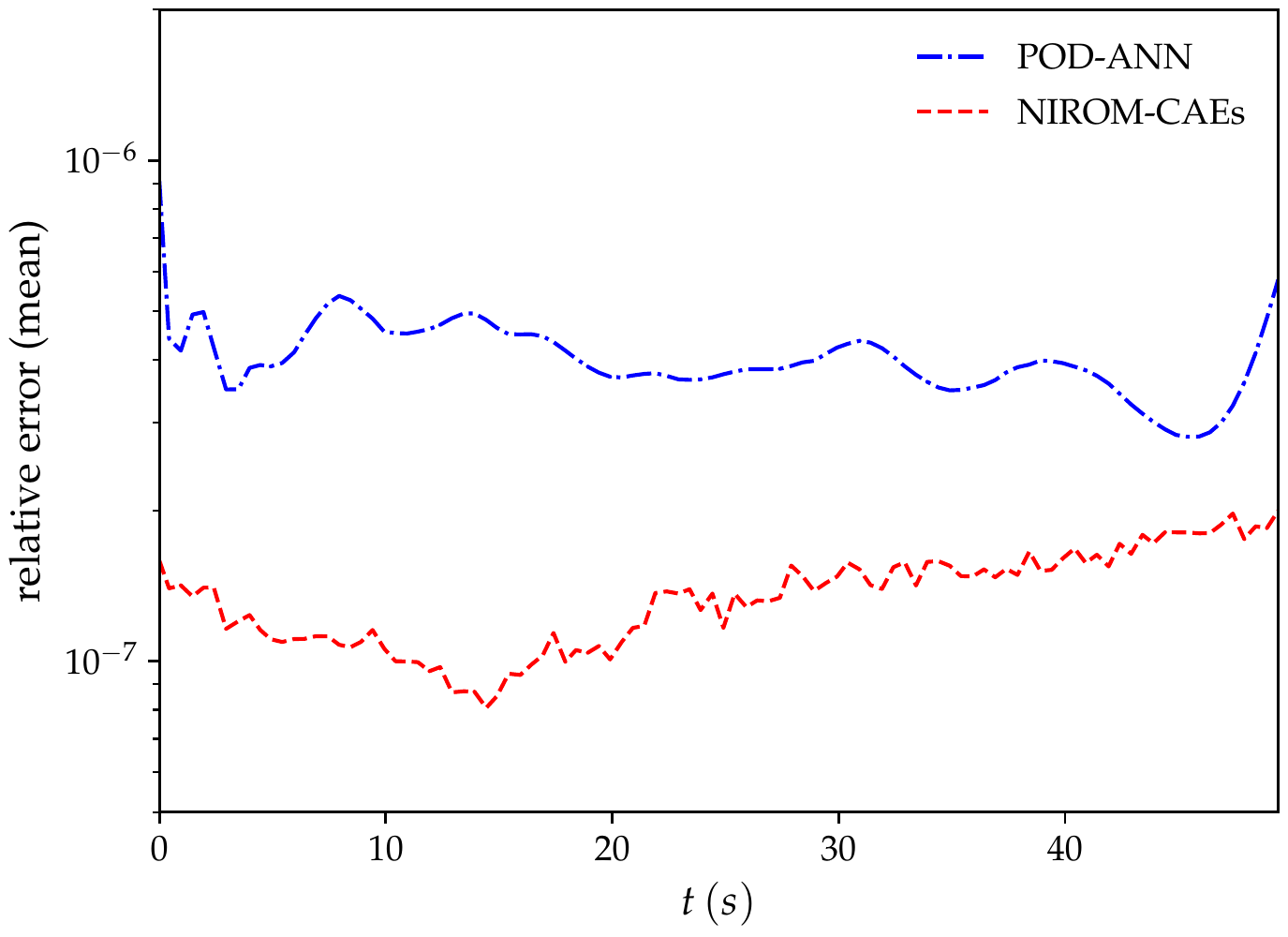}
         \caption{Mean}
         \label{fig:Error_Mean_Mille_Iles_compari}
    \end{subfigure}  
  \hfill
    \begin{subfigure}[b]{0.49\textwidth}
      \centering
        \includegraphics[width=\textwidth]{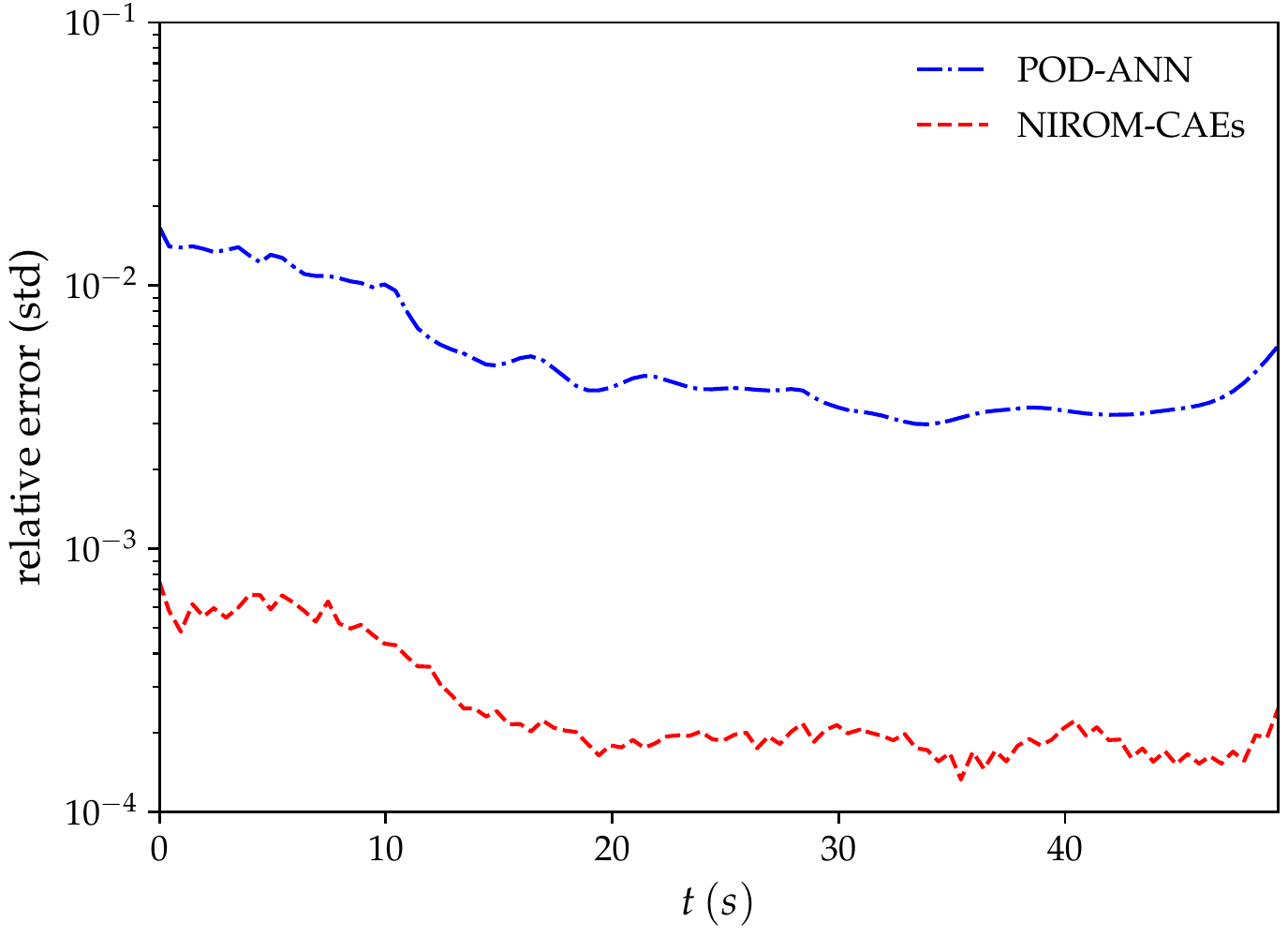}
         \caption{Std}
         \label{fig:Error_Std_Mille_Iles_compari}
     \end{subfigure}  
 %++++++++++++++++++++++++++++++ 
   %++++++++++++
   \caption{Comparison of the relative $L^{2}$-error of the mean and standard deviation of the water level as a function of time based on the POD-ANN technique and the proposed NIROM-CAEs approach. Errors are computed with respect to the reference LHS solution (with $N_{s}=2\,000$ realizations).}
   \label{fig:Error_MEan_Std_Mille_Iles_compari}
\end{figure}

For a better insight into the accuracy of both reduced order-model approaches, namely POD-ANN and NIROM-CAEs, a quantitative evaluation is performed by computing the relative $L^{2}$-error over the whole computational domain (the LHS reference sampling solution is obtained with $N_{s}=2\,000$ points). Thus, the relative error of the mean and standard deviation is represented as a function of time as shown in Fig.\ref{fig:Error_MEan_Std_Mille_Iles_compari}. One can observe from these plots that the proposed NIROM-CAEs have much lower values of the relative error for both mean and standard deviation. The effect of the training set sample size ($N_{s}$) on the maximum relative error over time is reported in table.\ref{tab:Tab_Max_Err_Compa_Mille_Iles}. Three values of the sample size $N_{s}=30,\,90$ and $300$ are tested, and for each value, the maximum $L^{2}$-error is computed over time for both the mean and the standard deviation. One can note that as the sample size increases, the relative error of the NIROM-CAEs decreases faster than that of POD-ANN, more particularly for the standard deviation where a difference in the error values can reach almost two orders of magnitude. This quantitative analysis supports the aforementioned observations concerning the predictive abilities of the proposed non-linear NIROM-CAEs. Another meaningful result from the table.\ref{tab:Tab_Max_Err_Compa_Mille_Iles} concerns the dimension of the generated latent space which is considerably reduced for the NIROM-CAEs of the order of $L_{x}=50$ and $L_{t}=10$ independently of the sample size, conversely to the POD-ANN whose dimension of the reduced basis increases from $L_{POD}=800$ for $N_{s}=30$ to reach $L_{POD}=3\,337$ for $N_{s}=300$, which may present a challenging task for the available computational capacities.

%========================================
%++++++++++++++++
%++++++++++++++++
%========================================

%+++++++++++++++++++++++++++++++++++++++++++++++++++++++++++
%% Max Error
\begin{table}[h!]
\caption{Effect of the sample size $N_{s}$ on the maximum relative error in the $L^{2}$-norm for the mean and standard deviation obtained of the  from the NIROM-CAEs ($L_{x}=50$, $L_{t}=10$) and POD-ANN ($\epsilon_{s}=\epsilon_{t}=10^{-8}$). Errors are computed with respect to the LHS reference solution (with $N_{s}=2\,000$ realizations).}
\centering
%\begin{tabular}{l l l  llll}  
\begin{tabular*}{0.85\textwidth}{@{\extracolsep{\fill}} ll  llll }
\hline        % inserting first horizontal line
& & \multicolumn{2}{c}{$Err_{L^{2},\,Mean}^{max}$}& \multicolumn{2}{c}{$Err_{L^{2},\,Std}^{max}$}\\%[-0.0ex]
\cline{3-4}
\cline{5-6}
\raisebox{2ex}{$ N_{s} $}& \raisebox{2ex}{$ L_{POD} $}&POD-ANN& NIROM-CAEs& POD-ANN& NIROM-CAEs\\%[-1ex] 
%+++++++++++++++++++++++
\hline        % insertinh second horizontal line
% 1st row 
$30$& $860$& $8.2855$E-07& $2.1896$E-06& $0.024537$& $0.010349$\\%[-1ex]
%% 3rd row
$90$& $2\,112$& $6.0781$E-07& $9.0018$E-07& $0.013303$& $0.005838$\\%[-1ex]
% 4th row
$300$& $3\,387$& $9.1037$E-07& $1.9986$E-07& $0.016687$& $7.4981$E-04\\%[-1ex]

\hline        % insertinh last horizontal line     
\end{tabular*}
\label{tab:Tab_Max_Err_Compa_Mille_Iles}
\end{table}
%+++++++++++++++++++++++++++++++++++++++++++++++++++++++++++

%========================================
\section{Conclusion}\label{conc}

This paper presents a non-intrusive reduced order model based on convolutional autoencoders (NIROM-CAEs) for parameter-varying and time-dependent fluid problems. The model is fully data-driven and exploits the nonlinear framework provided by the convolutional autoencoders to tackle physical problems presenting a high degree of complexity. The construction of the training offline stage of the proposed approach consists of two successive compression levels along the spatial and temporal dimensions, through the encoder part of the 1d CAEs. The first encoder (space-encoder) performs compression along the spatial dimension of the input snapshot matrix to generate the encoded low dimensional latent space. A second encoder (time-encoder), whose structure is similar to the former one, applies convolutional operations along the temporal dimension of the encoded space to generate a second spatio-temporal latent vector with a much lower dimension. The encoded latent space thus obtained is used as output data for a multilayer perceptron (MLP) which is used at the bottom level to train the mapping with the points from the parametric space. Once the training offline stage is constructed, the online predictive stage consists of a generation of a new set of unseen points in the design space to provide the trained MLP at the bottom level. The resulting spatio-temporal latent vectors are then decoded through successive 1d decoders to predict solutions describing the high-dimensional spatio-temporal dynamics of the output quantities of interest.\\

The performance and accuracy of the proposed technique are demonstrated on three non-linear examples parametrized by a random variable with an appropriate variability range. The two first test cases concern the Burgers and Stoker's solutions known as challenging test cases presenting a high hyperbolic behavior or a discontinuity accompanying the shock wave. It is shown from the numerical results that the NIROM-CAEs offer accurate approximations of the statistical moments in comparison to the reference solutions from the LHS sampling method, unlike the linear POD-ANN model which shows some limitations to reproduce the dynamics of the outputs where an oscillatory behavior is observed in the predicted profiles. The results also pointed out the low relative error that the proposed NIROM-CAEs which demonstrates further its abilities. The model is then applied to a third case to analyze uncertainty propagation in a dam-break flow over real terrain. The predictions from the NIROM-CAEs have shown an accurate reconstruction of the mean and the standard deviation profiles, where good concordances with those from the LHS reference solutions have been observed. The predictions from the POD-ANN show spurious oscillations, particularly for the temporal evolution at gauging points. Thus, the proposed non-intrusive reduced order model based on convolutional autoencoders presents a powerful tool for non-linear dimensionality reduction for parameter-varying, time-dependent and large-scale problems characterized by strong hyperbolic behavior or even with the presence of a discontinuity. The model offers the construction of accurate surrogate predictions with a high reduction ratio at a cheaper cost.

\clearpage
%+++++++++++++++++++++++++++++++++++
\section*{Acknowledgments}
This research was supported by the Natural Sciences and Engineering Research Council of Canada, the financial support is gratefully acknowledged.
\appendix
%\section{\textcolor{ForestGreen}{CAEs and MLP architecture}}\label{appe_A}

\section{Training convergence history and models' configurations}\label{appe_A}

%+=+=+=+=+=+=+=+=+=+=+=+=+
%+++++++++++++++++++++++++
\begin{table}[h!]
\caption{CAE-space artchitecture for Burger's and Stoker's test cases}
\centering  
\begin{tabular*}{0.95\textwidth}{@{\extracolsep{\fill}}l llll}
\hline        % inserting first horizontal line 
\textbf{Encoder-space}&  & & & \\
\hline
% 1st row 
Layer       & Nb of filters & Kernel size & Activation function & Output shape\\
\hline
Input       & -             & -           & -                   & $1000\times1$\\
%\hline
Conv-pool     & $32$        & $3-2$       & PReLU               & $500\times32$\\
Conv-pool     & $64$        & $3-2$       & PReLU               & $250\times64$\\
Conv-pool     & $128$       & $3-5$       & PReLU               & $50\times128$\\
%\hline
Flatten     & -               & -         & -                   & $6\,400$\\
%\hline
Dense       & -               & -         & PReLU               & $60$\\
%\hline
Dense (output) & -            & -         & PReLU               & $L_{x}=50$\\
\hline
\textbf{Decoder-space}&  & & & \\
\hline
Input       & -             & -           & -                   & $L_{x}=50$\\
Dense       & -               & -         & PReLU               & $60$\\
Dense       & -               & -         & PReLU               & $6\,400$\\
Reshape     & -               & -         & -                   & $50\times128$\\
Conv-Upsamp  & $128$           & $3-5$    & PReLU               & $250\times128$\\
Conv-Upsamp  & $64 $           & $3-2$    & PReLU               & $500\times64$\\
Conv-Upsamp  & $32 $           & $3-2$    & PReLU               & $1000\times32$\\
Conv (output)  & $1 $            & $3$      & PReLU               & $1000\times1$\\
\hline

\end{tabular*}
\label{tab:Tab_CAE_space_archit_Burg_Stok}
\end{table}
%++++++++++++++++++++++++
%+=+=+=+=+=+=+=+=+=+=+=+=+

%+=+=+=+=+=+=+=+=+=+=+=+=+
%+++++++++++++++++++++++++
\begin{table}[h!]
\caption{CAE-time artchitecture for Burger's and Stoker's test cases}
\centering  
\begin{tabular*}{0.95\textwidth}{@{\extracolsep{\fill}}l llll}
\hline        % inserting first horizontal line 
\textbf{Encoder-time}&  & & & \\
\hline
% 1st row 
Layer       & Nb of filters & Kernel size & Activation function & Output shape\\
\hline
Input       & -             & -           & -                   & $104\times L_{x}$\\
%\hline
Conv-pool     & $32$        & $3-2$       & PReLU               & $52\times32$\\
Conv-pool     & $64$        & $3-2$       & PReLU               & $26\times64$\\
Conv-pool     & $128$       & $3-2$       & PReLU               & $13\times128$\\
%\hline
Flatten     & -               & -         & -                   & $1\,664$\\
Dense (output) & -            & -         & PReLU               & $L_{t}=10$\\
\hline
\textbf{Decoder-time}&  & & & \\
\hline
Input       & -               & -           & -                 & $L_{t}=10$\\
Dense       & -               & -         & PReLU               & $1\,664$\\
Reshape     & -               & -         & -                   & $13\times128$\\
Conv-Upsamp  & $128$           & $3-2$    & PReLU               & $26\times128$\\
Conv-Upsamp  & $64 $           & $3-2$    & PReLU               & $52\times64$\\
Conv-Upsamp  & $32 $           & $3-2$    & PReLU               & $104\times32$\\
Conv (output) & $1 $            & $3$      & PReLU               & $104\times L_{x}$\\
\hline

\end{tabular*}
\label{tab:Tab_CAE_time_archit_Burg_Stok}
\end{table}
%++++++++++++++++++++++++
%+=+=+=+=+=+=+=+=+=+=+=+=+

%+=+=+=+=+=+=+=+=+=+=+=+=+
%+++++++++++++++++++++++++
\begin{table}[h!]
\caption{CAE-space architecture for hypothetical dam break of Miles-Iles river}
\centering  
\begin{tabular*}{0.95\textwidth}{@{\extracolsep{\fill}}l llll}
\hline        % inserting first horizontal line 
\textbf{Encoder-space}&  & & & \\
\hline
% 1st row 
Layer       & Nb of filters & Kernel size & Activation function & Output shape\\
\hline
Input       & -             & -           & -                   & $10200\times1$\\
%\hline
Conv-pool     & $32$        & $3-5$       & PReLU               & $2040\times32$\\
Conv-pool     & $64$        & $3-5$       & PReLU               & $408\times64$\\
%\hline
Flatten     & -               & -         & -                   & $26\,112$\\
%\hline
Dense       & -               & -         & PReLU               & $120$\\
%\hline
Dense (output) & -            & -         & PReLU               & $L_{x}=50$\\
\hline
\textbf{Decoder-space}&  & & & \\
\hline
Input       & -             & -           & -                   & $L_{x}=50$\\
Dense       & -               & -         & PReLU               & $120$\\
Dense       & -               & -         & PReLU               & $26\,112$\\
Reshape     & -               & -         & -                   & $408\times64$\\
Conv-Upsamp  & $64 $           & $3-5$    & PReLU               & $2040\times64$\\
Conv-Upsamp  & $32 $           & $3-5$    & PReLU               & $10200\times32$\\
Conv (output)  & $1 $            & $3$      & PReLU               & $10200\times1$\\
\hline

\end{tabular*}
\label{tab:Tab_CAE-space_archit_Miles_Iles}
\end{table}

%+=+=+=+=+=+=+=+=+=+=+=+=+
%+++++++++++++++++++++++++
\begin{table}[h!]
\caption{CAE-time artchitecture for hypothetical dam break of Miles-Iles river}
\centering  
\begin{tabular*}{0.95\textwidth}{@{\extracolsep{\fill}}l llll}
\hline        % inserting first horizontal line 
\textbf{Encoder-time}&  & & & \\
\hline
% 1st row 
Layer       & Nb of filters & Kernel size & Activation function & Output shape\\
\hline
Input       & -             & -           & -                   & $100\times L_{x}$\\
%\hline
Conv-pool     & $32$        & $3-2$       & PReLU               & $50\times32$\\
Conv-pool     & $64$        & $3-2$       & PReLU               & $25\times64$\\
Conv-pool     & $128$       & $3-5$       & PReLU               & $5\times128$\\
%\hline
Flatten     & -               & -         & -                   & $640$\\
Dense (output) & -            & -         & PReLU               & $L_{t}=10$\\
\hline
\textbf{Decoder-time}&  & & & \\
\hline
Input       & -               & -           & -                 & $L_{t}=10$\\
Dense       & -               & -         & PReLU               & $640$\\
Reshape     & -               & -         & -                   & $5\times128$\\
Conv-Upsamp  & $128$           & $3-5$    & PReLU               & $25\times128$\\
Conv-Upsamp  & $64 $           & $3-2$    & PReLU               & $50\times64$\\
Conv-Upsamp  & $32 $           & $3-2$    & PReLU               & $100\times32$\\
Conv (output) & $1 $            & $3$      & PReLU               & $100\times L_{x}$\\
\hline

\end{tabular*}
\label{tab:Tab_CAE-time_archit_Miles_Iles}
\end{table}
%++++++++++++++++++++++++
%+=+=+=+=+=+=+=+=+=+=+=+=+

%+=+=+=+=+=+=+=+=+=+=+=+=+
%+++++++++++++++++++++++++
\begin{table}[h!]
\caption{MLP artchitecture for Burgers, Stoker's test cases and hypothetical dam break}
\centering  
\begin{tabular*}{0.75\textwidth}{@{\extracolsep{\fill}}l ll}
\hline        % inserting first horizontal line 
\textbf{MLP}&  &\\
\hline
% 1st row 
Layer           & Activation function & Output shape\\
\hline
Input           & ReLU    & $1$         \\
Dense           & ReLU    & $128$       \\
Dense           & ReLU    & $128$        \\
Dense           & ReLU    & $128$        \\
Dense (output)  & Linear  & $L_{t}=10$   \\
\hline

\end{tabular*}
\label{tab:MLP_archit}
\end{table}
%++++++++++++++++++++++++
%+=+=+=+=+=+=+=+=+=+=+=+=+

%========================================
%+++++++++++++++++++++++++++++++
\begin{figure}[ht!]
 \centering
\includegraphics[width=1\textwidth]{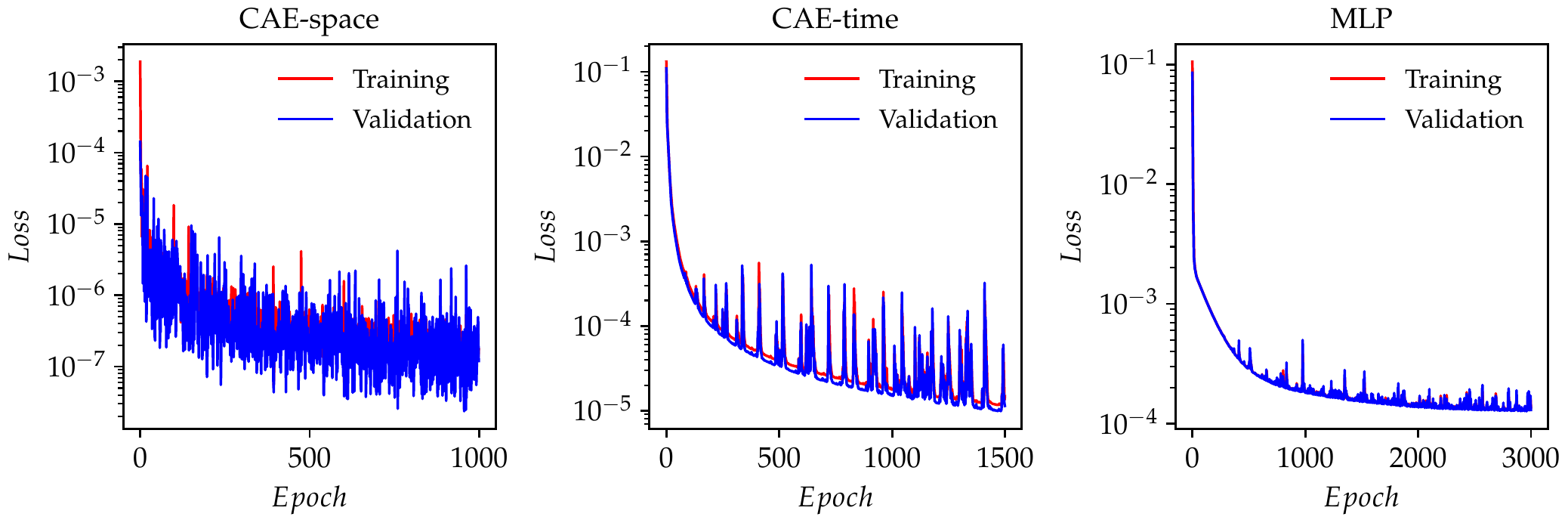}
 \caption{Evolution of the training and validation losses of CAEs and MLP for the Burgers equation test case.}
   \label{fig:Conv_hist_Burg}
\end{figure}
%+++++++++++++++++++++++++++++++
%========================================

%========================================
%+++++++++++++++++++++++++++++++
\begin{figure}[ht!]
 \centering
\includegraphics[width=1\textwidth]{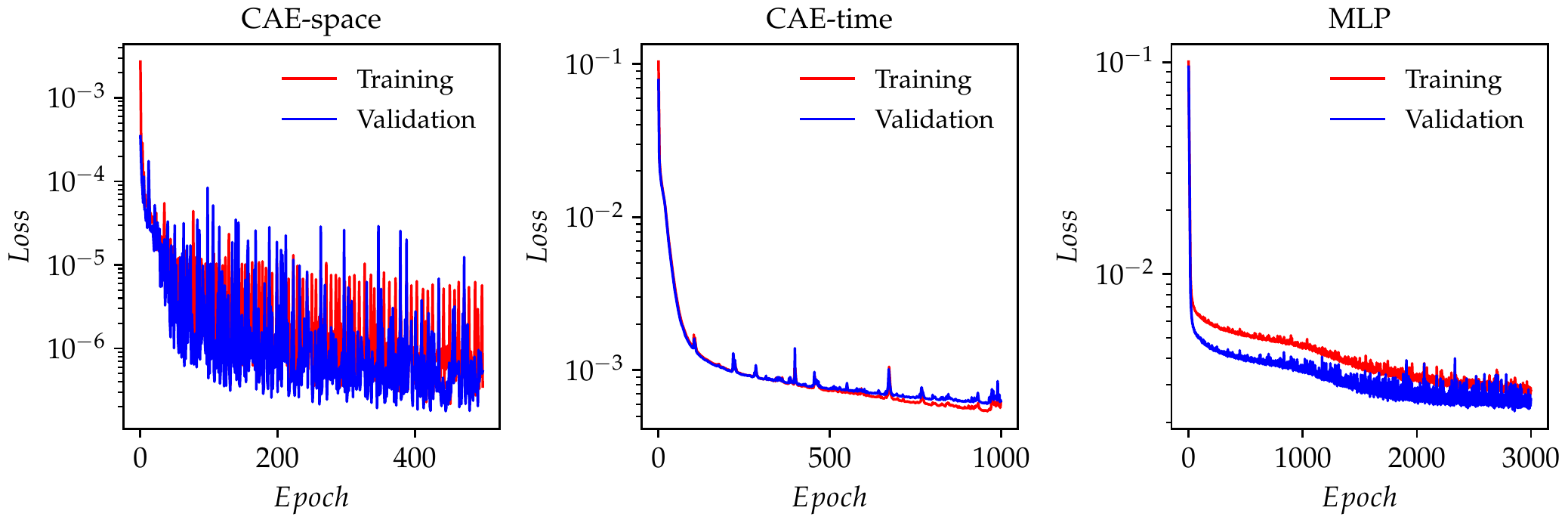}
 \caption{Evolution of the training and validation losses of CAEs and MLP for the Stoker's analytical solution test case.}
   \label{fig:Conv_hist_Stoker}
\end{figure}
%+++++++++++++++++++++++++++++++
%========================================

%========================================
%+++++++++++++++++++++++++++++++
\begin{figure}[ht!]
 \centering
\includegraphics[width=1\textwidth]{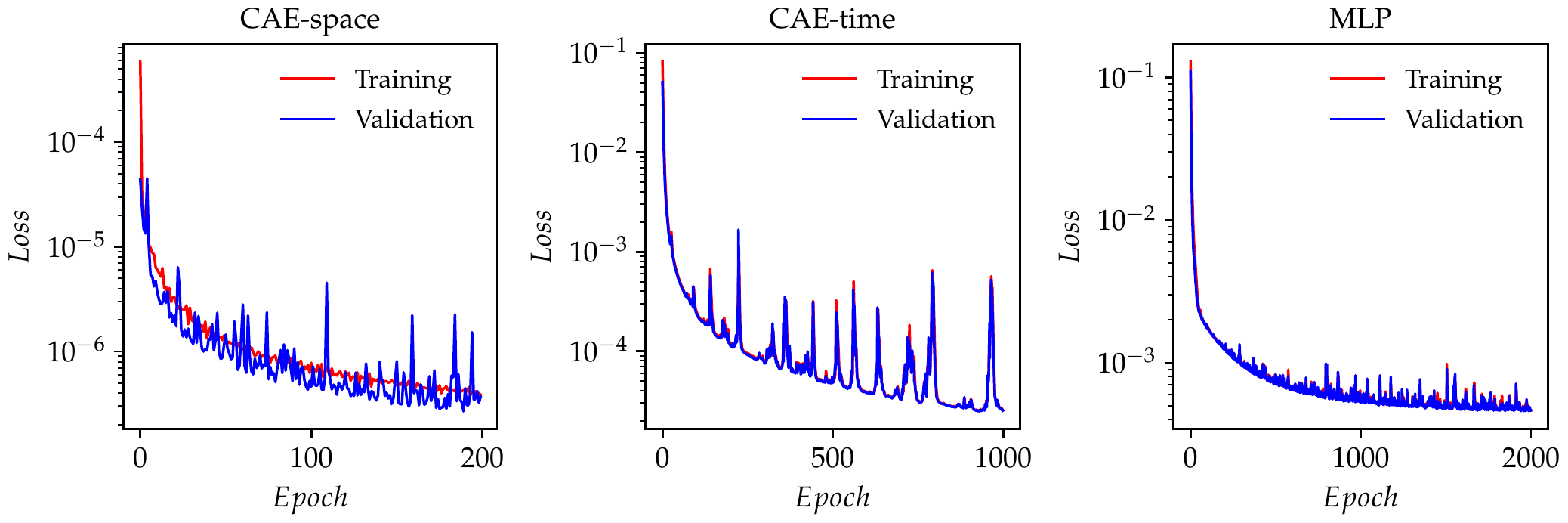}
 \caption{Evolution of the training and validation losses of CAEs and MLP for the hypothetical dam break of the Mille-Iles river.}
   \label{fig:Conv_hist_Mille_Iles}
\end{figure}
%+++++++++++++++++++++++++++++++
%========================================

 %%++++++++++++++++++++++++++++++++++++++++++++
%-------------
\clearpage
%\section*{References}
\bibliography{mybibfile}
\bibliographystyle{elsarticle-num}
%\bibliographystyle{elsarticle-harv}
%\bibliography{mybibfile}
%++++++++++++++++++++++++++++++++++++++++++++++++++++++
\end{document}